\documentclass[fleqn,usenatbib]{mnras}

\DeclareRobustCommand{\VAN}[3]{#2}
\let\VANthebibliography\thebibliography
\def\thebibliography{\DeclareRobustCommand{\VAN}[3]{##3}\VANthebibliography}

\usepackage{graphicx}	
\usepackage{amsmath}	
\usepackage{amssymb}	

\newcommand{\fthin}{f_{\rm thin\,disk,\,recent}}
\newcommand{\tcools}{t_{10^5\,{\rm K}}}
\newcommand{\tacc}{t_{\rm acc}}
\newcommand{\Mdot}{\dot{M}}
\newcommand{\Rvir}{r_{\rm vir}}
\newcommand{\nH}{n_{\rm H}}
\newcommand{\Tvir}{T_{\rm vir}}
\newcommand{\msun}{{\rm M}_\odot}
\newcommand{\vvir}{v_{\rm vir}}
\newcommand{\Nsample}{17}
\newcommand{\Rcirc}[0]{r_{\rm circ}}
\newcommand{\vc}[0]{v_{\rm c}}
\newcommand{\mturb}[0]{\mathcal{M}_{\rm turb}}

\title[Rotating cooling flows and thin galactic disks]{Hot-mode accretion and the physics of thin-disk~galaxy~formation}

\author[Hafen, Stern, Bullock et al.]{
\parbox{\textwidth}{
Zachary Hafen$^{1,2}$\thanks{E-mail: zhafen@uci.edu},
Jonathan Stern$^{3,2}$,
James Bullock$^{1}$,
Alexander B. Gurvich$^2$,
Sijie Yu$^1$,
Claude-Andr\'e Faucher-Gigu\`ere$^2$,
Drummond B. Fielding$^4$,
Daniel Angl\'es-Alc\'azar$^{5,4}$,
Eliot Quataert$^6$,
Andrew Wetzel$^7$,
Tjitske Starkenburg$^2$,
Michael Boylan-Kolchin$^8$,
Jorge Moreno$^{9,1,10}$,
Robert Feldmann$^{11}$,
Kareem El-Badry$^{12,13}$,
T. K. Chan$^{14}$,
Cameron Trapp$^{15}$,
Du\v{s}an Kere\v{s}$^{15}$,
Philip F. Hopkins$^{10}$
} \vspace{0.4cm}\\
\parbox{\textwidth}{
$^1$ Department of Physics and Astronomy, University of California Irvine, CA 92697, USA \\
$^2$ Department of Physics \& Astronomy and CIERA, Northwestern University, 1800 Sherman Ave, Evanston, IL 60201, USA \\
$^3$ School of Physics \& Astronomy, Tel Aviv University, Tel Aviv 69978, Israel \\
$^4$ Center for Computational Astrophysics, Flatiron Institute, 162 5th Ave, New York, NY 10010 \\
$^5$ Department of Physics, University of Connecticut, 196 Auditorium Road, U-3046, Storrs, CT 06269-3046, USA \\
$^6$ Department of Astrophysical Sciences, Princeton University, Princeton, NJ 08544, USA \\
$^7$ Department of Physics \& Astronomy, University of California, Davis, CA, USA 95616 \\
$^8$ Department of Astronomy, The University of Texas at Austin, 2515 Speedway, Stop C1400, Austin, TX 78712-1205, USA \\
$^9$ Department of Physics and Astronomy, Pomona College, Claremont, CA 91711, USA \\
$^{10}$ TAPIR, Mailcode 350-17, California Institute of Technology, Pasadena, CA 91125, USA \\
$^{11}$ Institute for Computational Science, University of Zurich, Zurich CH-8057, Switzerland \\
$^{12}$ Center for Astrophysics $|$ Harvard \& Smithsonian, 60 Garden Street, Cambridge, MA 02138, USA \\
$^{13}$ Harvard Society of Fellows, 78 Mount Auburn Street, Cambridge, MA 02138 \\
$^{14}$ Institute for Computational Cosmology, Department of Physics, Durham University, South Road, Durham DH1 3LE, UK \\
$^{15}$ Department of Physics, Center for Astrophysics and Space Sciences, University of California, San Diego, 9500 Gilman Drive, \\ La Jolla, CA 9209, USA \\
}
}

\date{Accepted XXX. Received YYY; in original form ZZZ}

\pubyear{2020}

\begin{document}
\label{firstpage}
\pagerange{\pageref{firstpage}--\pageref{lastpage}}
\maketitle

\begin{abstract}
We use FIRE simulations to study disk formation in $z\sim 0$, Milky Way-mass galaxies, and conclude that a key ingredient for the formation of thin stellar disks is the ability for accreting gas to develop an aligned angular momentum distribution via internal cancellation \textit{prior} to joining the galaxy.
Among galaxies with a high fraction ($>70\%$) of their young stars in a thin disk ($h/R \sim 0.1$), we find that:
(i) hot, virial-temperature gas dominates the inflowing gas mass on halo scales ($\gtrsim 20$ kpc), with radiative losses offset by compression heating;
(ii) this hot accretion proceeds until angular momentum support slows inward motion, at which point the gas cools to $\lesssim10^4\,{\rm K}$;
(iii) prior to cooling, the accreting gas develops an angular momentum distribution that is aligned with the galaxy disk, and while cooling transitions from a quasi-spherical spatial configuration to a more-flattened, disk-like configuration.
We show that the existence of this ``rotating cooling flow'' accretion mode is strongly correlated with the fraction of stars forming in a thin disk, using  a sample of \Nsample\ $z\sim0$ galaxies spanning a halo mass range of $10^{10.5} M_\odot \lesssim M_{\rm h} \lesssim 10^{12} M_\odot$ and stellar mass range of $10^8 M_\odot \lesssim M_\star \lesssim 10^{11} M_\odot$.
Notably, galaxies with a thick disk or irregular morphology do not undergo significant angular momentum alignment of gas prior to accretion and show no correspondence between halo gas cooling and flattening.
Our results suggest that rotating cooling flows (or, more generally, rotating subsonic flows) that become coherent and angular momentum-supported prior to accretion onto the galaxy are likely a necessary condition for the formation of thin, star-forming disk galaxies in a $\Lambda$CDM universe.
\end{abstract}




\section{Introduction}
\label{s: introduction}

\begin{figure*}
    \centering
    \includegraphics[width=\textwidth]{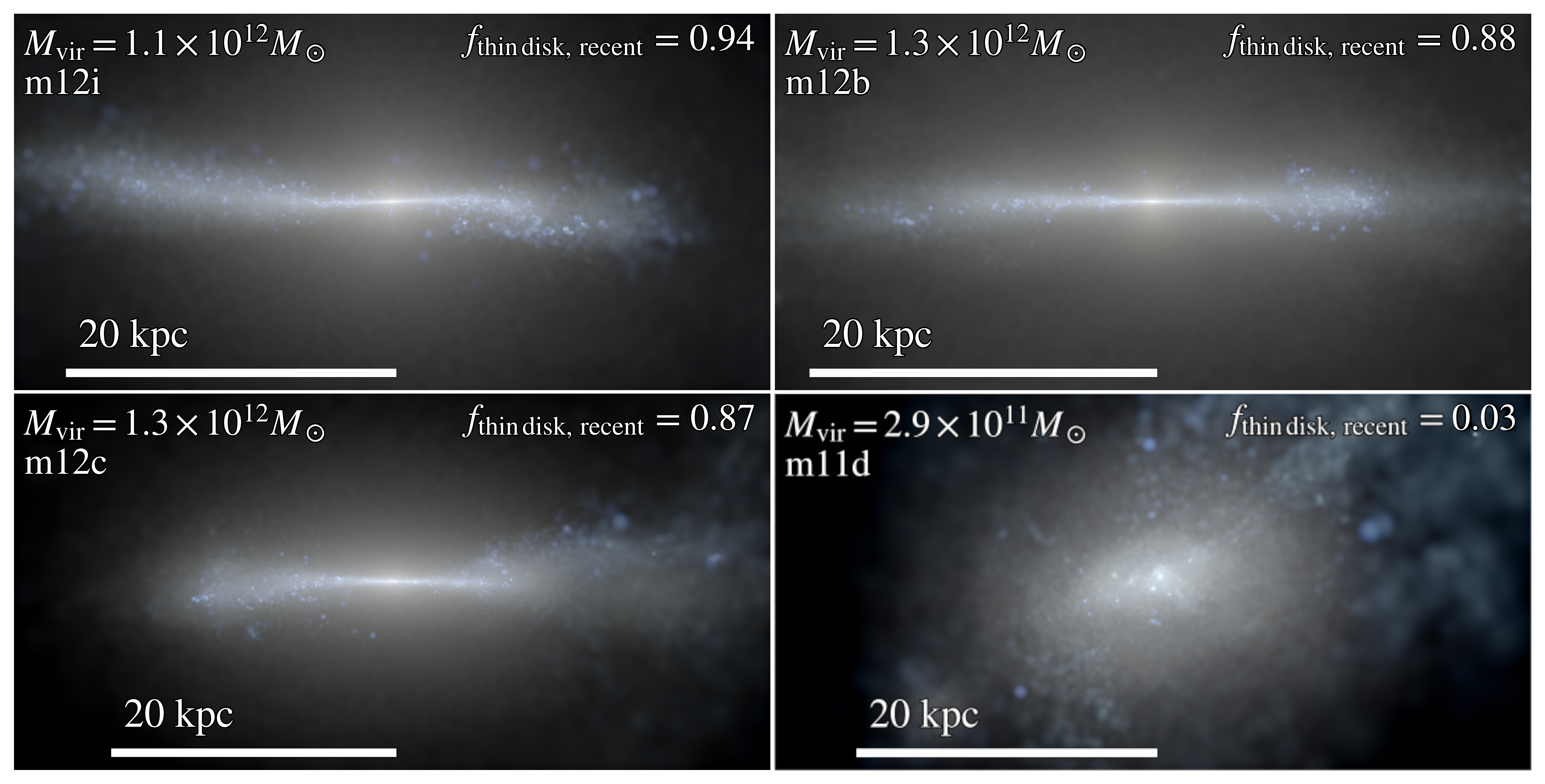}
    \caption{
    Mock Hubble images of four $z=0$ galaxies in the FIRE-2 simulations.
    Halo masses are noted in the panels, together with the mass fraction of young ($<1$ Gyr) stars which reside in a thin disk.
    The Milky Way-mass galaxies have thin disk morphologies (bottom-left, top), while the lower mass galaxy has an irregular morphology (bottom-right).
    }
    \label{f: stars}
\end{figure*}

Our present picture for the formation of galactic disks can be largely traced to analytic ideas first explored by \citet{fall1980}, where a galaxy's angular momentum is intimately tied to the corresponding properties of its host dark matter halo.
Collapsing structures in an otherwise expanding universe will be spun up by the large-scale matter field \citep{Peebles69};
this can deliver enough angular momentum to allow (at least some) galaxies to have significant angular-momentum support~\citep[e.g.][]{MMW98}. 
While these ideas have provided foundational insights into the origin of disk galaxies in a $\Lambda$CDM universe, our understanding of disk  formation, and thin-disk galaxy formation in particular, remains incomplete. 
While dark matter haloes of all masses are predicted to have similar angular momentum distributions \citep[e.g.][]{Barnes87}, disk fraction varies noticeably as a function of galaxy mass \citep[e.g.][]{Bernardi2010, Bluck2014, Moffett16}.
Moreover, dark matter spin alone is insufficient to explain the detailed properties of disks formed in cosmological simulations~\citep[e.g.][]{Sales2012, GK18, Rohr2021}.  

Given what we know about the angular momentum distribution in galactic haloes, it is somewhat surprising that so many galaxies are dominated by {\em thin} disks, with  scale heights and vertical velocity dispersions that are small compared to their scale radii and circular speeds, $h/R \sim \sigma_z/V_c \sim 0.1$~\citep[][]{Kregel2002}.
We know, for example, that the angular momentum distribution of dark matter \citep{Bullock2001,Lian2018} and gas \citep{Stewart2013,DeFelippis2020} in galactic haloes is quite \textit{broad}.
This means that in order for a tightly-ordered thin disk to emerge, the gas must become coherently aligned along a single plane \textit{after} being part of the quasi-spherical extended galactic halo and \textit{before} forming stars.
This suggests that the process by which gas is deposited from the galactic halo into the galaxy, and how this process affects the angular momentum distribution, is an essential aspect of thin disk formation.

The process by which gas is deposited into a galaxy from the circumgalactic medium (CGM) has been explored considerably~\citep[e.g.][]{Keres2005, Keres2009a, Dekel2006, Faucher-Giguere2011a, VanDeVoort2011a, Angles-Alcazar2017, Stevens2017, Martin2019a}.
Broadly speaking two paths of galaxy fueling have been identified: hot mode and cold mode.
In the cold-mode case, gas is deposited into the galaxy without ever virializing; this occurs typically in lower-mass haloes, where the gas cooling time is shorter than the infall time.
In hot-mode accretion, which dominates for massive haloes at late times ($M_{\rm vir} \gtrsim 10^{12} \Mdot$, $z \lesssim 1$; e.g. \citealt{Faucher-Giguere2011a, VandeVoort2011, VandeVoort2012a, Joung2012, Murante2012, Nelson2013, Fielding2017}), gas first shock-heats to the halo virial temperature, and then radiates away its gravitational and thermal energy prior to accreting onto the galaxy.
Our own Milky Way (MW) is one such galaxy for which hot-mode accretion is expected to dominate.
As discussed above, we expect the mode of gas delivery, and the precise means by which gas mixes, cools, and accretes to have a substantial bearing on the ability to form thin, coherently rotating disks.

In cold mode accretion, cool ($T \sim 10^4$ K) gas travels from cosmological scales through the CGM and intergalactic medium (IGM), and into the galaxy in filaments~\cite[e.g.][]{Keres2005, Keres2009, Dekel2006, Faucher-Giguere2011a, Martin2019a}, often along with embedded satellite galaxies~\citep[e.g.][]{Faucher-Giguere2015, Faucher-Giguere2016, Hafen2019, Esmerian2021}.
This mode is expected to dominate the mass inflow rate at high redshifts \citep[$z\gtrsim2$; e.g.][]{Keres2009a, Dekel2009, Huscher2021}.
It is unclear, however, if the cool filaments remain intact down to the galaxy, or rather heat up and dissolve into the surrounding hot phase~(e.g. \citealt{Keres2009b, Nelson2016, Mandelker2016, Mandelker2018, Mandelker2020a}), in which case hot accretion onto the galaxy would also be important at high redshift).
Cool filamentary inflow typically carries more specific angular momentum (on average) than either hot gas or dark matter~\citep[e.g.][]{Stewart2017}, and systems fed by cold accretion often have extended, messy, ``cold flow disks'' orbiting the galaxy~\citep[e.g.][]{Stewart2011a, Stewart2013, Danovich2015, Dekel2019}.
However, the tendency for such gas to have a wide range of trajectories as it approaches the galactic region may hinder the ability to develop a thin, coherently aligned structure prior to star formation.
Also, the fact that thin disk galaxies are common only among fairly massive systems ($L \sim 0.1 L_\star - L_\star$) at lower redshift~\citep[e.g.][]{Kranz2003, Kassin2006, Bizyaev2021, Kassin2012a, Simons2015, Simons2017}, suggests that cold-mode delivery may not be conducive to thin disk formation.

An alternative possibility is that hot-mode accretion, believed to dominate in more massive haloes at lower redshift, is more favorable to thin-disk formation.
Indeed, some cosmological simulations show a correlation between hot accretion and disk formation, albeit with wide scatter~\citep{Sales2012}.
The connection depends on the specific mechanics of this accretion mode, and on whether the hot gas manages to cool and accrete rather than being reheated by galactic feedback processes.
One possible subset of hot-mode accretion is instability-driven accretion, wherein gas precipitates out of a hydrostatic hot halo due to thermal instabilities, forming cool clouds that lose buoyancy and accrete onto the galaxy~\citep[e.g.][]{Fall85, Maller2004, Mccourt2012, Voit2015, Armillotta2016, Gronke2020b,Fielding2020, Voit2021}.
Alternatively, radiative cooling in the hot CGM can cause the entire hot halo to flow radially inward. 
In this latter scenario, compression heating of the hot gas due to the inflow roughly balances radiative losses, so the hot gas stays at an approximately constant temperature.
This type of hot accretion has been termed a `cooling flow' in the context of galaxy cluster studies~\citep{Mathews1978, Cowie1980, Fabian1984, Balbus1988, Bertschinger1989,McNamara2007}, and was recently revisited in the context of galaxy-scale haloes by \cite{Stern2019}.
A hot inflow is qualitatively distinct from precipitation, since the entire hot medium inflows on a cooling timescale, which implies that thermal instabilities do not have time to grow substantially (\citealt{Balbus1989}, see also Fig.~10 in \citealt{ Stern2019}).
The angular momentum content of such hot inflows was considered by \cite{Stern2020}, who showed that angular momentum sets a maximum accretion rate in which the inflow remains hot and subsonic down to the galaxy scale. 
In the present paper we demonstrate that such cooling flows with angular momentum (i.e. `rotating cooling flows')
are the primary mode of gas accretion onto disky Milky Way-mass galaxies at $z \sim 0$ in the FIRE-2 cosmological simulations \citep{Hopkins2018}, and may be a necessary condition for the formation of thin star-forming disks.

The analysis here follows \cite{Stern2021}, which showed that the formation of disks is closely connected to the formation of a virialized and stable hot CGM.
\cite{Yu2021} subsequently showed that the emergence of an inner virialized CGM correlates specifically with a transition from {\em thick}-disk to {\em thin}-disk formation.
The conditions necessary for thin disk formation thus appear to correlate with the conditions necessary for the onset of hot accretion modes, and cooling flows in particular.
In contrast, it seems unlikely that the transition to a thin disk in FIRE-2 is driven by increased ejection of low angular momentum gas by feedback as suggested by analysis of other simulations \citep[e.g.][]{Brook2011, Ubler2014, Genel2015, DeFelippis2017}, since in FIRE-2 the mass loss due to feedback drops when the galaxy becomes disky ~\citep{Hafen2020,Pandya2021,Stern2021a}.
These findings motivate the exploration that follows.

Central to our analysis, and the analyzes of \citeauthor{Stern2021} and \citeauthor{Yu2021}, are the FIRE simulations~(Feedback in Realistic Environments; \citealt{Hopkins2014, Hopkins2018})\footnote{\url{https://fire.northwestern.edu/}} a set of ``zoom-in'' simulations that resolve stellar feedback on the scale of giant molecular clouds in the interstellar medium (ISM)~\citep{Guszejnov2020b, Benincasa2020}, producing winds that expand into the CGM and interact with accreting gas~\citep{Muratov2015, Muratov2017, Angles-Alcazar2017, Hafen2019, Hafen2020, Pandya2021}.
The resultant galaxies are broadly consistent with the stellar mass-halo mass relation~\citep{Hopkins2017}, satellite galaxy populations~\citep{Wetzel2016, Garrison-Kimmel2019a, Samuel2020}, and can have thin-disks consistent with Milky Way-like galaxies~\citep{Ma2017a, Garrison-Kimmel2018, El-Badry2018, Yu2021}, albeit potentially with hotter kinematics~\citep{Sanderson2020}. 
Here we use the FIRE-2 simulations to study gas accretion onto $z\sim0$ galaxies and its relation to thin disk morphology.
Our approach uses the particle-tracking methodology developed in \citet{Hafen2019, Hafen2020} to explore the mechanics of rotating cooling flows near the disk-halo interface where angular momentum support is substantial.
Our analysis goes beyond 1D steady-state solutions for rotating cooling flows developed in classic ICM studies~\citep[e.g.][]{Cowie1980}, and extends the idealised 3D simulations of rotating cooling flows in \cite{Stern2020} to a more realistic cosmological setting. 
Our work is complementary to \citet{Trapp2021}, who characterized the phenomenological properties of accretion onto MW-like FIRE galaxies, and the particle tracking analysis of \cite{Angles-Alcazar2017}, who provided an overview of the connection between the cosmic baryon cycle and galaxy mass assembly. 

This paper is structured as follows. 
In \S\ref{s: methods} we describe our sample of FIRE simulations and the sample of accreting particles selected from the simulations.
In \S\ref{s: results} we analyze the characteristics and mechanics of gas accretion onto $z\sim0$ galaxies in FIRE, and their relation to thin disk morphology in the central galaxy.
We discuss our results in \S\ref{s: discussion} and conclude in \S\ref{s: conclusions}.

\section{Methods}
\label{s: methods}

\subsection{Simulations}
\label{s: methods -- simulations}

\begin{table*}
\caption{Simulation parameters.}
\begin{tabular}{cccccccc}
\hline
Name  &  $\fthin$  & $M_{\rm vir}$  &  $M_\star$  &  $R_{\textrm{vir}}$  &  $\Delta f_{\rm aligned}$  &  Metal diffusion?  &  Reference  \\
  &   & $M_\odot$  & $M_\odot$  &  kpc  &  &  &  \\
 \hline
\texttt{m12i}  &  0.94  &  $1.1\times10^{12}$  &  $7.3\times10^{10}$  &  268  &  0.34  &  \checkmark  &  \cite{Wetzel2016}    \\
\texttt{m12i\_core}  &  0.89  &  $1.1\times10^{12}$  &  $8.0\times10^{10}$  &  274  &  0.34  &    &  \cite{Hopkins2018}    \\
\texttt{m12b}  &  0.88  &  $1.3\times10^{12}$  &  $1.0\times10^{11}$  &  286  &  0.35  &  \checkmark  &  \cite{Garrison-Kimmel2019a}    \\
\texttt{m12c}  &  0.87  &  $1.3\times10^{12}$  &  $6.8\times10^{10}$  &  283  &  0.25  &  \checkmark  &  \cite{Garrison-Kimmel2019a}    \\
\texttt{m12f}  &  0.87  &  $1.5\times10^{12}$  &  $9.7\times10^{10}$  &  302  &  0.24  &  \checkmark  &  \cite{Garrison-Kimmel2017}    \\
\texttt{m11h}  &  0.52  &  $1.8\times10^{11}$  &  $3.9\times10^{9}$  &  146  &  0.047  &  \checkmark  &  \cite{El-Badry2018a}    \\
\texttt{m12w}  &  0.49  &  $9.5\times10^{11}$  &  $6.5\times10^{10}$  &  253  &  0.15  &  \checkmark  &  \cite{Samuel2020}    \\
\texttt{m12m}  &  $^*$0.46  &  $1.5\times10^{12}$  &  $1.4\times10^{11}$  &  298  &  0.26  &    &  \cite{Hopkins2018}    \\
\texttt{m11b}  &  0.44  &  $4.4\times10^{10}$  &  $1.2\times10^{8}$  &  92.4  &  0.12  &    &  \cite{Chan2018}    \\
\texttt{m12r}  &  0.31  &  $1.0\times10^{12}$  &  $2.4\times10^{10}$  &  257  &  0.11  &  \checkmark  &  \cite{Samuel2020}    \\
\texttt{m12z}  &  0.11  &  $8.0\times10^{11}$  &  $2.5\times10^{10}$  &  242  &  0.052  &  \checkmark  &  \cite{Garrison-Kimmel2019a}    \\
\texttt{m11i}  &  0.046  &  $7.0\times10^{10}$  &  $1.0\times10^{9}$  &  106  &  0.013  &  \checkmark  &  \cite{El-Badry2018a}    \\
\texttt{m11c}  &  0.035  &  $1.4\times10^{11}$  &  $9.5\times10^{8}$  &  137  &  0.0068  &    &  \cite{Hopkins2018}    \\
\texttt{m11e}  &  0.031  &  $1.5\times10^{11}$  &  $1.6\times10^{9}$  &  136  &  0.044  &  \checkmark  &  \cite{El-Badry2018a}    \\
\texttt{m11d}  &  0.03  &  $2.9\times10^{11}$  &  $4.9\times10^{9}$  &  169  &  0.0026  &  \checkmark  &  \cite{El-Badry2018a}    \\
\texttt{m11a}  &  0.022  &  $4.1\times10^{10}$  &  $1.3\times10^{8}$  &  90.3  &  -0.022  &    &  \cite{Chan2018}    \\
\texttt{m11q}  &  0.0066  &  $1.5\times10^{11}$  &  $7.4\times10^{8}$  &  138  &  0.0045  &  \checkmark  &  \cite{Hopkins2018}    \\ 
\end{tabular}
\\
\begin{flushleft}
Properties at $z=0$ of the sample of FIRE galaxies analyzed in this work.
The value of $\fthin$ is the fraction of stars formed in the last Gyr prior to $z=0$ that are in a thin disk configuration ($j/j_c(E) > 0.8$; e.g. \citealt{Yu2021}).
The \texttt{m12m} galaxy has a sizable bar, which drives the thin-disk fraction lower by our adopted definition (marked by $^*$).
The value of $\Delta f_{\rm aligned}$ is a measure of the relation between cooling and flattening in accreted gas (\S\ref{s: characteristics -- aligns}).
The ``metal diffusion'' column marks whether or not the simulation includes a subgrid prescription for metal diffusion.
\end{flushleft}
\label{table: simulations_used}
\end{table*}

We analyze hydrodynamical cosmological zoom-in simulations produced as part of the FIRE project~\citep{Hopkins2014}.
The simulation sample, listed in Table~\ref{table: simulations_used}, were run with the FIRE-2 version~\citep{Hopkins2018} of the gravity and hydrodynamics code \textsc{GIZMO}\footnote{\url{http://www.tapir.caltech.edu/\~phopkins/Site/GIZMO.html}}~\citep{Hopkins2015}.
The simulations were produced using the meshless finite-mass (``MFM'') mode of \textsc{GIZMO}, a Lagrangian method with no inter-element mass flux.
This enables us to track the history of each resolution element.
The full details of simulations produced with the FIRE-2 code are available in~\cite{Hopkins2018}.
The FIRE simulations include detailed prescriptions for star formation and stellar feedback.
Each star particle contributes to the simulation momentum from radiation pressure; energy, momentum, mass, and metals from Type Ia and II supernovae and stellar winds; and photo-ionization and photo-electric heating.
The mass of the resolution elements ranges from 2100-7100 $M_\odot$.
Star formation is limited to self-gravitating, self-shielding (molecular) gas with a density of at least $n_{\rm SF} = 1000$ cm$^{-3}$.
In addition to stellar radiation, the simulations include a uniform meta-galactic radiation background that ionizes gas in the intergalactic and circumgalactic medium~\citep{Faucher-Giguere2009}.
In the simulations and throughout our analysis we use a standard flat $\Lambda$CDM cosmology with $\Omega_{\rm m }\approx 0.32$, $\Omega_{\Lambda}=1-\Omega_{\rm m}$, $\Omega_{\rm b} \approx 0.049$, and $H_{0} \approx 67$ km s$^{-1}$ Mpc$^{-1}$~\citep[e.g.,][]{PlanckCollaboration2018}.

Fig.~\ref{f: stars} shows edge-on mock Hubble images for the $z=0$ snapshots of four of our simulated galaxies, neglecting dust attenuation to more clearly illustrate the stellar distribution. 
The bottom-left and top panels show three MW-mass galaxies (\texttt{m12i}, \texttt{m12b}, and \texttt{m12c}) while the bottom-right panel shows a $M_\star \sim 5 \times 10^9 M_\odot$ dwarf galaxy, \texttt{m11d}.
As noted in previous studies~\citep{Garrison-Kimmel2018, El-Badry2018} MW-mass galaxies in FIRE typically have thin disk morphologies, while lower mass galaxies show a thick disk / irregular morphology.
We quantify this trend using $\fthin$, defined as the mass fraction of stars with age $<1$ Gyr (at $z=0$) that have $j_z/j_c(E) > 0.8$.
Here $j_z$ is the specific angular momentum in the $z$ direction and $j_c(E)$ is the angular momentum that the star would have if it had the same energy but was in a circular orbit.
The $z$ axis is defined as the direction of the total angular momentum vector of stars inside the galaxy.
This definition of the thin disk is the same definition used in~\cite{Yu2021}.
Values of $j_z/j_c(E)>0.8$ correspond to height to size ratios $h/R \sim 0.1$ (confirmed across a number of choices of $h$ and $R$) and rotation to dispersion ratios $V_{\rm rot}/\sigma_z \sim10$, and correlate strongly with the observed thin disk fraction in the $r$ band (Appendix \ref{s: appendix-sloan thin disk fraction}).
The values of $\fthin$ are noted in Fig.~\ref{f: stars} and listed in Table~\ref{table: simulations_used}.
Throughout we will refer to galaxies with $\fthin > 0.6$ as thin disk galaxies and galaxies with $\fthin < 0.6$ as irregular or thick disk galaxies.

\subsection{Analysis}
\label{s: methods -- analysis}

For each galaxy we use the particle tracking method described in \cite{Hafen2019, Hafen2020}, which in turn applies insight from previous particle-tracking applied to the FIRE simulations~\citep{Angles-Alcazar2017}.
We select an unbiased sample of resolution elements (particles) that have accreted onto the central galaxy during the last Gyr prior to $z=0$.
To select such particles, we require that they are
(1) within the galaxy radius $r_{\rm gal}$ at $z=0$, either in stars or in gas with density  $n_{\rm H} > 0.13$ cm$^{-3}$, and
(2) within the CGM (gas at $0.1 - 1 \Rvir$) in the snapshot corresponding to 1 Gyr prior to $z=0$.
Throughout the paper we use $r$ for the 3D distance. 
We define $r_{\rm gal} = 4 r_{\star,0.5}$ following \cite{Hafen2019, Hafen2020}, where $r_{\star, 0.5}$ is the stellar half-mass radius.
When selected this way, the number of particles that meet these conditions ranges from $\sim 1000$ for the low-mass irregular galaxies to $\sim 10^5$ for the thin-disk galaxies.
For each selected particle we retrieve the full history of the particle, including temperature, density, and metallicity. 
We discard a small fraction ($<2\%$) of the particles whose mass increases by a factor two as a consequence of mass deposition by stellar feedback\footnote{
These particles pose a problem for our tracking method because the history of the additional mass is not recorded, and because they are split into multiple particles after gaining sufficient mass.}.
In Appendix~\ref{s: appendix-sample validation} we show that the accretion histories of selected particles are representative of the accretion histories of all stars formed within $1$ Gyr of $z=0$, and that around thin disk galaxies our selected particles are drawn from the angular momentum distribution of the CGM without bias.

A key time for our analysis is the accretion time $\tacc$, the time at which a particle first accretes onto the galaxy after being identified as part of the CGM 1 Gyr prior to $z=0$.
Explicitly, $\tacc$ is:
\begin{equation}
    \tacc \equiv t \bigg( {\rm last\,snapshot\,outside\,galaxy} \bigg) ~,
\label{e: tacc}
\end{equation}
i.e. the last snapshot prior to the particle matching the above criteria for being within a galaxy.

Another key time for our analysis is the last time a particle cools prior to entering the galaxy.
For a given particle, we define this time as:
\begin{equation}
    \tcools \equiv t \bigg( {\rm last\,snapshot\,outside\,galaxy\,with\,}T>10^5\,{\rm K} \bigg) ~,
\label{e: tcools}
\end{equation}
i.e., the last snapshot the particle was hot prior to $\tacc$.
For gas that cools as it accretes, $\tcools$ occurs as the gas passes through the galaxy-halo interface.
In all halos we find that $\gtrsim99\%$ of tracked particles heat above $T > 10^5$ K at some point prior to accretion, even in our low mass halos where the virial temperature is $T \approx 2 \times 10^5$ K and  accretion onto the galaxy is dominated by cold inflows. 
In such halos heating is typically temporary due to the short cooling times, but hot gas still comprises $\gtrsim 30\%$ of the halo's gas mass at any given time~\citep{Hafen2019,Hafen2020}.
Below we analyze the statistical properties of all accreting particles as a function of time relative to their cooling time ($t-\tcools$) or relative to their accretion time ($t-\tacc$).

\section{Results}
\label{s: results}

\begin{figure*}
    \centering
    \includegraphics[width=\textwidth]{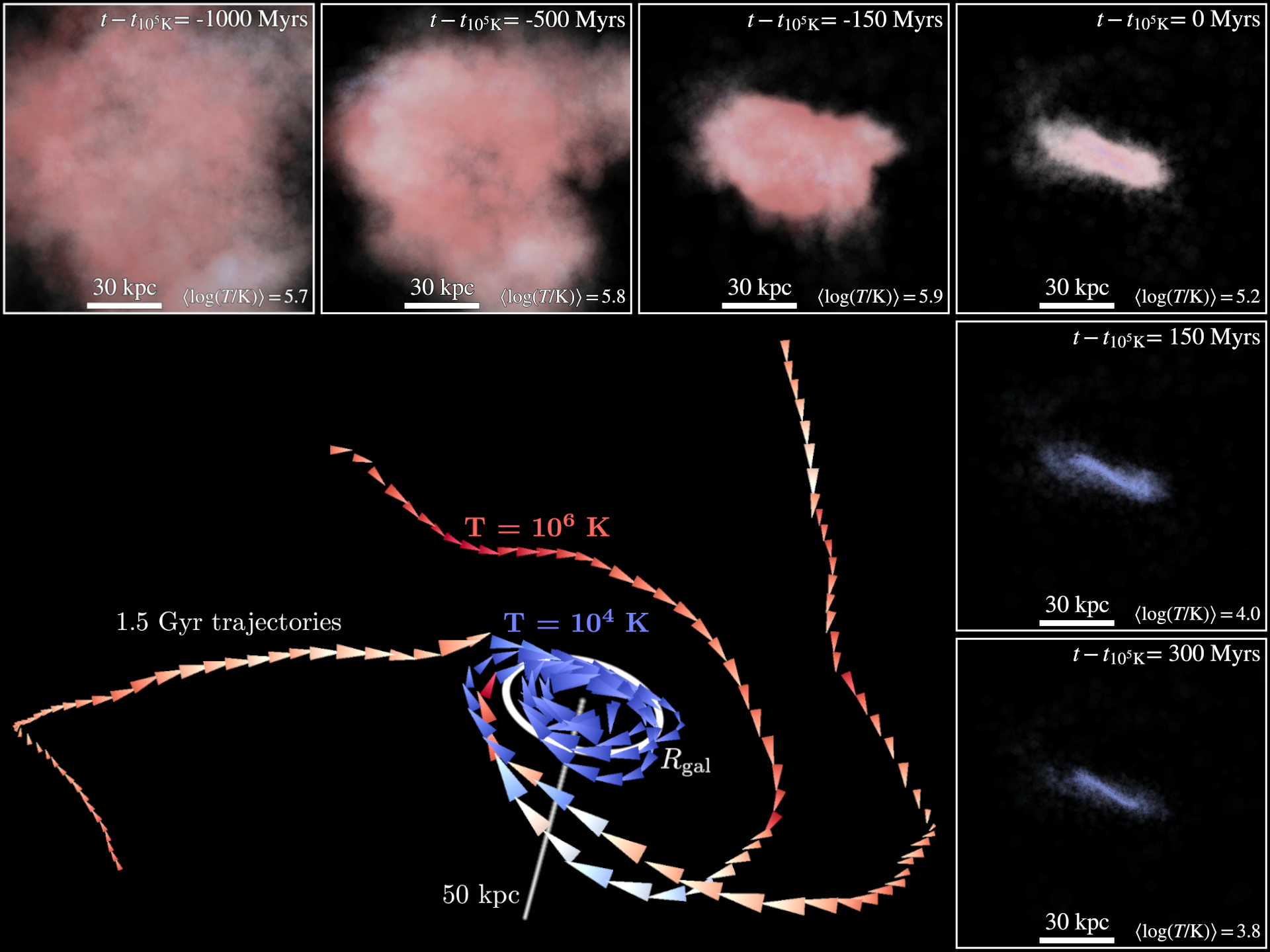}
    \caption{
Gas accretion onto a Milky Way-mass disk galaxy in FIRE, \texttt{m12i}, near $z\approx0$.
\textbf{Top and right panels:}
Temperature and spatial evolution of accreting gas with respect to $\tcools$, defined as the last time at which the gas cools below $10^5$ K prior to accreting onto the galaxy.
Red, white, and blue indicates $T=10^6$ K, $10^5$ K, and $10^4$ K respectively. 
\textbf{Bottom-left panel:}
Three representative trajectories for accreting gas elements.
The panels show that accretion is hot ($\approx 10^6$ K) and contracts quasi-spherically at early times relative to the time of cooling.
At the time of cooling the geometry of accreting gas transitions from a quasi-spherical distribution at $t-\tcools < -150$ Myr to a cool disk at $t-\tcools > 150$ Myr.
    }
    \label{f: overview}
\end{figure*}

To characterize gas accretion we analyze the central galaxy and its CGM from $z=0$ to one Gyr prior.
Fig.~\ref{f: overview} shows a visual overview of how gas accretes onto \texttt{m12i} -- a MW-mass galaxy that forms a substantial thin disk ($\fthin = 0.94$). 
The top and right panels plot the temperature and spatial evolution of accreting gas versus time prior to cooling ($t-\tcools$), while the bottom-left panel plots three representative trajectories for accreting gas elements.
The trajectories were visualized using the Firefly visualization software~\citep{Geller2018}.\footnote{
A 3D version of these trajectories is available online at \url{zhafen.github.io/rotating-cooling-flows}. The Firefly homepage is at \url{alexbgurvi.ch/Firefly}.}
The trajectory color scales with temperature, with red, white, and blue indicating $T=10^6$ K, $10^5$ K, and $10^4$ K, respectively.
The figure shows that at early times relative to cooling ($t-\tcools \sim -1000$ Myr) the accretion is hot ($\approx10^6$ K) and contracts quasi-spherically.
Then, around the time of cooling the geometry of accreting gas transitions from quasi-spherical at $t - \tcools=-150$ Myr to a cool disk aligned with the galaxy at $t - \tcools=+ 150$ Myr.
A quantitative analysis of this transition follows in \S\S\ref{s: characteristics -- inflowing gas phase}--\ref{s: mechanics -- energy balance}.

\subsection{Gas inflow onto thin-disk MW analogs is hot through the CGM}
\label{s: characteristics -- inflowing gas phase}

\begin{figure*}
\includegraphics[width=\textwidth]{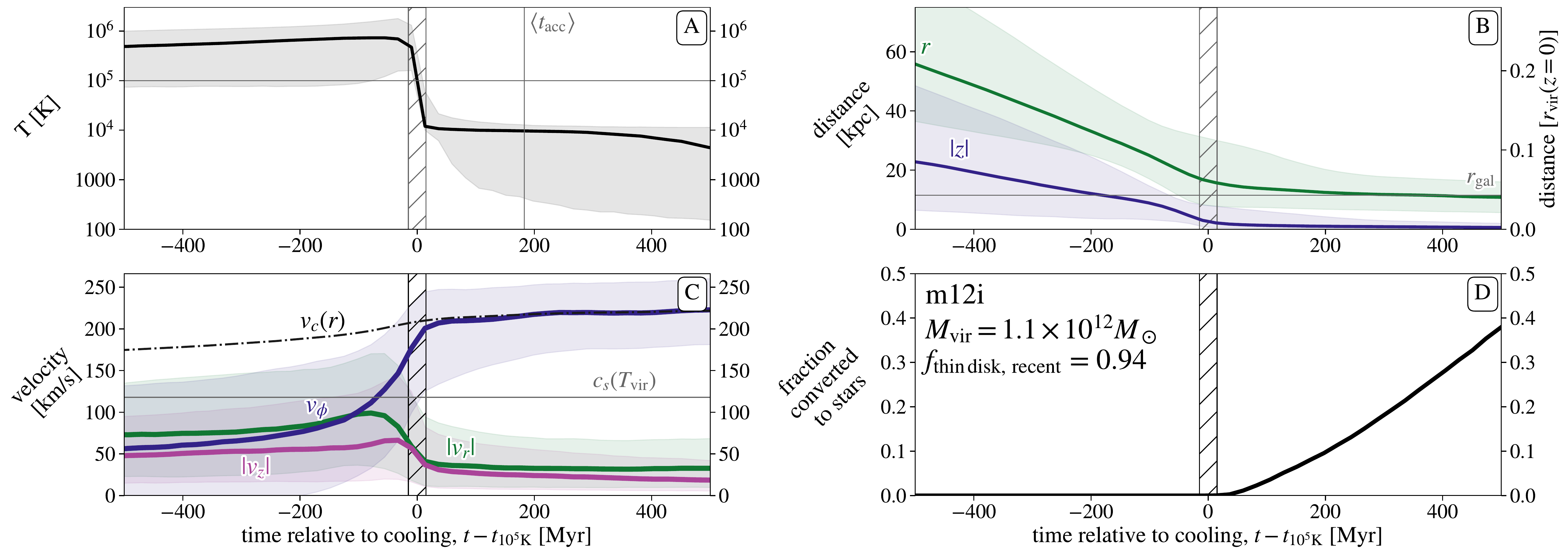}
\caption{
Properties of gas accretion onto a $z\sim0$ thin disk galaxy in FIRE (\texttt{m12i}), versus time relative to the final cooling time ($t - \tcools$).
In each panel solid lines and shaded regions mark the medians and 16th to 84th percentile ranges of all particles accreted within 1 Gyr prior to $z=0$. 
\textbf{A:}
Temperature.
The inflow is predominantly hot ($\gtrsim 10^5$ K) prior to cooling, with a median temperature approaching $\Tvir \sim 10^6$ K.
At $t = \tcools$ the gas cools (by definition), and achieves temperatures $T\lesssim10^4$ K or forms stars.
\textbf{B:}
3D distance from halo center (green) and absolute height from the disk plane (blue).
Cooling occurs at $r=10-30$ kpc, corresponding to $0.7-2.5\,r_{\rm gal}$.
Prior to cooling the gas forms an inflow while after cooling the inflow decreases speed.
Most of the gas collapses into a disk upon cooling, with a median $\vert z \vert \approx 2$ kpc at $t=\tcools$.
\textbf{C:}
Velocity components of accretion (colored lines and bands), relative to the median circular velocity (dash-dotted line).
The gas reaches full rotational support upon cooling.
The gray line marks the sound speed of virial temperature gas.
\textbf{D:}
Fraction of gas converted into stars.
Star formation starts after cooling, at a rate of $\sim10\%$ per $100$ Myr.
}
\label{f: before and after A}
\end{figure*}

Fig.~\ref{f: before and after A} plots various characteristics of accreting gas on the \texttt{m12i} thin disk galaxy, as a function of time relative to the accretion's last cooling time ($t - \tcools$).
In each panel, solid lines and shaded regions mark the medians and 16th to 84th percentile ranges of all particles accreted within $0.5-1$ Gyr prior to $z=0$.
The lower limit on the time range applied in this figure is to ensure that particles are present for most of the time displayed after $\tcools$, although removing the limit does not change the qualitative results.
In the temperature panel (A) we exclude from the distribution particles that have converted into stars. 

Panel (A) demonstrates that during the 500 Myr prior to cooling for a final time, the inflow is predominantly hot ($\gtrsim 10^5$ K), with a median temperature of $4-8\times10^5$ K, similar to the halo virial temperature of $\Tvir(z=0)=6.5\times10^5$ K.
This is not true by construction --- gas could maintain a temperature $T>10^5$ K for only a short time prior to cooling.
In Appendix~\ref{s: appendix-mass flow} we verify that the inflow is  predominantly hot also in terms of mass inflow rate.
During the time prior to cooling the accreting gas is inflowing toward the galaxy (panel B), from a median $r \approx 55$ kpc at $t-\tcools=-500$ Myr, to a median $r\approx18\,{\rm kpc}\approx1.4 r_{\rm gal}$ at $t=\tcools$, after which time the inflow decelerates.
The characteristic inflow velocity of $v_r \approx-70$ km s$^{-1}$ (Panel C) is substantially lower than the circular velocity of $170-200$ km s$^{-1}$.
This radial velocity is also smaller than the median sound speed of $100-130$ km s$^{-1}$, indicating a subsonic flow. 

We show below that a similar hot inflow is seen in other thin disk galaxies in our sample, and that the accretion occurs on a cooling timescale. 
Thus, accretion from the CGM onto $z=0$ thin disks in FIRE is dominated by the inflow of hot, virial temperature gas, as in classic cooling flow solutions \citep[e.g.][]{Fabian1984}.
This hot accretion mode, in which the entire hot phase inflows, is qualitatively different from `precipitation', in which gas clumps cool out of the hot phase at halo scales, lose buoyancy and subsequently accrete (see introduction). 
The dominance of a hot inflow over precipitation in MW-like galaxies simulated in FIRE has previously been noted by \cite{Esmerian2021}. 

\subsection{Accretion cools and decelerates at the galaxy-halo interface}
\label{s: characteristics -- cools}

Panel (A) in Fig.~\ref{f: before and after A} demonstrates that at $t = \tcools$ the temperature drops quickly from ${\sim}10^6$ K to ${\sim}10^4$ K.
Panel (B) shows that this cooling occurs at the galaxy edge or shortly beyond, at $r(\tcools) \approx 10-30$ kpc, equivalent to $\approx0.7 - 2.5 r_{\rm gal}$\footnote{Gas can have $r(\tcools) < r_{\rm gal}$ if it is under-dense relative to the galaxy when it crosses $r_{\rm gal}$ ($n_{\rm H} < 0.13$ cm$^{-3}$; see section \ref{s: methods -- analysis}).}.
Less than $10\%$ of particles cool beyond $\sim 40$ kpc.
After cooling to $T \sim 10^4$ K, the temperature further drops to cool ISM temperatures of $100-10^4$ K and stars begin to form at a rate of $\approx10\%$ per $100$ Myr (panel D), roughly equal to the average rate in the galaxy ISM. 
The vertical grey line in panel (A) marks the median time at which the gas accretes, though $\tacc - \tcools$ spans $\sim0-500$ Myr depending on the particle.

Panel (C) in Fig.~\ref{f: before and after A} demonstrates that cooling at the galaxy scale is further associated with a change in kinematics.
The radial inflow velocity $\vert v_r \vert $ starts decelerating $\approx40$ Myr prior to cooling, finishing at $v_{r} \lesssim 50$ km s$^{-1}$ after cooling.
The deceleration is associated with $v_\phi$ reaching $v_c$, where $\phi$ is defined with respect to the total angular momentum of stars in the galaxy at $z=0$, indicating a transition from pressure-support in the hot CGM to rotational support in the cool ISM.
Note that deceleration prior to accretion is possible due to the subsonic nature of the radial hot flow --- the gas pressure can adjust close to the galaxy so the transition to a rotating disk happens smoothly rather than in a single shock.
Similar results are seen for accretion onto other thin-disk galaxies in our sample (see below), indicating that the majority of accretion onto $z\sim0$ thin disk galaxies in FIRE is a hot inflow which cools and decelerates just outside the galaxy.
As shown below (\S\ref{s: thin vs irregular}; Appendix~\ref{s: appendix-individual}), this is {\em not} the case for galaxies that lack thin disks in our sample. 
Galaxies that are dominated by thick/irregular morphology demonstrate no such deceleration at the galaxy-halo interface.

\subsection{Cooling of accreted gas is concurrent with flattening}
\label{s: characteristics -- aligns}

\begin{figure*}
    \centering
    \includegraphics[width=\textwidth]{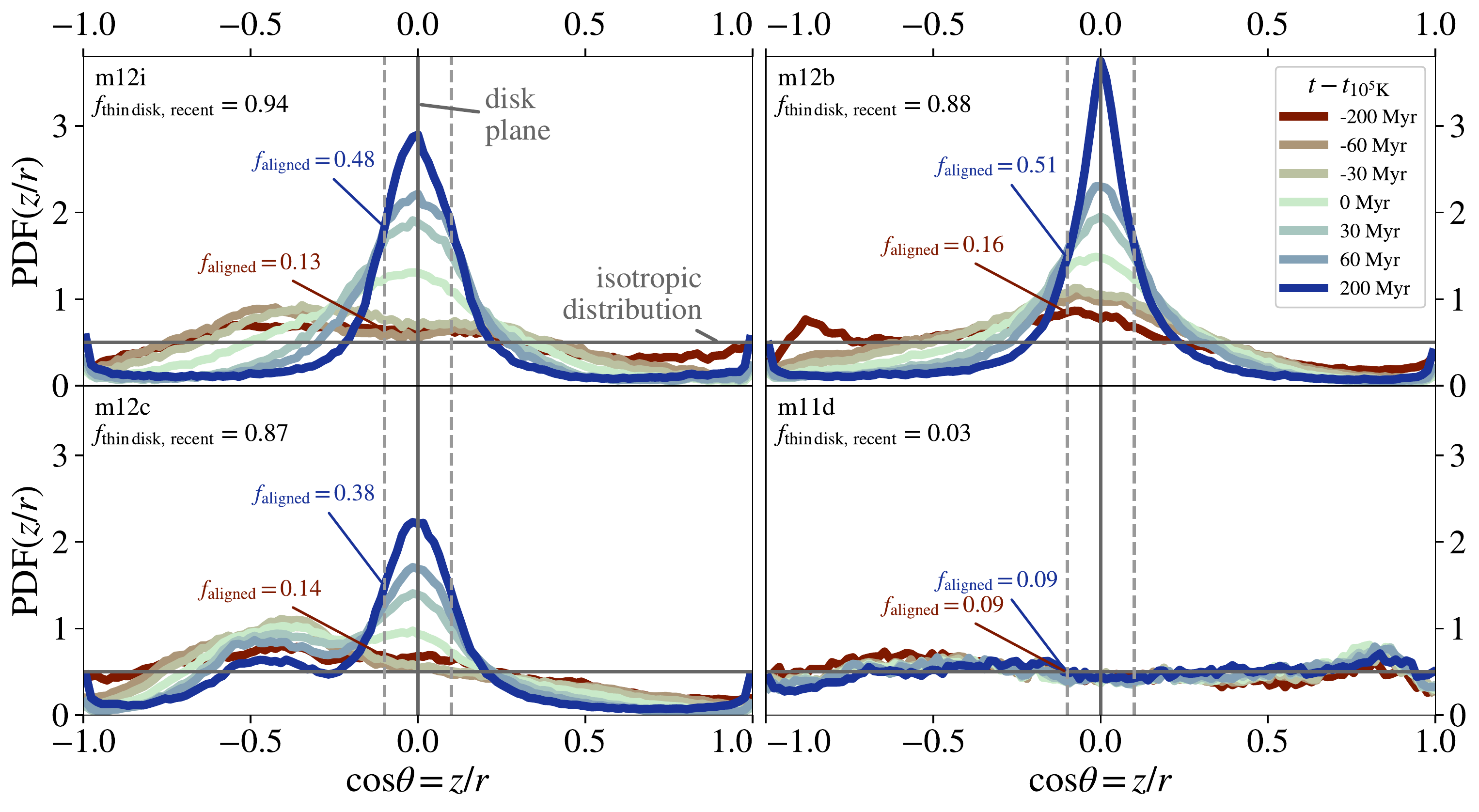}
    \caption{
    Geometry (ratio of height to radius) of accreting gas, as a function of time relative to the last cooling time.
    The galaxies displayed are the same as in Fig.~\ref{f: stars}.
    In thin disk galaxies (top and bottom-left panels) the geometry of accreting gas evolves significantly around the time of cooling, from a quasi-spherical distribution prior to cooling (red curves) to a disk-like configuration after cooling (blue curves). 
    In contrast, in the irregular galaxy (bottom-right) accreting gas is roughly spherical both prior to and after cooling. The quantity $f_{\rm aligned}$ measures the fraction of accreting gas that has $-0.1<z/r<0.1$ at a given $t-t_{10^5{\rm K}}$. 
    }
    \label{f: theta vs t}
\end{figure*}

Panel (B) in Fig.~\ref{f: before and after A} also plots the distance $\vert z \vert$ above or below the disk plane as a function of time.
The panel shows that as the gas collapses, it becomes increasingly flattened in the disk plane.
At the time of cooling, the gas has a median height of $\approx 2$ kpc, indicating a disk geometry with height to radius ratio of $\vert z\vert/r_{\rm gal}\approx0.2$, consistent with the transition to rotation support indicated by panel (C).
Panel (D) in Fig.~\ref{f: before and after A} emphasizes that all of the star formation occurs after rotation support is achieved and the disk geometry is in place.

This correspondence between a transition to disk geometry and cooling is further explored in Fig.~\ref{f: theta vs t}, which plots the geometry of accreting gas at different times relative to $\tcools$ for the four galaxies shown in Fig.~\ref{f: stars}. 
Times prior to cooling are colored in red, while times after cooling are plotted in blue.
All particles with $t-\tcools$ within $\pm$30 Myr of the value noted in the legend are included in the distribution. 
The curves show the PDF of $\cos \theta = z/r$, i.e.\ $\theta$ is the angle between the gas element position and the total angular momentum vector of stars in the galaxy at $z=0$.
A spherical distribution of accreted gas would have a flat PDF with a value of 0.5, while 
the PDF of an infinitely thin disk would be a $\delta$-function centered at $z/r = 0$.
The figure shows that gas accreting onto the three thin-disk galaxies transitions from being distributed quasi-spherically at $t = \tcools-200$ Myr to being distributed in the galaxy plane at $t = \tcools+200$ Myr.
This indicates that the cooling and flattening of the accreting gas occurs simultaneously in these galaxies, consistent with the conclusion from Figs.~\ref{f: overview} and \ref{f: before and after A}.
In contrast, in the irregular galaxy shown in the bottom-right there is no association between cooling and flattening.
Rather, the geometry of accreting gas is quasi-spherical both before and after cooling.

\begin{figure*}
    \centering
    \includegraphics[width=\textwidth]{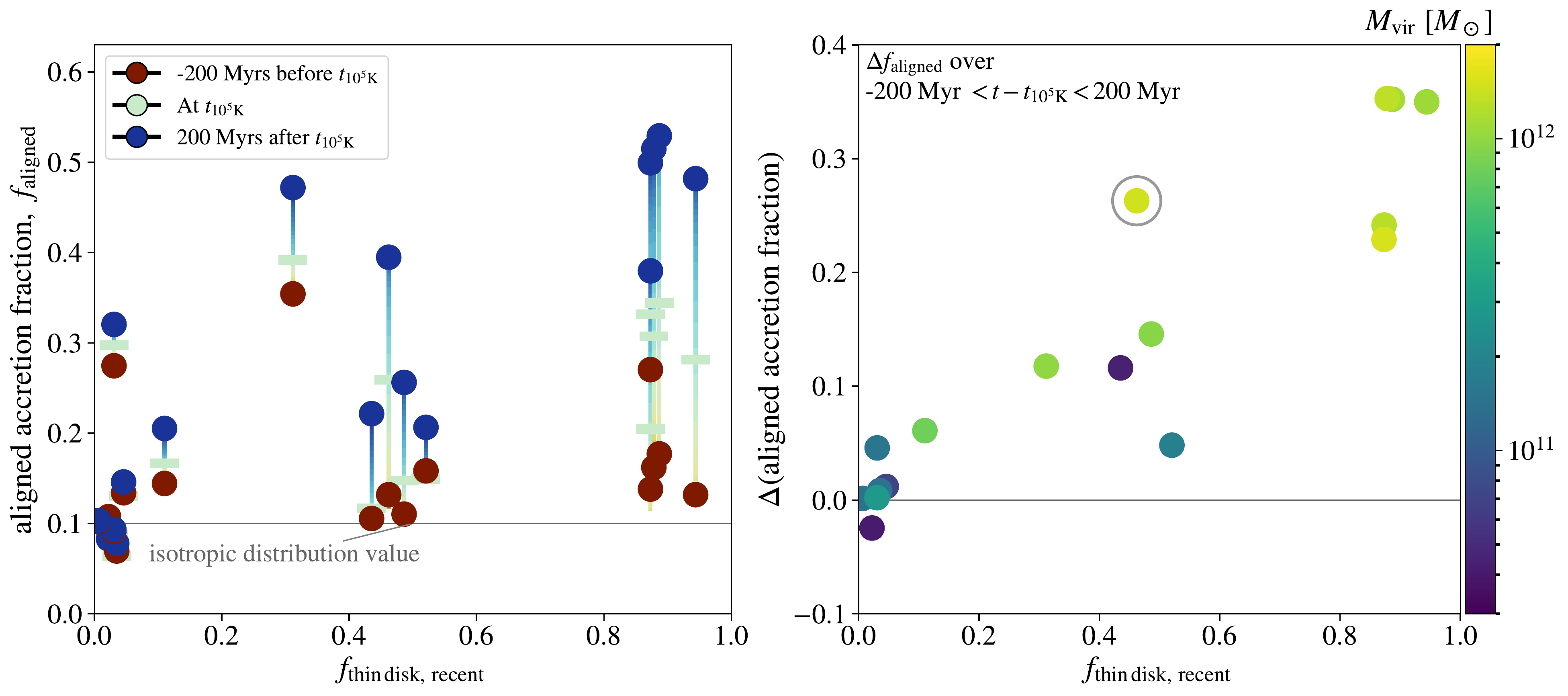}
    \caption{
    \textbf{Left:}
    Mass fraction of accreting gas aligned with the disk ($\vert z/r < 0.1 \vert$, see Fig.~\ref{f: theta vs t}) before and after cooling, for our sample of \Nsample~FIRE haloes.
    The horizontal axis plots the fraction of young stars in the central galaxy that are in a thin disk.
    \textbf{Right:}
    Change in aligned mass fraction during the $\pm200$ Myr around cooling time shown in the left panel.
    A large $\Delta f_{\rm aligned}$ indicates that flattening is concurrent with cooling.
    Color indicates virial mass.
    The value of $\Delta f_{\rm aligned}$ is strongly correlated with the fraction of young stars in a thin disk. 
    The point circled on the right is \texttt{m12m}, which has developed a sizable bar at late times.
    }
    \label{f: prevalence}
\end{figure*}

Fig.~\ref{f: prevalence} extends the analysis in Fig.~\ref{f: theta vs t} to the full sample.
We parametrize the extent of flattening via a parameter $f_{\rm aligned}$ (for ``aligned accretion''), which corresponds to the fraction of accreting gas mass aligned with the disk plane ($\vert z/r \vert < 0.1$, marked by dashed vertical lines in Fig.~\ref{f: theta vs t}).
The left panel of Fig.~\ref{f: prevalence} shows the evolution of $f_{\rm aligned}$ from 200 Myr before $\tcools$ (red) to 200 Myr after $\tcools$ (blue), while the right panel shows the change in $f_{\rm aligned}$ between these two epochs.
The horizontal axes plot the fraction of young stars in a thin disk $\fthin$. 
For most haloes the accreting gas is largely unaligned prior to cooling, with $f_{\rm aligned}\sim 0.1 - 0.2$ comparable to $f_{\rm aligned} = 0.1$ expected for an isotropic distribution.
Upon cooling, the alignment of accreting gas sharply increases in haloes with $\fthin > 0.6$ --- in most cases $\gtrsim 50\%$ of mass collapses to $\vert z/r \vert < 0.1$ during this time.
In contrast, in haloes with $\fthin \approx 0$ there is practically no change in $f_{\rm aligned}$ upon cooling.
Intermediate cases with $\fthin \approx 0.2-0.6$ typically show a modest increase in $f_{\rm aligned}$.
The right panel of Fig.~\ref{f: prevalence} demonstrates the strong correlation between the change in $f_{\rm aligned}$ when the gas cools and the fraction of recent stars in a thin disk.
That is, this panel shows that flattening is concurrent with cooling in accretion onto thin-disk galaxies, while no such association exists for accretion onto irregular or thick disk galaxies.

The galaxy circled in the right panel, \texttt{m12m}, has a stellar bar~\citep{Debattista2019}, which tends to decrease $\fthin$. 
This bar may help explain why this galaxy is somewhat offset to low $\fthin$ relative to other galaxies with a similar large change in  $f_{\rm aligned}$.

\begin{figure*}
\includegraphics[width=\textwidth]{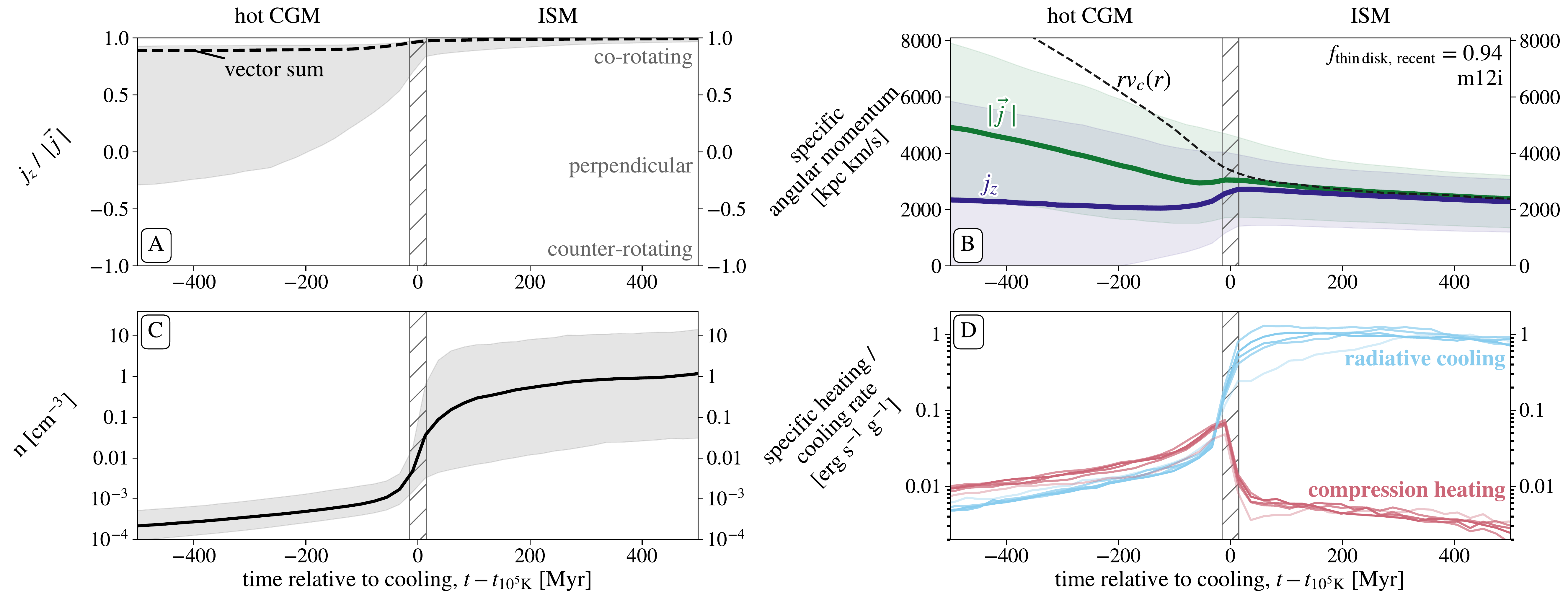}
\caption{
Angular momentum and energetics of gas accreting onto the $z\sim0$ thin disk galaxy (\texttt{m12i}) shown in Fig.~\ref{f: before and after A}, versus time relative to the final cooling time.
\textbf{A:}
The ratio $j_z / \vert \vec j \vert$.
The shaded regions plots the 16th to 84th percentile range of the accreted particles, while the dashed line plots the vector sum.
The shaded region shrinks as gas flows inward, indicating the flow becomes more coherent and that angular momentum unaligned with the total angular momentum is cancelled out. 
\textbf{B:}
The magnitude of the specific angular momentum of particles ($\vert\vec{j}\vert$, green) and the component of angular momentum aligned with the galaxy disk ($j_z$; purple).
Solid lines and shaded regions mark the medians and 16th to 84th percentile ranges, respectively. 
The dashed line shows the angular momentum necessary for rotational support.
Cooling occurs when angular momentum support becomes significant, as expected in subsonic cooling flows \citep{Cowie1980, Stern2020}.
\textbf{C:}
Baryon number density.
Prior to cooling the gas density increases due to the inflow.
As the gas cools the density sharply increases, due to the deceleration of the inflow at the galaxy scale and the collapse into a disk geometry. 
\textbf{D:}
Radiative cooling (blue) and heating from $PdV$ work on accreting gas particles (red).
Only gas with $T>10^{4.5}$ K is included. 
In the CGM compression heating offsets radiative cooling.
This yields the mostly flat temperature profile seen in Fig.~\ref{f: before and after A}, as expected in a cooling flow.
}
\label{f: before and after B}
\end{figure*}

\subsection{Angular momentum and energetics of accreting gas}
\label{s: mechanics -- energy balance}

The mechanics of concurrent cooling and flattening in rotating cooling flows is explored in Fig.~\ref{f: before and after B}, for the example case of \texttt{m12i}.
The shaded region in panel (A) shows the 16th--84th percentiles in $j_z/\vert\vec j\vert$, the ratio of angular momentum in the $z$ direction to the magnitude of total angular momentum, for individual particles as a function of $t-\tcools$.
The dashed line plots $(\Sigma \vec{j})_z/\vert\Sigma \vec{j}\vert$, i.e. the same ratio but for the total angular momentum of accreting gas at a given $t-\tcools$.
Note that this latter ratio is not identical to unity because the $z$ direction is with respect to the central galaxy rather than the gas itself.
At $t-\tcools=-500$ Myr the ratio $j_z/\vert\vec j\vert$ spans a large range of $\approx -0.3 - 0.9$, while by $t=\tcools$ nearly all the accreting gas has $j_z\approx\vert\vec j\vert$, indicating all hot accreting gas particles are co-rotating.
On the other hand, the alignment of the total angular momentum is nearly constant with time prior to accretion. 
These trends indicate that components unaligned with the net angular momentum are canceling out due to interaction in the hot halo.

Fig.~\ref{f: before and after B} panel (B) shows the magnitude of the specific angular momentum ($\vert \vec j \vert$; green) and the z-component of the specific angular momentum ($j_z$; purple).
The median value of $\vert \vec j \vert$ decreases prior to cooling, from $\approx 5000$ kpc km s$^{-1}$ to $\approx 3000\,{\rm kpc\,km\,s}^{-1}$, in contrast with  $j_z$ which remains constant.
The nearly constant $j_{\rm z}$ indicates the inflow roughly conserves its net angular momentum.
At the same time the decreasing $\vert \vec j \vert$ indicates the inflow is cancelling out unaligned components, consistent with the conclusion from panel (A).
We note that this result appears contradictory to the result of \cite{Stevens2017} based on the EAGLE simulations, which found a $\gtrsim50\%$ decrease in net angular momentum when accreting hot gas cools onto the galaxy.

The value of $j_z\approx 2500$ kpc km s$^{-1}$ is comparable to the average specific angular momentum of $j_{\rm DM} \simeq \sqrt{2}\lambda \Rvir \vvir$ expected in dark matter haloes due to tidal torques \citep[e.g.][]{Bullock2001}.
Using a typical dimensionless spin parameter $\lambda \simeq 0.035$, and the virial radius $\Rvir=270$ kpc and virial velocity $\vvir=130$ km s$^{-1}$ of this halo, we get $j_{\rm DM}\approx1750$ kpc km s$^{-1}$, i.e.~the expected average value of the dark matter is within $30\%$ of the net angular momentum of accreting hot gas shown in Fig.~\ref{f: before and after B}. 
The fact that the gas has slightly more angular momentum than naively expected for the dark matter is consistent with previous findings that gas typically has a slightly higher spin than dark matter \citep[e.g.][]{Stewart2017}.

The dashed line in panel (B) of Fig.~\ref{f: before and after B} plots the median specific angular momentum necessary for gas to be fully supported by angular momentum at a given radius, i.e. $rv_c(r)$.
The accretion attains significant angular momentum support as it proceeds through the inner CGM, consistent with other analyzes of CGM rotational support which find that rotation support is more prominent at smaller CGM radii~\citep{Oppenheimer2018, Trapp2021}.
The values of $j_z$ and $v_c(r)r$ converge shortly after $t=\tcools$, indicating that cooling and a transition to rotational support occur almost simultaneously, as indicated also by Fig.~\ref{f: before and after A}.
This result is consistent with 1D steady-state cooling flow solutions which include angular momentum -- these solutions demonstrate that the hot inflow cools to $\sim10^4$ K at the radius where $j_z=r v_c(r)$, as long as the flow remains subsonic down to this radius \citep{Cowie1980, Stern2020}.

There is a small but noticeable increase in $j_z$ at $t \approx \tcools$, i.e., the angular momentum of accreting gas increases somewhat when it transitions from the hot CGM to the ISM. This increase is  similar to the observed $30\%$ difference between the rotation velocity of the Milky Way disk and its hot CGM \citep{Hodges-Kluck2016}. While this increase is not the focus of our analysis,  we note that it may be a result of a difference in orientation between the angular momentum of the galaxy and the angular momentum of the accreting gas~\citep[e.g.][]{Danovich2012, DeFelippis2017}, which forces the accreting gas to co-rotate with the galaxy upon accretion.
Any angular momentum gained is expected to be lost by other particles, driving an  inflow in the disk or disk-halo interface~\citep{Mayor1981, Pezzulli2016a}.

Fig.~\ref{f: before and after B} panel (C) shows the distribution of the accreting gas density versus time relative to cooling.
Prior to cooling the gas density increases steadily due to the inflow, reaching $\approx10^{-3}$ cm$^{-3}$ just before cooling at $t\lesssim\tcools$.
This density is comparable to observational estimates of the hot gas density just outside the Milky Way disk (e.g.,~\citealt{Li2017a}).
At $t\approx\tcools$, the gas density sharply increases, reaching $0.1$ cm$^{-3}$ within $\approx50$ Myr, due to the decreasing radial velocity of the inflow upon accretion (Fig.~\ref{f: before and after A}), and due to the collapse from a quasi-spherical geometry into a disk geometry.
The gas then starts forming stars shortly after cooling (Fig.~\ref{f: before and after A}).
Note that star formation occurs despite that the 84th density percentile after cooling is $\approx10$ cm$^{-3}$, well below the minimum density for star formation of $n_{\rm SF} = 1000$ cm$^{-3}$ in FIRE.
This is because gas remains at densities approaching $n_{\rm SF}$ only for a short time before forming stars.

In Fig.~\ref{f: before and after B} panel (D) we assess the energetics of the hot gas to determine why it cools.
We study the energetics through two types of change in specific energy: radiative cooling (blue lines) and compression heating (red lines).
Radiative cooling per unit mass for an individual particle is calculated as $\nH^2 \Lambda / \rho$, where $\nH$ is the hydrogen density, $\rho$ is the mass density, and $\Lambda$ is the cooling function.
We include only fluid elements with $T>10^{4.5}$ K, to avoid optical thickness effects on cooling, and to avoid large fluctuations when $\Lambda$ approaches zero at $T\approx10^4$ K.
As our focus is the energetics of the hot $\sim\Tvir$ gas, this cut does not affect our conclusions. 
Compression heating per unit mass for an individual resolution element is calculated as $P \frac{dV}{dt} \approx \frac{ P }{ \rho^2 } \frac{ \Delta \rho }{ \Delta t }$, where $P$ is the thermal pressure, $V$ is the specific volume, $\Delta \rho$ is the change in density from one snapshot to the next, and $\Delta t$ is the snapshot time spacing.
Because accreting gas elements interact with other accreting gas elements thermodynamically we show the mean specific energy tracks of all gas elements binned into 100 Myr bins of $\tcools$. 
Also, to focus on the behavior of the majority of the particles we do not bin the $1\%$ of particles with the highest change in specific energy.
Some $\tcools$ bins contain much more accreting gas than others, and to reflect this we set the darkness of the lines proportional to the number of particles in the bin.

At $t-\tcools=-500$ Myr the radiative cooling rate is $\approx0.006$ erg s$^{-1}$ g$^{-1}$, corresponding to a cooling time $t_{\rm cool}=400$ Myr for the median temperature of $T=4\times 10^5$ K at this epoch (see Fig.~\ref{f: before and after A}, panel A). 
At later $t-\tcools$ but still prior to cooling, the radiative cooling rate increases due to the increase in gas density.
The panel shows that this energy loss to radiation is followed closely by compressive heating, explaining the roughly flat temperature profile at $t<\tcools$ (Fig.~\ref{f: before and after A}). 
This approximate equality between radiative cooling and compressive heating is a defining characteristic of classic cooling flows in which angular momentum support can be neglected \citep{Mathews1978, McNamara2007, Stern2019}. 
This balance between radiative cooling and compressive heating is not possible in a perfectly hydrostatic halo, since without inward movement compression will not occur.

Around $\tcools$ the cooling rate and compressive heating diverge.
This can be understood by noting that the radiative cooling rate per unit mass scales as $\propto\rho\Lambda$, while the compressive heating rate scales as $\propto T d\log\rho/d t$.
The deceleration of the hot inflow prior to accretion onto the galaxy causes $\rho$ to increase faster than $d\log\rho/d t$, causing radiative cooling to exceed compressive heating and the temperature to decrease.
This in turn increases the cooling rate which further accelerates the drop in temperature.
The result is gas that cools from $\approx10^6$ K to $\approx10^4$ K over the course of $\lesssim 50$ Myr.

Panel D also shows that  after $\tcools$ when the gas is part of the ISM, radiative cooling greatly exceeds compression heating, indicating that hot gas in the ISM is short lived. 

\begin{figure*}
    \centering
    \includegraphics[width=\textwidth]{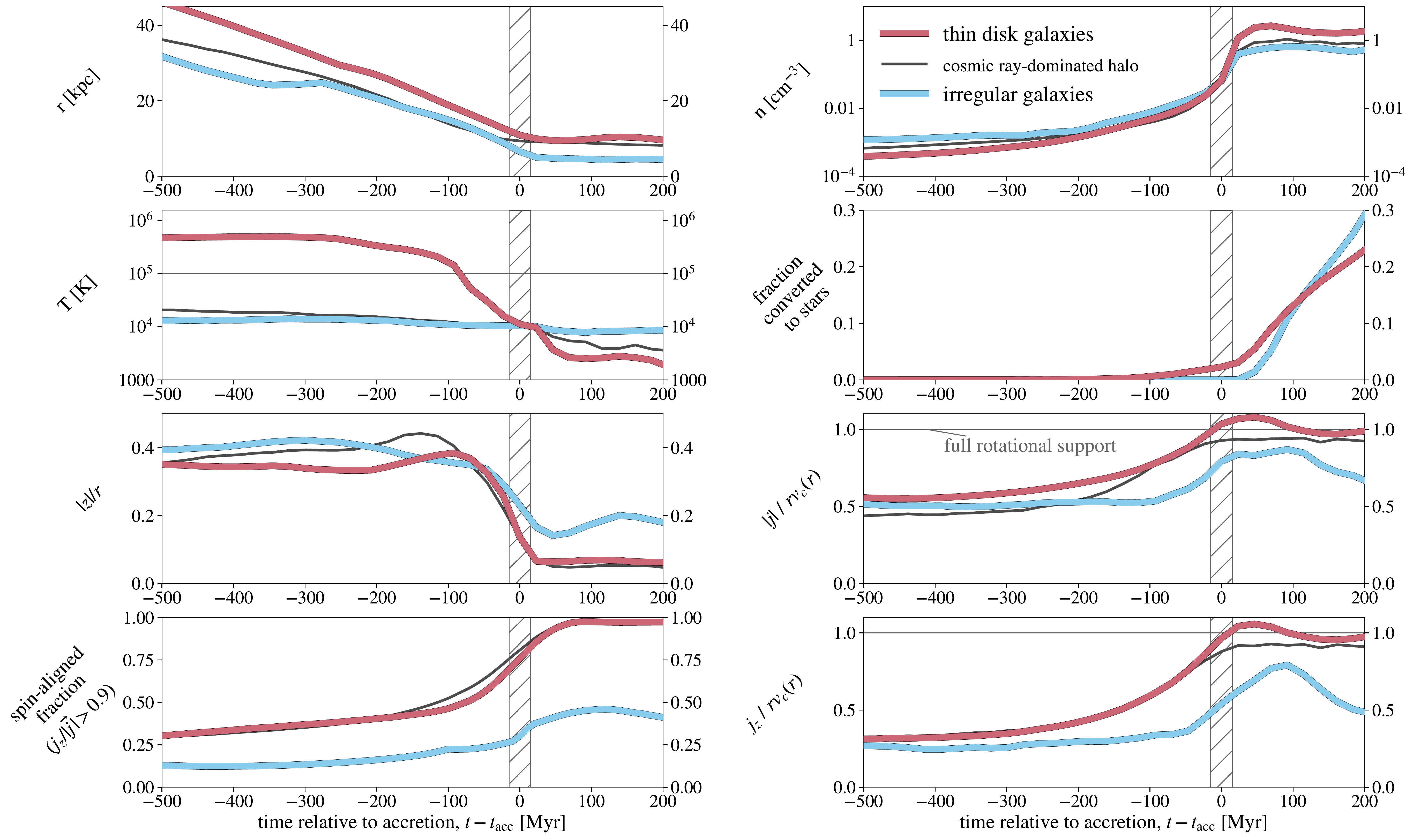}
    \caption{
    Hot accretion onto thin disk galaxies versus cool accretion onto irregular galaxies.
    Lines plot the median properties of the accreting gas relative to time of accretion, for thin disk galaxies (red; $\fthin>0.6$) and for irregular galaxies (blue; $\fthin<0.6$).
    The temperature panel (second row on the left) shows that accretion onto thin disk galaxies is hot ($T\gtrsim10^{5.5}$~K), in contrast with cool $T\approx10^4$ K accretion onto irregular galaxies.
    The bottom panels show that accretion onto thin disk galaxies also has an angular momentum distribution that is more aligned and is more rotationally supported than in irregular galaxies, suggesting that the coherence achieved in hot accretion is conducive to the formation of galaxy disks. 
    The dark grey lines show the median accretion properties in a cosmic ray-dominated halo.
    The accretion in this halo is cool, but similar in its angular momentum properties to hot accretion (see \S\ref{s: crs}). 
    }
    \label{f: before and after combined}
\end{figure*}

\subsection{Comparison of accretion onto thin disk galaxies and onto irregular galaxies}
\label{s: thin vs irregular}

Figure~\ref{f: before and after combined} compares the properties of accretion onto galaxies with $\fthin > 0.6$ (``thin disk galaxies'') versus accretion onto galaxies with $\fthin < 0.6$ (``irregular galaxies'').
For each property the colored lines show the median of the medians of the individual simulations in the group at a given $t - \tacc$.
The properties are plotted as a function of time relative to the time at which gas accretes onto the main galaxy ($\tacc$), defined as the first time gas is within $r_{\rm gal}$ and with a density $n_{\rm H} > 0.13$ cm$^{-3}$.
This is in contrast with plotting properties versus $\tcools$ as done above, since in irregular galaxies where accretion is generally cool ($T\approx10^4$K) the last cooling time is not  associated with accretion onto the galaxy. 
We also exclude \texttt{m12f} from the thin disk group since the gas cools and becomes rotationally supported at significantly larger radii than $r_{\rm gal}$, due to the exceptionally large $j_z\approx5000$ kpc km s$^{-1}$ of hot gas in this galaxy.
This galaxy is further discussed in Appendix~\ref{s: appendix-individual}. 

The red lines in Fig.~\ref{f: before and after combined} demonstrate that the conclusions from Figs.~\ref{f: before and after A} and~\ref{f: before and after B} hold on average for all thin disk galaxies in our sample:
gas inflow is hot through the CGM with a median $T \approx 10^{5.5}$ K,
and cooling and deceleration occurs at the galaxy edge,
at which time the gas distribution also flattens.
In contrast, accretion onto irregular galaxies is primarily cold, with no change in median temperature before and after accretion.
Gas becomes more spatially aligned as it accretes, but still has a median $ \vert z \vert / r \sim 0.2$, and never becomes fully rotationally supported.
Accretion onto thin disks and irregulars also differ significantly in their spin-aligned fraction  prior to $\tacc$ (lower-left panel) -- at $\tacc$ the fraction of spin-aligned accretion onto thin disk galaxies is $0.75$, a factor of three higher than the fraction of $0.25$ in irregular galaxies.
This suggests that achieving angular momentum coherence in the CGM is conducive to the formation of thin disks. 

\begin{figure*}
    \centering
    \includegraphics[width=\textwidth]{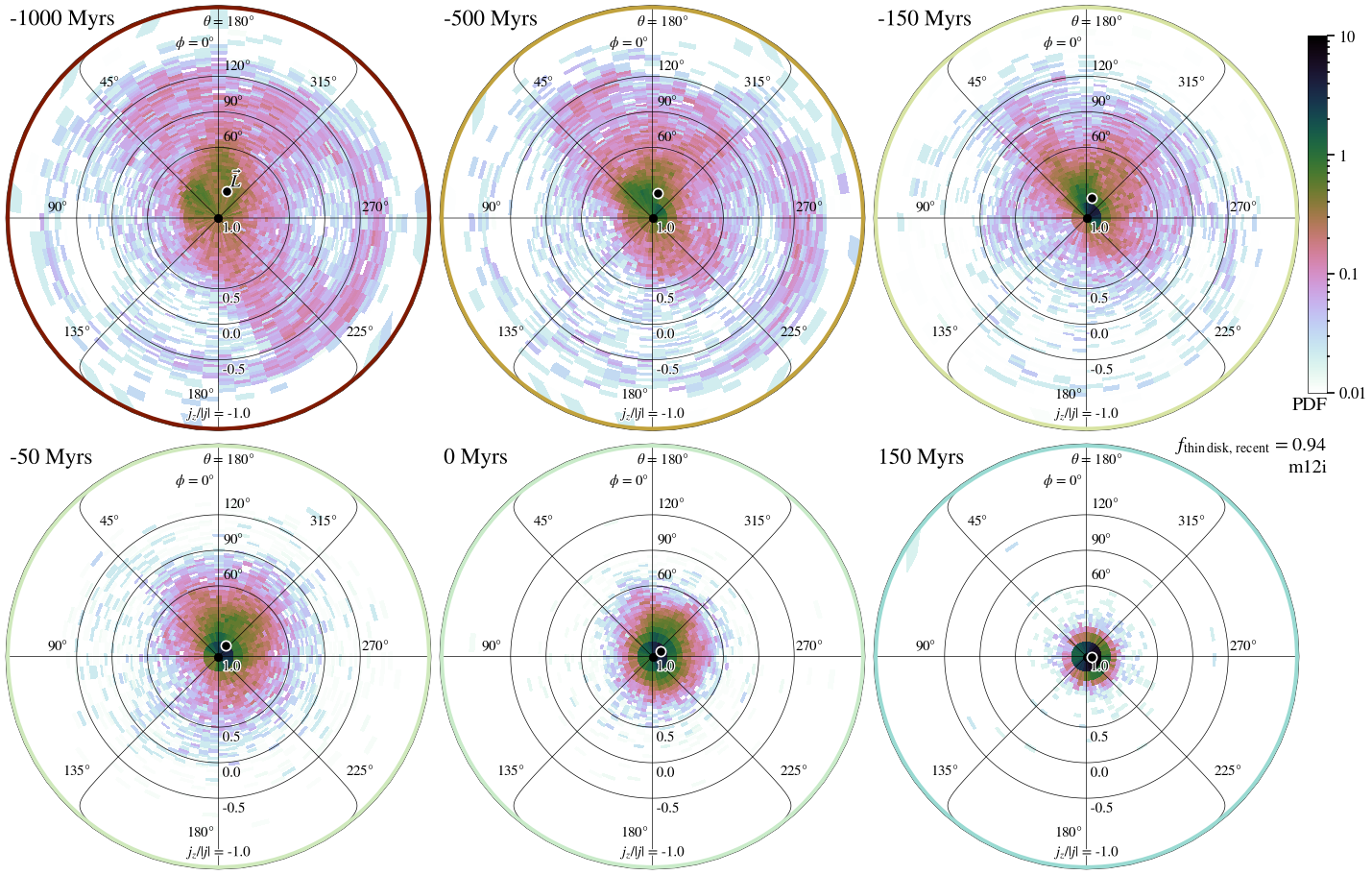}
    \caption{
    The evolution of the direction of angular momentum for gas accreting onto a MW-like galaxy, \texttt{m12i}, in an `on-sky' projection.
    Different panels show the distribution for different times relative to $\tacc$.
    The bin color is logarithmically proportional to the fraction of accreting mass with an angular momentum pointing in that direction.
    The angles $\theta$ and $\phi$ are spherical coordinates where $\theta=0^\circ$ is the direction of the galaxy angular momentum at $z=0$ (shown in the center of the projection).
    The black point outlined in white is the direction of the net angular momentum vector of accreting gas, $\Sigma \vec{j}$.
    Over the course of $1$ Gyr prior to accreting the angular momentum distribution narrows significantly (i.e. becomes more coherent), while the direction of the net vector changes only slightly.
    }
    \label{f: coherence -- on sky}
\end{figure*}

\begin{figure*}
    \centering
    \includegraphics[width=\textwidth]{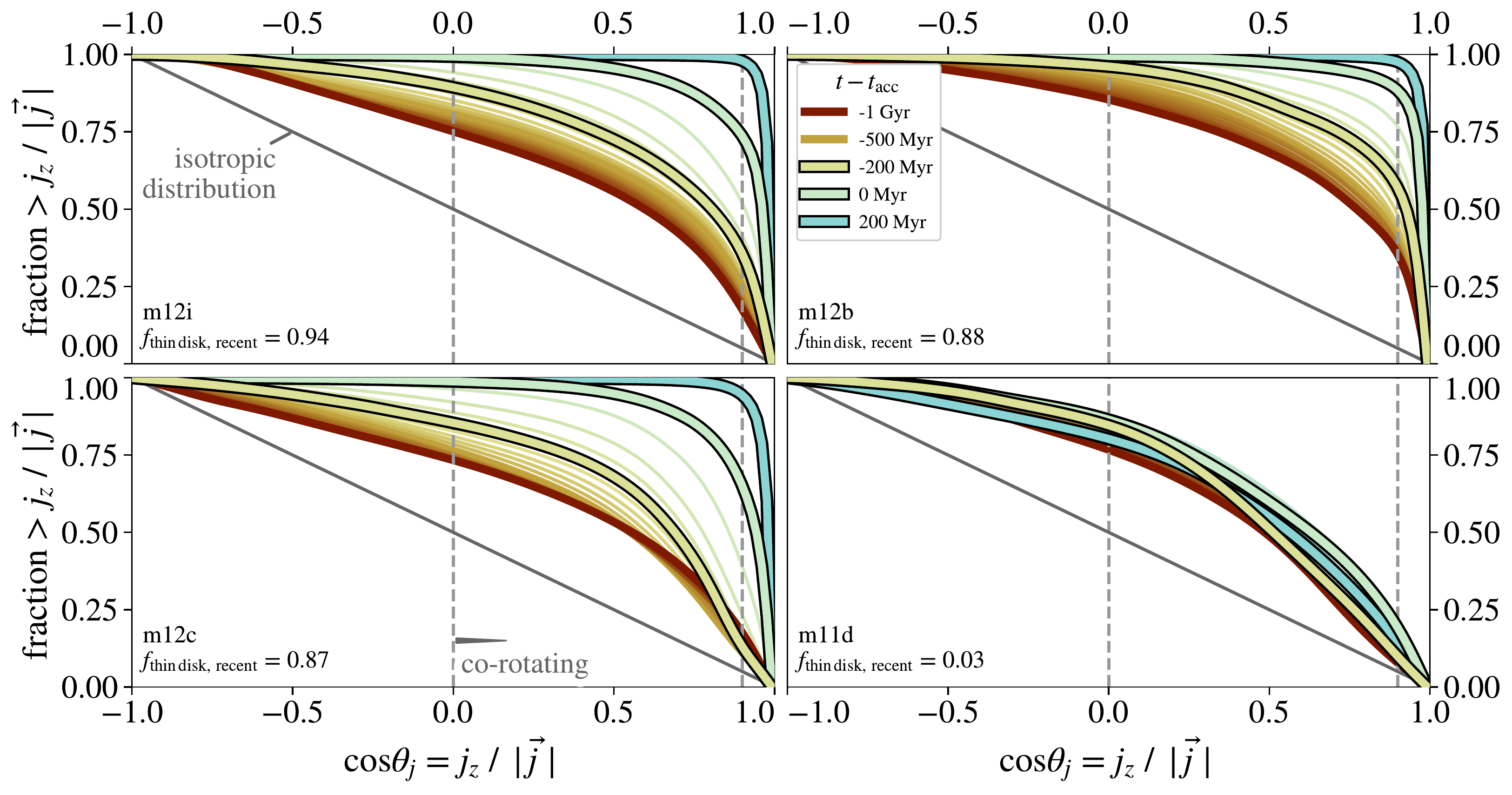}
    \caption{
    Angular momentum distribution of accreting gas, as a function of time relative to accretion time.
    The galaxies displayed are the same as in Figs.~\ref{f: stars} and~\ref{f: theta vs t}.
    Curves show the fraction of mass above a given $j_z / \vert \vec j \vert$. 
    In the thin-disk galaxies (top and bottom-left panels) the angular momentum distribution becomes more coherent and aligned with the central galaxy with time, especially during the 200 Myr prior to accretion.
    In the irregular galaxy (bottom right) the angular momentum distribution is only mildly co-rotating  both before and after accreting. 
    }
    \label{f: coherence}
\end{figure*}

\begin{figure}
    \centering
    \includegraphics[width=\columnwidth]{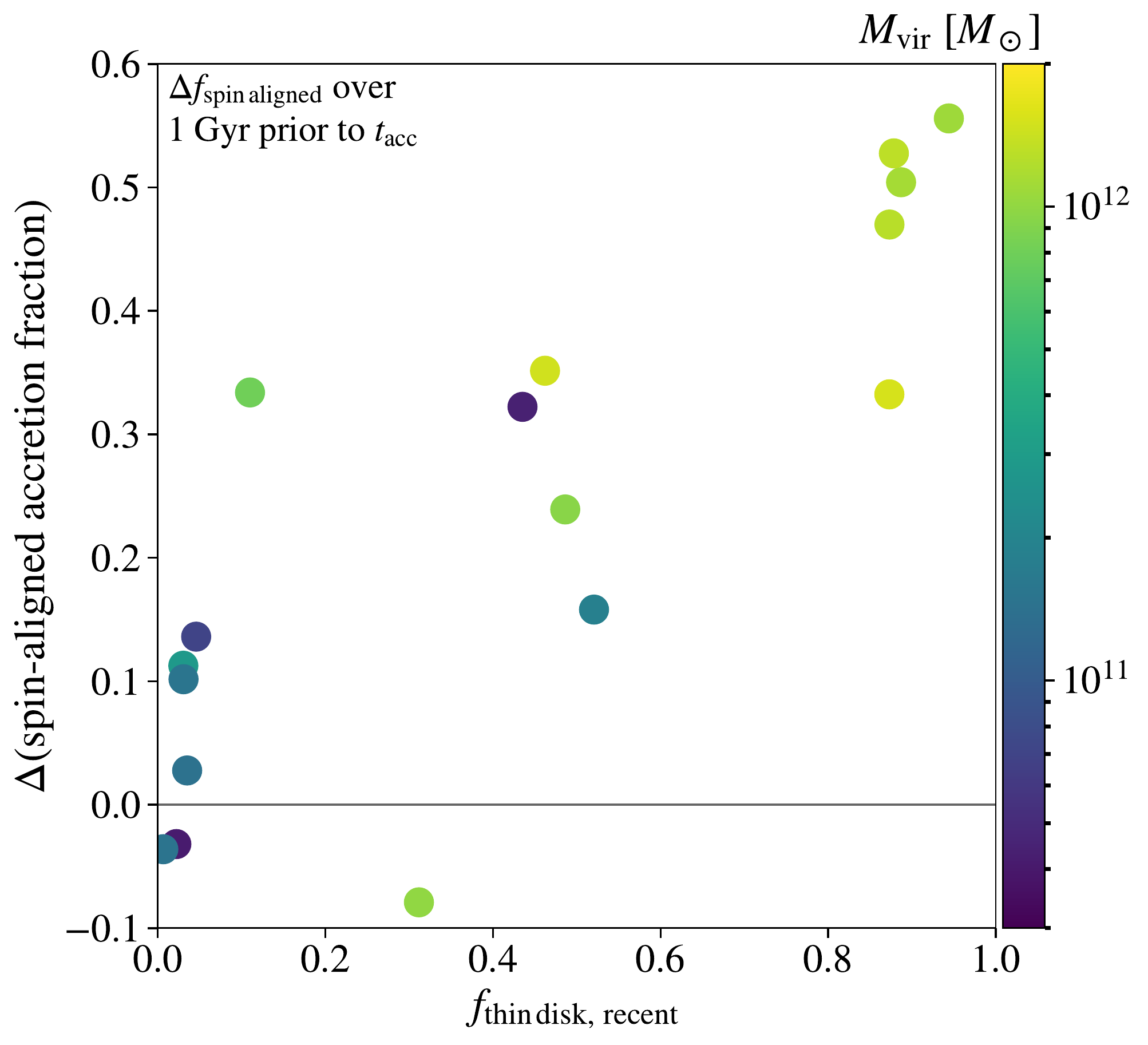}
    \caption{
    The change in the fraction of accreting gas mass which is spin-aligned with the galaxy ($j_z/\vert \vec j \vert > 0.9$, right dashed lines in Fig.~\ref{f: coherence}) over 1 Gyr prior to accretion, versus thin disk fraction for the full sample of simulations.
    This Figure is an angular momentum-space equivalent to the change in geometry shown in the right panel of Fig.~\ref{f: prevalence}. 
    The strong correlation between change in spin-alignment and thin disk fraction suggests that achieving angular momentum coherence in the CGM is conducive to the formation of thin disks.
    }
    \label{f: prevalence - angular momentum}
\end{figure}

\subsection{Angular momentum coherence in accreted gas}
\label{s: mechanics -- coherence}

Figure~\ref{f: coherence -- on sky} shows the evolution of the angular momentum distribution for \texttt{m12i} versus time relative to $\tacc$. 
We use an ``on-sky'' azimuthal equidistant projection of the angular momentum direction for \texttt{m12i}.
Azimuthal equidistant projections preserve the distance to the central point (the angular momentum direction of the galaxy at $z=0$) as well as the orientation ($\phi$) relative to the central point.
Each histogram bin covers an equal area.
The total angular momentum of the accreting gas at each time $\Sigma{\vec{j}}$ is shown via a black point outlined in white.
The Figure shows that prior to $t - \tacc = -500$ Myrs accretion is composed of gas rotating in a variety of directions, including counter-rotating ($j_z/\vert j \vert < 0$ or equivalently $\theta > 90^\circ$).
The net angular momentum $\Sigma{\vec{j}}$ is however roughly centered on the angular momentum of the galaxy $\theta = 0^\circ$.
By $t - \tacc = -50$ Myrs the angular momentum distribution is significantly more coherent/narrower, while only a minor change in seen in the direction of net angular momentum vector.
At and shortly after $\tcools$ the angular momentum distribution further narrows and the net vector nearly fully aligns itself with the galaxy.

Fig.~\ref{f: coherence} shows the cumulative distribution function of $j_z / \vert \vec j \vert$ in the accreted gas, weighted by mass, for the four simulations shown in Fig.~\ref{f: theta vs t}.
This distribution quantifies the level of alignment of angular momentum in the accreted gas with respect to the rotation axis of the stars.
For reference, perfectly co-rotating gas has $j_z / \vert \vec j \vert = 1$, while perpendicular and counter-rotating gas have $j_z / \vert \vec j \vert = 0$ and $-1$, respectively.
An isotropic distribution of angular momentum would appear as a diagonal line in this plot. 
Each curve corresponds to a different $t - \tacc$ as noted in the legend.
Outlined in black are the CDFs for $t - \tacc =$ -200, 0, 200 Myr.

In gas accreting onto thin disk galaxies (top and bottom-left panels of Fig.~\ref{f: coherence}) the angular momentum distribution becomes increasingly coherent with time relative to $\tacc$.
At $t-\tacc=-1$ Gyr the angular momentum distribution is marginally coherent, with $\approx50-70\%$ of accreting gas co-rotating with $j_z/\vert \vec j \vert > 0.5$.
In contrast at $t-\tacc=+200$ Myr the distribution is highly coherent, with $j_z/\vert \vec j \vert > 0.9$ for $\gtrsim 90\%$ of accretion.
The majority of the evolution in coherence occurs over $\lesssim 200$ Myr prior to $\tacc$, as seen by the differences between the distributions for $t-\tacc=-200$ Myr and $t-\tacc=0$ Myr.
This increase in coherence allows the accreting gas to collapse into a thin disk as shown in Figs.~\ref{f: theta vs t}--\ref{f: prevalence}.
Furthermore, this result shows that accreting gas is almost entirely co-rotating with the galaxy {\em prior} to accreting, i.e., while the accretion is still part of the galactic `hot corona'. 

In stark contrast with thin disk galaxies, the irregular galaxy \texttt{m11d} shown in the bottom-right panel of Fig.~\ref{f: coherence} experiences only a very mild evolution in angular momentum coherence --- angular momentum remains only marginally co-rotating both before and after accreting.

In Figure~\ref{f: prevalence - angular momentum} we show the change in angular momentum alignment across the full sample, similar to Fig.~\ref{f: prevalence} for spatial alignment. 
In this case, spin alignment is defined as the fraction of gas with $j_z/j > 0.9$.
The change in spin alignment is calculated focusing on the time prior to accretion, $-1$ Gyr $< t - \tacc < 0$ Gyr.
The figure shows a correlation between thin disk fraction and the increase in angular momentum alignment prior to accretion, suggesting that achieving angular momentum coherence in the CGM is conducive to the formation of thin disks.

\section{Discussion}
\label{s: discussion}

In this paper we analyze the properties of gas accreting onto $z\sim0$ galaxies simulated in FIRE, focusing on Milky-Way mass galaxies in which new stars form in a thin disk. 
We find that thin disk galaxies in FIRE accrete via `rotating cooling flows', which is a type of `hot accretion'. In this accretion mode the quasi-spherical $T\sim\Tvir$ CGM phase inflows towards the galaxy, remaining hot down to the radius where its angular momentum is sufficient to provide rotational support.
At this radius ($\gtrsim 4 r_{\star,0.5}$, just outside the galaxy radius) the hot inflow both decelerates and becomes coherently rotating, and then simultaneously cools and collapses into a rotating cool disk.
Our results thus extend classic cooling flow theory by demonstrating their applicability in realistic cosmological simulations, and by exploring the mechanics of cooling flows with angular momentum, a subject which has not yet been studied extensively~\citep[c.f.][]{Cowie1980, Stern2020}.
Moreover, we find a strong correlation between the prevalence of this accretion mode and the fraction of stars in the central galaxies that form in a thin disk, potentially indicating that a rotating cooling flow is a necessary condition for the formation of a thin star-forming disk.
In this section we discuss several interpretations, caveats, and implications of our results. 

\subsection{Why are rotating cooling flows conducive to thin disk formation?}
\label{s: why CFs thin disks}

A main result of our analysis is that when accretion occurs via rotating cooling flows, the accreted gas forms a coherently rotating disk \textit{prior} to cooling and accreting onto the ISM.
This is due to the decrease in angular momentum dispersion prior to cooling (Figs.~\ref{f: coherence -- on sky}--\ref{f: prevalence - angular momentum}, and panels A--B in Fig.~\ref{f: before and after B}) and due to the deceleration of the flow prior to accretion (panel D in Fig.~\ref{f: before and after A}).
Combined, these two properties indicate that upon accretion the flow is already in a coherent disk with $v_\phi\approx v_{\rm c}$ and $v_r, v_z \ll v_{\rm c}$.
In contrast, in other accretion modes (such as cold streams and precipitation) the inflow may reach the ISM with a large dispersion in angular momentum and substantial radial momentum, so equilibration will start only after accretion onto the ISM. That stars in FIRE galaxies which are fed by cool accretion form in irregular distributions for many Gyrs (Figures~\ref{f: before and after combined} and~\ref{f: m11d}) suggests that equilibration in the ISM alone may be insufficient to form a thin disk, and hence equilibriation in the CGM as we find in galaxies fed by rotating cooling flows is conducive to the formation of thin disks. 

Why is equilibration in the ISM alone insufficient to form a thin disk in FIRE? Equilibration is expected to proceed on a dynamical timescale ($\sim 100$ Myr), which is short relative to the star formation timescale of $\sim0.5-2$ Gyr (e.g. Fig.~\ref{f: before and after A}; \citealt{Bigiel2008}). However, some stars may succeed in forming while the accreting gas morphology is still a thick disk or irregular.
Such stars inject feedback momentum and energy into their surroundings which could further delay the equilibration process, allowing even more stars to form outside of a thin disk.

The tendency of hot inflows to decelerate and rotate coherently prior to accretion is potentially a result of the subsonic nature of the flow. 
In a subsonic flow gradual deceleration prior to accretion is expected since the inflow is `forewarned' (via changes in pressure) of the transition to rotational support at the galaxy scale.
This is in contrast with supersonic accretion modes where the accreting gas is expected to shock and halt abruptly at the galaxy scale.
Also, subsonic accretion makes it easier for the flow to reach angular momentum coherence, since the inflow timescale is longer than the sound-crossing timescale on which coherence can be achieved. In contrast, in supersonic, free-falling accretion flows as we find in low mass galaxies, the accretion and coherence timescales are both comparable to the dynamical time and hence comparable to each other, so coherence is not achieved in the CGM.

Panels A--B in Fig.~\ref{f: before and after B} and Figs.~\ref{f: coherence -- on sky}--\ref{f: prevalence - angular momentum} demonstrate that angular momentum coherence is achieved in hot inflows just before cooling and accretion, roughly at the rotational support radius $\Rcirc$ (defined via $\Sigma\vec{j}=\vc(\Rcirc)\Rcirc$, where $\Sigma\vec{j}$ is the net angular momentum in the flow), rather than farther out in the halo.
Since the hot inflow is subsonic at all radii, this result suggests that a dynamical timescale longer than the sound-crossing timescale is not, on its own, a sufficient condition for the flow to achieve angular momentum coherence.
We suspect that coherence is achieved specifically near $\Rcirc$ because of the relation between average rotation speed and subsonic turbulence in the hot CGM, where the latter is driven by e.g., clumpy cosmological accretion, stirring by subhalos, and feedback from the galaxy.
To understand this, we can look at the relative magnitude of angular momentum fluctuations due to turbulence $\sigma_{\rm turb} r / j_z$, where $\sigma_{\rm turb}$ is the turbulent velocity.
For subsonic turbulence with a mach number of say $\mturb=1/3$ we have $\sigma_{\rm turb}\approx \mturb\vc$, since the sound speed in hot inflows is roughly equal to $\vc$.
Combining this with the definition of $\Rcirc$ and the conservation of $j_z$ in the inflow (Fig.~\ref{f: before and after B}, panel B), we get that angular momentum fluctuations are of order $r \sigma_{\rm turb}/j_z \approx \mturb r/ \Rcirc$.
Thus, for subsonic turbulence the relative magnitude of angular momentum fluctuations will be small at $r\approx\Rcirc$, but could be large at larger radii, as found above.
We defer further exploration of this potential relation between turbulence and angular momentum coherence to future work. 

We note that accretion via a rotating cooling flow does not ensure a thin disk forms. 
A merger, a strong feedback event, or other disruptions may take a thin gaseous disk and disturb it, preventing stars from forming in a thin disk.
Simulation \texttt{m12m} provides an example of this:
it has rotating cooling flow accretion comparable to thin disk galaxies ($\Delta f_{\rm aligned} \approx 0.25$), but has a relatively small $\fthin \approx 0.5$ (circled point in Fig.~\ref{f: prevalence}) due to the presence of a bar.
However, the strong correlation seen in Fig.~\ref{f: prevalence} between the prevalence of rotating cooling flow accretion (quantified via $\Delta f_{\rm aligned}$) and the thin disk fraction $\fthin$ suggests that such events are relatively rare.

\subsection{When and where do we expect thin disks?}

\begin{figure*}
    \centering
    \includegraphics[width=\textwidth]{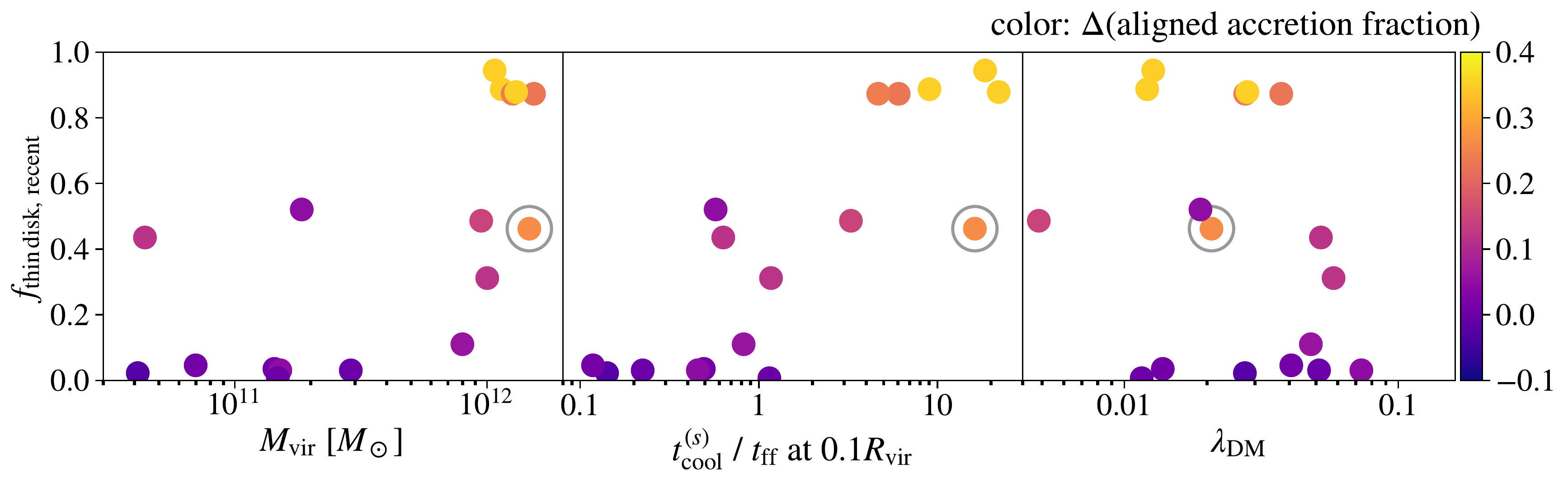}
    \caption{
    Fraction of young stars in a thin disk versus $M_{\rm vir}$ (left), versus $t_{\rm cool}^{(s)}/t_{\rm ff}$ evaluated at $0.1 R_{\rm vir}$ (middle), and versus dark-matter spin parameter (right).
    Markers are colored by $\Delta f_{\rm aligned}$, a measure of the dominance of rotating cooling flows in accreting gas (Fig.~\ref{f: prevalence}). The middle panel shows that thin disks appear in halos with $t_{\rm cool}^{(s)}\gg t_{\rm ff}$, while irregular galaxies appear in halos with $t_{\rm cool}^{(s)}\lesssim t_{\rm ff}$. This relation further supports our conclusion that rotating cooling flows are conducive to the formation of thin disks, since a rotating cooling flow is expected after the inner CGM virializes, which occurs when $t_{\rm cool}^{(s)}$ exceeds $t_{\rm ff}$ at $0.1 R_{\rm vir}$ \citep{Stern2021}.
    }
    \label{f: prevalence vs galaxy properties}
\end{figure*}

\newcommand{\tcoolsh}{t_{\rm cool}^{(s)}}
\newcommand{\tff}{t_{\rm ff}}
\newcommand{\Mvir}{M_{\rm vir}}

The scenario suggested by this paper, that thin star forming disks are a result of accretion via rotating cooling flows, allows us to predict at which halo masses and redshifts we expect thin disks to form.
At low redshift, cooling flows are expected if a time-steady and pressure-supported `virialized CGM' has formed, which occurs when the cooling time of hot shocked CGM gas $\tcoolsh$ exceeds the free-fall time $\tff$, typically at halo masses above a threshold of $\sim10^{11}-10^{12}\msun$ \citep[e.g.,][]{White1978, White1991, Birnboim2003}.
The existence of cooling flows also requires that radiative cooling in this hot gas is not balanced by feedback. While there is evidence for feedback balancing radiative cooling in group and cluster halos, at which the cooling rate implied by X-ray observations greatly exceeds the central galaxy SFR (i.e., halos with a `cooling flow problem', see \citealt{McDonald18} for a recent study), there is no similar evidence for a cooling flow problem in the halos of disk galaxies \citep{Li14b}, though uncertainties are still large due to the weak X-ray emission from the hot gas at this mass scale. 
Cooling flows in the low redshift universe are thus possible at intermediate masses, where on one hand the halo is massive enough so the CGM has virialized, but on the other hand the halo is not so massive that X-ray observations rule them out. 

\cite{Stern2020} recently refined the classic condition for the existence of cooling flows, by showing that even if a virialized CGM exists in the outer halo the resulting cooling flow may go through a sonic point and reach the galaxy as cool supersonic accretion.
Thus, to produce cooling flows that remain hot and subsonic down to the galaxy scale as found above for thin disk galaxies, a more specific condition should be satisfied, namely that $\tcoolsh$ exceeds $\tff$ at an inner CGM radius of $\approx0.1 r_{\rm vir}$, i.e.~subsonic cooling flow requires that the \textit{inner} CGM is virialized. 
In our simulation sample, the condition $\tcoolsh\gtrsim\tff$ at $\approx0.1 r_{\rm vir}$ is met at $\Mvir\approx10^{12}\msun$ (\citealt{Stern2021}). 
Fig.~\ref{f: prevalence vs galaxy properties} shows the relationship between thin disk fraction and virial mass (left panel), $\tcoolsh/\tff$ at $0.1 r_{\rm vir}$ (middle panel), and dark-matter spin parameter in our sample of $z\sim0$ FIRE galaxies.
Markers are colored by $\Delta f_{\rm aligned}$, a measure of the dominance of rotating cooling flows in accreting gas (Fig.~\ref{f: prevalence}).
The figure shows that thin disks and rotating cooling flows appear at $M_{\rm vir}\approx 10^{12}\msun$ and $\tcoolsh/\tff(0.1\Rvir)>1$, consistent with the expectation noted above that cooling flows and hence thin disks commence only when the inner CGM virializes.
This mass scale for thin disk formation is however somewhat larger than in $z\sim0$ observations, as further discussed in section~\ref{s: caveats}.
While thin disk fraction does correlate with mass scale, the dark-matter mass and angular momentum properties alone are likely insufficient to predict galaxy morphology, as indicated by the lack of correlation between the dark-matter spin parameter and $\fthin$.

How does thin disk formation depend on redshift?
While the threshold of inner CGM virialization in FIRE is roughly independent of redshift \citep{Stern2021,Stern2021a}, at $z>1$
cold streams may penetrate the hot halo and dominate the accretion \cite[e.g.,][]{Keres2005,Dekel2006, Dekel2009}, and thus if thin disks form as a result of hot accretion as suggested in this work they would be rarer at high redshift.
Alternatively, thin disks could be rarer at high redshift due to evolving gas fractions, since in equilibrium disk models the thickness of the disk scales with gas fraction \citep{Thompson2005, Faucher-Giguere2013, Krumholz2018}.
These predictions are consistent with morphological and kinematic observations of galaxy samples, where thin disks are found mainly at $z<1$ and intermediate galaxy masses \citep[e.g.,][]{Kassin2012, vanderWel14}. 

\subsection{Comparison to other models for the hot CGM}

Our result that the inner CGM of Milky-Way mass galaxies forms a rotating cooling flow can be compared to other models for the hot CGM of Milky Way-like galaxies.
In both the precipitation-regulated model of \cite{Sharma2012} and \cite{Voit2017} and the isentropic model of \cite{Faerman2020}, the expected inflow induced by radiation is suppressed due to a balance between radiative cooling and heating by feedback (`thermal balance'), and the hot gas is in hydrostatic equilibrium.
Our results based on the FIRE simulations suggest a different scenario, where the hot CGM is not in thermal balance but rather inflows on a cooling timescale as expected from classic cooling flow solutions.
This implies that stellar feedback in Milky-Way mass galaxies in FIRE is insufficient to disrupt the cooling flow, consistent with the weak outflows at this mass and redshift scale in the FIRE simulations~\citep{Muratov2015, Muratov2017, Angles-Alcazar2017, Pandya2021} and in observations \citep[e.g.,][]{Heckman2019}.
The potential effects of AGN feedback, not implemented in the FIRE sample used in this work, are further discussed in section~\ref{s: caveats}.
We note though that even in cooling flows the pressure profile of the hot CGM is consistent with hydrostatic equilibrium to zeroth-order, with relatively small deviations of order $\approx(t_{\rm ff}/t_{\rm cool})^2$ \citep{Stern2019}. 

Our results also differ from precipitation models in that the hot inflow \textit{dominates} the accretion onto disk galaxies, rather than accretion by cool clouds condensed from the hot medium (Figures.~\ref{f: before and after combined} and~\ref{f: Mdot}, see also \citealt{Esmerian2021}).
We expect this dominance of hot inflows to be robust to resolution effects despite that the formation of cool clouds in simulations depends on resolution \citep[e.g.][]{McCourt2018, Mandelker2019a,Mandelker2021, Fielding2020}. This follows since the hot gas accretes on the cooling timescale (\S\ref{s: mechanics -- energy balance}), i.e., the same timescale on which thermal instabilities grow. Thermal instabilities in an inflowing hot CGM thus do not have time to grow substantially, as indicated by idealized calculations of cooling flows \citep{Balbus1989,Stern2019}.
A systematic resolution test of this result is however beyond the scope of this paper. 

It is also interesting to compare our results to the calculations of \cite{Pezzulli2017} and \cite{Sormani2018}, who considered the effect of angular momentum support on the structure of a hot CGM which is hydrostatic except for the rotation component. Since cooling flows are also hydrostatic to zeroth-order as mentioned above, they are expected to satisfy similar constraints on their pressure profiles as in these previous models (see eqn.~1 in \citealt{Sormani2018}). Our results however suggest another constraint on the hot CGM structure which was not considered by these previous studies -- the radial distribution of the net angular momentum is flat, since the hot CGM conserves angular momentum as it inflows (Fig.~\ref{f: before and after B}, panel B). We defer building idealized hot CGM models which satisfy this additional constraint to future work.

\subsection{Analogous accretion in cosmic ray-dominated haloes}
\label{s: crs}

The dark grey line in Fig.~\ref{f: before and after combined} shows the median properties of accreting gas in a simulation that includes cosmic ray physics~\citep[\texttt{m12i\_cr;}][]{Chan2019, Hopkins2020a}, thus providing a window into the differences in gas accretion modes in simulations with cosmic-ray (CR) support in the CGM.
The accretion onto the central galaxy in the cosmic ray-dominated halo is similar to rotating cooling flows in many aspects:
accreting gas gains coherence in the halo, becomes rotationally supported at the galaxy edge (with a decreased radial velocity), and subsequently collapses into a disk.
These properties were highlighted also by \cite{Trapp2021}, who performed an analysis on the accretion of gas onto MW-mass disk galaxies with CR-dominated haloes.
Consistent with our results, \citeauthor{Trapp2021} found that accretion onto MW-mass disks has the same qualitative behavior, regardless of whether or not the halo is CR-dominated.
However, the top left panel of Fig.~\ref{f: before and after combined} demonstrates that accreting gas in CR-dominated haloes never shocks to a temperature $T \sim T_{\rm vir} \sim 10^{5.5}$ K --- instead the median temperature for the CR-dominated halo is $T\approx10^4$ K.
This is because in \texttt{m12i\_cr} the CGM is not supported against gravity by thermal pressure, but rather by CR pressure \citep{Ji2020}.

Increased coherence and decreased radial velocity prior to accretion are the two properties of rotating cooling flows we identify as promoting thin disk formation (\S\ref{s: why CFs thin disks}), and are clearly present in \texttt{m12i\_cr} (Fig.~\ref{f: before and after combined}).
Consistent with this, \texttt{m12i\_cr} has a high thin disk fraction, $\fthin=0.9$.
In \S\ref{s: why CFs thin disks} we argue that the coherent co-rotation and deceleration of rotating cooling flows is a result of its subsonic nature.
The accreting gas in the CR-dominated halo is also effectively subsonic --- gas velocities are below the effective sound speed $c_{s,{\rm eff}}$ and therefore have properties characteristic of subsonic gas.
The effective sound speed $c_{s,{\rm eff}}$ is defined as part of the local strong-coupling approximation used for the subgrid cosmic ray physics prescription, wherein the total pressure used to determine gas trajectories is a combination of hydrodynamic and CR pressure ($P = P_{\rm gas} + P_{\rm cr}$) and the sound speed is $c^2_{s,{\rm eff}}=c^2_{s}+\gamma_{\rm cr} P_{\rm cr}/\rho$~\citep{Hopkins2020a}.
Cool $\sim10^4$ K accreting gas can therefore act subsonic, and thus be conducive to thin disk formation, as long as $P_{\rm cr}$ is sufficiently large.
The qualitatively similar behavior between both rotating cooling flows and accretion in CR-dominated haloes may suggest that these accretion modes are a subset of a more general form of \textit{disk-conducive subsonic accretion}.

\subsection{Caveats}
\label{s: caveats}

In our MW-mass haloes the average SFR over $z=0-0.5$ is SFR $\approx 3-10\,M_\odot/$yr, while the observationally-based average SFR for $M_{\rm vir} \sim 10^{12} M_\odot$ haloes over the same redshift range is a lower SFR $\approx 0.7-6\,M_\odot/$yr~\citep{Behroozi2013}. 
Since one component that regulates the SFR is the gas accretion rate onto the galaxy, this may suggest that our simulations have higher accretion rates from the CGM onto the galaxy $\Mdot_{\rm CGM}$ than in real galaxies at the same mass scale. 
The CGM accretion rate in FIRE may be reduced if the cooling flow is disrupted by additional physics implemented in some other suites of FIRE simulations, including cosmic rays~\citep{Chan2019, Hopkins2020a,Hopkins2021d,Hopkins2021e} and AGN feedback \citep{Wellons22}.
It is however unclear whether additional processes such as these can disrupt cooling flows around MW-way mass galaxies.
For example, CR transport models remain highly uncertain, especially in the CGM~\citep{Hopkins2021,Quataert2021a, Quataert2021}, and it is not yet known how AGN feedback couples to halo gas.
Moreover, if AGN feedback is intermittent, it is possible that a rotating cooling flow could reform between bursts of AGN feedback, especially at small CGM radii as analyzed above where cooling and dynamical times are relatively short. 

An alternative solution to the elevated accretion rate problem is that the cooling flow is not disrupted, but just weaker than in FIRE, as expected if the CGM mass or metallicity are overpredicted in FIRE at $z\sim0$.
This follows since in a cooling flow $\Mdot_{\rm CGM}\propto M_{\rm CGM}^2 \Lambda(Z_{\rm CGM})$, where $M_{\rm CGM}$ and $Z_{\rm CGM}$ are the CGM mass and metallicity, respectively, while $\Lambda$ is the cooling function.
Thus, if the CGM mass is overpredicted by a factor of two then $\Mdot$ and the SFR would be overestimated by a factor of four.
Both the CGM mass and metallicity are a result of the integrated enrichment and depletion of the CGM by outflows over cosmic time, and thus are uncertain~\citep[e.g.,][]{Davies2021, Kelly2021}.

A lower CGM mass or metallicity may also help address the difference between the mass threshold for rotational support in FIRE and in observations. 
At $z\sim0$, FIRE galaxies have significant rotational support ($V_{\rm rot}/\sigma_z \gg 1$) above $M_\star\sim10^{10}\msun$, higher than in observations which find rotationally-supported galaxies above $M_\star\sim10^9\msun$ (\citealt{Wheeler16, El-Badry2018a, El-Badry2018}, see also \citealt{Peebles2020}).
A lower $M_{\rm CGM}$ or $Z_{\rm CGM}$ would imply a higher $\tcoolsh$ and $\tcoolsh/\tff$ for a given halo mass, thus decreasing the mass threshold for inner CGM virialization. The latter effect would cause the onset of cooling flows and the formation of thin disks to occur at lower halo masses than suggested by Fig.~\ref{f: prevalence vs galaxy properties}.
At $z\sim0$, a factor of a few increase in $\tcoolsh$ relative to that in FIRE would decrease the threshold halo mass in which the inner CGM virializes to $M_{\rm vir}\sim10^{11}\,\msun$ \citep{Stern2021}, corresponding to a threshold stellar mass for thin disk formation of $M_\star\sim10^9\,\msun$  similar to the observed value. 

This difference between observed and simulated CGM mass or metallicity, as well as the general picture of rotating cooling flows, can be tested by comparison with CGM observations. 
The analysis of \cite{Stern2019} suggests that a cooling flow model with $\Mdot_{\rm CGM}\approx1\msun$ yr$^{-1}$, roughly equal to the accretion rate needed to sustain star formation in the Galaxy, is consistent with X-ray absorption and emission observations in the Milky-Way CGM.
Also, in line with the above scenario where FIRE somewhat overpredicts $M_{\rm CGM}$, some estimates of the CGM mass based on X-ray observations find $M_{\rm CGM}/f_bM_{\rm vir} \approx 0.1$~\citep[][]{Li2018, Bregman2018}, where $f_b M_{\rm halo}$ is the cosmological baryon fraction multiplied by the halo mass.
This is lower than  $M_{\rm CGM}/f_bM_{\rm vir} \approx 0.2-0.4$ for similar-mass FIRE haloes~\citep{Hafen2019}.
However, other studies deduce a higher $M_{\rm CGM}/f_bM_{\rm halo}=0.3$ based on the same X-ray data~\citep{Faerman2020}, while \cite{Chan2021} find X-ray emission in FIRE is at the low end of the observed distribution when matching simulated and observed galaxies by their SFR.
We defer a more direct comparison of the predictions of the rotating cooling flow model realized in FIRE with observations to future work.

\section{Summary}
\label{s: conclusions}

In this paper we use the particle-tracking method developed in \cite{Hafen2019,Hafen2020} to study how gas accretes onto $z\sim0$ galaxies in the FIRE-2 cosmological zoom simulations~\citep{Hopkins2018}, focusing on Milky-Way mass galaxies in which stars form in a thin disk. 
Our main results are as follows.
\begin{enumerate}
    \item \textbf{Mechanics of rotating cooling flows at $z \sim 0$:}
    We find that gas accretion onto thin disk galaxies in FIRE is dominated by rotating cooling flows, wherein the hot $T\approx T_{\rm vir}$ CGM forms a subsonic and angular momentum-conserving inflow down to the galaxy-halo interface, at which it cools to $T\lesssim10^4$ K (Figs.~\ref{f: before and after A} and \ref{f: before and after B}).
    Cooling occurs at the radius where rotational support in the flow becomes substantial, and is concurrent with flattening of the flow, i.e.~the flow transitions rapidly from a quasi-spherical hot medium into a cool thin disk (Figs. \ref{f: overview} and \ref{f: theta vs t}).
    Prior to cooling and flattening, the hot flow decelerates and becomes coherently rotating with a narrow angular momentum distribution (Figs.~\ref{f: coherence -- on sky}-- ~\ref{f: prevalence - angular momentum}), properties which we attribute to the subsonic nature of the flow (\S~\ref{s: why CFs thin disks}).
    Our results thus extend classic cooling flow theory by demonstrating rotating cooling flows are a primary accretion mode for many cosmologically-simulated galaxies, and by further exploring the mechanics of cooling flows with angular momentum beyond previous studies \citep{Cowie1980, Stern2020}.
    Also, the inflowing nature of the hot CGM phase surrounding disk galaxies in FIRE is in contrast with the common assumption that the hot CGM is static  \citep[e.g.,][]{Maller2004,Sharma2012, Voit2017, Faerman2017,Faerman2020}.\\ 
    \item \textbf{Relation between rotating cooling flows and thin stellar disks:}
    We find that across a sample of \Nsample~galaxies spanning $10^8 M_\odot < M_\star < 10^{11} M_\odot$, there is strong correlation between the dominance of cooling flow accretion and the fraction of stars formed in a thin disk in the central galaxy (Figs.~\ref{f: theta vs t}--\ref{f: prevalence}, Fig~\ref{f: prevalence - angular momentum}).
    This expands the result of \cite{Stern2021}, which found a relation between the formation of disks and the virialization of the inner CGM in FIRE, where the latter is a necessary condition for the onset of cooling flows. 
    We theorize that rotating cooling flows are conducive to the formation of thin disks since these flows achieve coherent, purely rotating kinematics prior to cooling to $\lesssim10^4$ K, and thus before any star-formation in the accreting gas takes place (Fig.~\ref{f: before and after A}, \S\ref{s: why CFs thin disks}).
    If rotating cooling flows are indeed a requirement for forming thin disks, then we expect thin disks mainly in relatively massive star-forming galaxies in which the CGM has virialized or is composed of subsonic gas, and at low redshift where cold streams are not expected to dominate the accretion.
    These trends are qualitatively consistent with observations which shows that thin disk morphologies are prevalent mainly in high mass and low-redshift star forming galaxies \citep{Kassin2012, Simons2017}. 
\end{enumerate}

\section*{Acknowledgements}

ZH was supported by a Gary A. McCue postdoctoral fellowship at UC Irvine.
JS was supported by the Israel Science Foundation (grant No. 2584/21) and by the German Science Foundation via DIP grant STE 1869/2-1 GE 625/17-1. 
JSB was supported by NSF grant AST-1910346.
ABG was supported by an NSF-GRFP under grant DGE-1842165 and was additionally supported by NSF grants DGE-0948017 and DGE-145000.
SY was supported by NSF grant AST-1910346.
CAFG was supported by NSF through grants AST-1715216, AST-2108230,  and CAREER award AST-1652522; by NASA through grant 17-ATP17-0067; by STScI through grant HST-AR-16124.001-A; and by the Research Corporation for Science Advancement through a Cottrell Scholar Award.
DBF is supported by the Simons Foundation through the Flatiron Institute.
DAA was supported in part by NSF grants AST-2009687 and AST-2108944 and by the Flatiron Institute, which is supported by the Simons Foundation.
EQ was supported in part by a Simons Investigator grant from the Simons Foundation and NSF grant 2107872.
AW received support from: NSF grants CAREER 2045928 and 2107772; NASA ATP grant 80NSSC20K0513; HST grants AR-15809 and GO-15902 from STScI; a Scialog Award from the Heising-Simons Foundation; and a Hellman Fellowship.
MBK acknowledges support from NSF CAREER award AST-1752913, NSF grants AST-1910346 and AST-2108962, NASA grant NNX17AG29G, and HST-AR-15006, HST-AR-15809, HST-GO-15658, HST-GO-15901, HST-GO-15902, HST-AR-16159, and HST-GO-16226 from the Space Telescope Science Institute, which is operated by AURA, Inc., under NASA contract NAS5-26555.
JM gratefully acknowledges sabbatical leave support from Pomona College and the Harry and Grace Steele Foundation.
RF acknowledges financial support from the Swiss National Science Foundation (grant no PP00P2\_194814 and 200021\_188552).
TKC is supported by Science and Technology Facilities Council (STFC) astronomy consolidated grant ST/P000541/1 and ST/T000244/1.
CT and DK were supported by NSF grants AST-1715101 and  AST-2108314.
Numerical calculations were performed on the Quest computing cluster at Northwestern University, the Wheeler computing cluster at Caltech, XSEDE allocations TG-AST130039, TG-AST120025, TG-AST140064, and TG-AST140023, Blue Waters PRAC allocation NSF.1713353, NASA HEC allocation SMD16-7592, and allocations AST21010 and AST20016 supported by the NSF and TACC.
This research benefited from the Halo21 KITP workshop which was supported in part by the National Science Foundation under Grant No.~NSF PHY-1748958.

This research used the Python programming language and the following modules:
Firefly~\citep{Geller2018},
Numpy~\citep{Harris2020},
Matplotlib~\citep{Hunter2007},
pytest~\citep{pytest3.4},
Jug~\citep{Coelho2017},
h5py~\citep{h5py},
SciPy~\citep{Virtanen2020},
pandas~\citep{McKinney2010,Reback2020},
palettable~(\url{https://github.com/jiffyclub/palettable}),
and Numba~\citep{Lam2015}.

\section*{Data availability}
The data underlying this article will be shared on reasonable request to the corresponding author (ZH). The simulation initial conditions, snapshot files, and visualization can be found in \href{https://fire.northwestern.edu/data/}{https://fire.northwestern.edu/data/}.
A public version of the GIZMO simulation code is available \href{http://www.tapir.caltech.edu/~phopkins/Site/GIZMO.html}{http://www.tapir.caltech.edu/~phopkins/Site/GIZMO.html}.




\bibliographystyle{mnras}
\bibliography{references,jsbref} 

\begin{thebibliography}{}
\makeatletter
\relax
\def\mn@urlcharsother{\let\do\@makeother \do\$\do\&\do\#\do\^\do\_\do\%\do\~}
\def\mn@doi{\begingroup\mn@urlcharsother \@ifnextchar [ {\mn@doi@}
  {\mn@doi@[]}}
\def\mn@doi@[#1]#2{\def\@tempa{#1}\ifx\@tempa\@empty \href
  {http://dx.doi.org/#2} {doi:#2}\else \href {http://dx.doi.org/#2} {#1}\fi
  \endgroup}
\def\mn@eprint#1#2{\mn@eprint@#1:#2::\@nil}
\def\mn@eprint@arXiv#1{\href {http://arxiv.org/abs/#1} {{\tt arXiv:#1}}}
\def\mn@eprint@dblp#1{\href {http://dblp.uni-trier.de/rec/bibtex/#1.xml}
  {dblp:#1}}
\def\mn@eprint@#1:#2:#3:#4\@nil{\def\@tempa {#1}\def\@tempb {#2}\def\@tempc
  {#3}\ifx \@tempc \@empty \let \@tempc \@tempb \let \@tempb \@tempa \fi \ifx
  \@tempb \@empty \def\@tempb {arXiv}\fi \@ifundefined
  {mn@eprint@\@tempb}{\@tempb:\@tempc}{\expandafter \expandafter \csname
  mn@eprint@\@tempb\endcsname \expandafter{\@tempc}}}

\bibitem[\protect\citeauthoryear{{Angl{\'e}s-Alc{\'a}zar},
  {Faucher-Gigu{\`e}re}, Kere{\v s}, Hopkins, Quataert  \&
  Murray}{{Angl{\'e}s-Alc{\'a}zar} et~al.}{2017}]{Angles-Alcazar2017}
{Angl{\'e}s-Alc{\'a}zar} D.,  {Faucher-Gigu{\`e}re} C.-A.,  Kere{\v s} D.,
  Hopkins P.~F.,  Quataert E.,   Murray N.,  2017, \mn@doi [Monthly Notices of
  the Royal Astronomical Society] {10.1093/mnras/stx1517}, 470, 4698

\bibitem[\protect\citeauthoryear{Armillotta, Fraternali  \&
  Marinacci}{Armillotta et~al.}{2016}]{Armillotta2016}
Armillotta L.,  Fraternali F.,   Marinacci F.,  2016, \mn@doi [Monthly Notices
  of the Royal Astronomical Society] {10.1093/mnras/stw1930}, 462, 4157

\bibitem[\protect\citeauthoryear{Balbus}{Balbus}{1988}]{Balbus1988}
Balbus S.~A.,  1988, \mn@doi [The Astrophysical Journal] {10.1086/166301}, 328,
  395

\bibitem[\protect\citeauthoryear{Balbus \& Soker}{Balbus \&
  Soker}{1989}]{Balbus1989}
Balbus S.~A.,  Soker N.,  1989, The Astrophysical Journal, 341, 611

\bibitem[\protect\citeauthoryear{{Barnes} \& {Efstathiou}}{{Barnes} \&
  {Efstathiou}}{1987}]{Barnes87}
{Barnes} J.,  {Efstathiou} G.,  1987, \mn@doi [\apj] {10.1086/165480}, \href
  {https://ui.adsabs.harvard.edu/abs/1987ApJ...319..575B} {319, 575}

\bibitem[\protect\citeauthoryear{Behroozi, Wechsler  \& Conroy}{Behroozi
  et~al.}{2013}]{Behroozi2013}
Behroozi P.~S.,  Wechsler R.~H.,   Conroy C.,  2013, \mn@doi [The Astrophysical
  Journal] {10.1088/0004-637X/770/1/57}, 770, 57

\bibitem[\protect\citeauthoryear{Benincasa et~al.,}{Benincasa
  et~al.}{2020}]{Benincasa2020}
Benincasa S.~M.,  et~al., 2020, \mn@doi [Monthly Notices of the Royal
  Astronomical Society] {10.1093/mnras/staa2116}, 497, 3993

\bibitem[\protect\citeauthoryear{{Bernardi}, {Shankar}, {Hyde}, {Mei},
  {Marulli}  \& {Sheth}}{{Bernardi} et~al.}{2010}]{Bernardi2010}
{Bernardi} M.,  {Shankar} F.,  {Hyde} J.~B.,  {Mei} S.,  {Marulli} F.,
  {Sheth} R.~K.,  2010, \mn@doi [\mnras] {10.1111/j.1365-2966.2010.16425.x},
  \href {https://ui.adsabs.harvard.edu/abs/2010MNRAS.404.2087B} {404, 2087}

\bibitem[\protect\citeauthoryear{Bertschinger}{Bertschinger}{1989}]{Bertschinger1989}
Bertschinger E.,  1989, \mn@doi [The Astrophysical Journal] {10.1086/167428},
  340, 666

\bibitem[\protect\citeauthoryear{Bigiel, Leroy, Walter, Brinks, {de Blok},
  Madore  \& Thornley}{Bigiel et~al.}{2008}]{Bigiel2008}
Bigiel F.,  Leroy A.,  Walter F.,  Brinks E.,  {de Blok} W. J.~G.,  Madore B.,
   Thornley M.~D.,  2008, \mn@doi [The Astronomical Journal]
  {10.1088/0004-6256/136/6/2846}, 136, 2846

\bibitem[\protect\citeauthoryear{Birnboim \& Dekel}{Birnboim \&
  Dekel}{2003}]{Birnboim2003}
Birnboim Y.,  Dekel A.,  2003, \mn@doi [Monthly Notices of the Royal
  Astronomical Society] {10.1046/j.1365-8711.2003.06955.x}, 345, 349

\bibitem[\protect\citeauthoryear{Bizyaev, Makarov, Reshetnikov, Mosenkov,
  Kautsch  \& Antipova}{Bizyaev et~al.}{2021}]{Bizyaev2021}
Bizyaev D.,  Makarov D.~I.,  Reshetnikov V.~P.,  Mosenkov A.~V.,  Kautsch
  S.~J.,   Antipova A.~V.,  2021, arXiv:2105.11855 [astro-ph]

\bibitem[\protect\citeauthoryear{Bluck, Mendel, Ellison, Moreno, Simard, Patton
   \& Starkenburg}{Bluck et~al.}{2014}]{Bluck2014}
Bluck A. F.~L.,  Mendel J.~T.,  Ellison S.~L.,  Moreno J.,  Simard L.,  Patton
  D.~R.,   Starkenburg E.,  2014, \mn@doi [Monthly Notices of the Royal
  Astronomical Society] {10.1093/mnras/stu594}, 441, 599

\bibitem[\protect\citeauthoryear{Bregman, Anderson, Miller, {Hodges-Kluck},
  Dai, Li, Li  \& Qu}{Bregman et~al.}{2018}]{Bregman2018}
Bregman J.~N.,  Anderson M.~E.,  Miller M.~J.,  {Hodges-Kluck} E.,  Dai X.,  Li
  J.-T.,  Li Y.,   Qu Z.,  2018, \mn@doi [The Astrophysical Journal]
  {10.3847/1538-4357/aacafe}, 862, 3

\bibitem[\protect\citeauthoryear{Brook et~al.,}{Brook et~al.}{2011}]{Brook2011}
Brook C.~B.,  et~al., 2011, \mn@doi [Monthly Notices of the Royal Astronomical
  Society] {10.1111/j.1365-2966.2011.18545.x}, 415, 1051

\bibitem[\protect\citeauthoryear{Bullock, Dekel, Kolatt, Kravtsov, Klypin,
  Porciani  \& Primack}{Bullock et~al.}{2001}]{Bullock2001}
Bullock J.~S.,  Dekel A.,  Kolatt T.~S.,  Kravtsov A.~V.,  Klypin A.~A.,
  Porciani C.,   Primack J.~R.,  2001, \mn@doi [The Astrophysical Journal]
  {10.1086/321477}, 555, 240

\bibitem[\protect\citeauthoryear{Chan, Kere{\v s}, Wetzel, Hopkins,
  {Faucher-Gigu{\`e}re}, {El-Badry}, {Garrison-Kimmel}  \&
  {Boylan-Kolchin}}{Chan et~al.}{2018}]{Chan2018}
Chan T.~K.,  Kere{\v s} D.,  Wetzel A.,  Hopkins P.~F.,  {Faucher-Gigu{\`e}re}
  C.-A.,  {El-Badry} K.,  {Garrison-Kimmel} S.,   {Boylan-Kolchin} M.,  2018,
  \mn@doi [Monthly Notices of the Royal Astronomical Society]
  {10.1093/mnras/sty1153}, 478, 906

\bibitem[\protect\citeauthoryear{Chan, Kere{\v s}, Hopkins, Quataert, Su,
  Hayward  \& {Faucher-Gigu{\`e}re}}{Chan et~al.}{2019}]{Chan2019}
Chan T.~K.,  Kere{\v s} D.,  Hopkins P.~F.,  Quataert E.,  Su K.-Y.,  Hayward
  C.~C.,   {Faucher-Gigu{\`e}re} C.-A.,  2019, \mn@doi [Monthly Notices of the
  Royal Astronomical Society] {10.1093/mnras/stz1895}, 488, 3716

\bibitem[\protect\citeauthoryear{Chan, Keres, Gurvich, Hopkins, Trapp, Ji  \&
  {Faucher-Giguere}}{Chan et~al.}{2021}]{Chan2021}
Chan T.~K.,  Keres D.,  Gurvich A.~B.,  Hopkins P.,  Trapp C.,  Ji S.,
  {Faucher-Giguere} C.-A.,  2021, arXiv:2110.06231 [astro-ph]

\bibitem[\protect\citeauthoryear{Coelho}{Coelho}{2017}]{Coelho2017}
Coelho L.~P.,  2017, \mn@doi [Journal of Open Research Software]
  {10.5334/jors.161}, 5, 30

\bibitem[\protect\citeauthoryear{Collette}{Collette}{2013}]{h5py}
Collette A.,  2013, Python and HDF5.
O'Reilly

\bibitem[\protect\citeauthoryear{Cowie, Fabian  \& Nulsen}{Cowie
  et~al.}{1980}]{Cowie1980}
Cowie L.~L.,  Fabian A.~C.,   Nulsen P. E.~J.,  1980, \mn@doi [Monthly Notices
  of the Royal Astronomical Society] {10.1093/mnras/191.2.399}, 191, 399

\bibitem[\protect\citeauthoryear{Danovich, Dekel, Hahn  \& Teyssier}{Danovich
  et~al.}{2012}]{Danovich2012}
Danovich M.,  Dekel A.,  Hahn O.,   Teyssier R.,  2012, \mn@doi [Monthly
  Notices of the Royal Astronomical Society]
  {10.1111/j.1365-2966.2012.20751.x}, 422, 1732

\bibitem[\protect\citeauthoryear{Danovich, Dekel, Hahn, Ceverino  \&
  Primack}{Danovich et~al.}{2015}]{Danovich2015}
Danovich M.,  Dekel A.,  Hahn O.,  Ceverino D.,   Primack J.,  2015, \mn@doi
  [Monthly Notices of the Royal Astronomical Society] {10.1093/mnras/stv270},
  449, 2087

\bibitem[\protect\citeauthoryear{Davies, Crain  \& Pontzen}{Davies
  et~al.}{2021}]{Davies2021}
Davies J.~J.,  Crain R.~A.,   Pontzen A.,  2021, \mn@doi [Monthly Notices of
  the Royal Astronomical Society] {10.1093/mnras/staa3643}, 501, 236

\bibitem[\protect\citeauthoryear{DeFelippis, Genel, Bryan  \& Fall}{DeFelippis
  et~al.}{2017}]{DeFelippis2017}
DeFelippis D.,  Genel S.,  Bryan G.~L.,   Fall S.~M.,  2017, \mn@doi [The
  Astrophysical Journal] {10.3847/1538-4357/aa6dfc}, 841, 16

\bibitem[\protect\citeauthoryear{DeFelippis, Genel, Bryan, Nelson, Pillepich
  \& Hernquist}{DeFelippis et~al.}{2020}]{DeFelippis2020}
DeFelippis D.,  Genel S.,  Bryan G.~L.,  Nelson D.,  Pillepich A.,   Hernquist
  L.,  2020, \mn@doi [ApJ] {10.3847/1538-4357/ab8a4a}, 895, 17

\bibitem[\protect\citeauthoryear{Debattista, Gonzalez, Sanderson, {El-Badry},
  {Garrison-Kimmel}, Wetzel, {Faucher-Gigu{\`e}re}  \& Hopkins}{Debattista
  et~al.}{2019}]{Debattista2019}
Debattista V.~P.,  Gonzalez O.~A.,  Sanderson R.~E.,  {El-Badry} K.,
  {Garrison-Kimmel} S.,  Wetzel A.,  {Faucher-Gigu{\`e}re} C.-A.,   Hopkins
  P.~F.,  2019, \mn@doi [Monthly Notices of the Royal Astronomical Society]
  {10.1093/mnras/stz746}, 485, 5073

\bibitem[\protect\citeauthoryear{Dekel \& Birnboim}{Dekel \&
  Birnboim}{2006}]{Dekel2006}
Dekel A.,  Birnboim Y.,  2006, \mn@doi [Monthly Notices of the Royal
  Astronomical Society] {10.1111/j.1365-2966.2006.10145.x}, 368, 2

\bibitem[\protect\citeauthoryear{Dekel et~al.,}{Dekel et~al.}{2009}]{Dekel2009}
Dekel A.,  et~al., 2009, \mn@doi [Nature] {10.1038/nature07648}, 457, 451

\bibitem[\protect\citeauthoryear{Dekel, Lapiner  \& Dubois}{Dekel
  et~al.}{2019}]{Dekel2019}
Dekel A.,  Lapiner S.,   Dubois Y.,  2019, arXiv:1904.08431

\bibitem[\protect\citeauthoryear{{El-Badry} et~al.,}{{El-Badry}
  et~al.}{2018a}]{El-Badry2018a}
{El-Badry} K.,  et~al., 2018a, \mn@doi [Monthly Notices of the Royal
  Astronomical Society] {10.1093/mnras/stx2482}, 473, 1930

\bibitem[\protect\citeauthoryear{{El-Badry} et~al.,}{{El-Badry}
  et~al.}{2018b}]{El-Badry2018}
{El-Badry} K.,  et~al., 2018b, \mn@doi [Monthly Notices of the Royal
  Astronomical Society] {10.1093/mnras/sty730}, 477, 1536

\bibitem[\protect\citeauthoryear{Esmerian, Kravtsov, Hafen,
  {Faucher-Gigu{\`e}re}, Quataert, Stern, Kere{\v s}  \& Wetzel}{Esmerian
  et~al.}{2021}]{Esmerian2021}
Esmerian C.~J.,  Kravtsov A.~V.,  Hafen Z.,  {Faucher-Gigu{\`e}re} C.-A.,
  Quataert E.,  Stern J.,  Kere{\v s} D.,   Wetzel A.,  2021, \mn@doi [Monthly
  Notices of the Royal Astronomical Society] {10.1093/mnras/stab1281}, 505,
  1841

\bibitem[\protect\citeauthoryear{Fabian, Nulsen  \& Canizares}{Fabian
  et~al.}{1984}]{Fabian1984}
Fabian A.~C.,  Nulsen P. E.~J.,   Canizares C.~R.,  1984, \mn@doi [Nature]
  {10.1038/310733a0}, 310, 733

\bibitem[\protect\citeauthoryear{Faerman, Sternberg  \& McKee}{Faerman
  et~al.}{2017}]{Faerman2017}
Faerman Y.,  Sternberg A.,   McKee C.~F.,  2017, \mn@doi [The Astrophysical
  Journal] {10.3847/1538-4357/835/1/52}, 835, 52

\bibitem[\protect\citeauthoryear{Faerman, Sternberg  \& McKee}{Faerman
  et~al.}{2020}]{Faerman2020}
Faerman Y.,  Sternberg A.,   McKee C.~F.,  2020, \mn@doi [ApJ]
  {10.3847/1538-4357/ab7ffc}, 893, 82

\bibitem[\protect\citeauthoryear{Fall \& Efstathiou}{Fall \&
  Efstathiou}{1980}]{fall1980}
Fall S.~M.,  Efstathiou G.,  1980, \mn@doi [Monthly Notices of the Royal
  Astronomical Society] {10.1093/mnras/193.2.189}, 193, 189

\bibitem[\protect\citeauthoryear{{Fall} \& {Rees}}{{Fall} \&
  {Rees}}{1985}]{Fall85}
{Fall} S.~M.,  {Rees} M.~J.,  1985, \mn@doi [\apj] {10.1086/163585}, \href
  {https://ui.adsabs.harvard.edu/abs/1985ApJ...298...18F} {298, 18}

\bibitem[\protect\citeauthoryear{{Faucher-Gigu{\`e}re}, Lidz, Zaldarriaga  \&
  Hernquist}{{Faucher-Gigu{\`e}re} et~al.}{2009}]{Faucher-Giguere2009}
{Faucher-Gigu{\`e}re} C.,  Lidz A.,  Zaldarriaga M.,   Hernquist L.,  2009,
  \mn@doi [The Astrophysical Journal] {10.1088/0004-637X/703/2/1416}, 703, 1416

\bibitem[\protect\citeauthoryear{{Faucher-Gigu{\`e}re}, Kere{\v s}  \&
  Ma}{{Faucher-Gigu{\`e}re} et~al.}{2011}]{Faucher-Giguere2011a}
{Faucher-Gigu{\`e}re} C.~A.,  Kere{\v s} D.,   Ma C.~P.,  2011, \mn@doi
  [Monthly Notices of the Royal Astronomical Society]
  {10.1111/j.1365-2966.2011.19457.x}, 417, 2982

\bibitem[\protect\citeauthoryear{{Faucher-Gigu{\`e}re}, Quataert  \&
  Hopkins}{{Faucher-Gigu{\`e}re} et~al.}{2013}]{Faucher-Giguere2013}
{Faucher-Gigu{\`e}re} C.-A.,  Quataert E.,   Hopkins P.~F.,  2013, \mn@doi
  [Monthly Notices of the Royal Astronomical Society] {10.1093/mnras/stt866},
  433, 1970

\bibitem[\protect\citeauthoryear{{Faucher-Gigu{\`e}re}, Hopkins, Kere, Muratov,
  Quataert  \& Murray}{{Faucher-Gigu{\`e}re}
  et~al.}{2015}]{Faucher-Giguere2015}
{Faucher-Gigu{\`e}re} C.-A.,  Hopkins P.~F.,  Kere D.,  Muratov A.~L.,
  Quataert E.,   Murray N.,  2015, \mn@doi [Monthly Notices of the Royal
  Astronomical Society] {10.1093/mnras/stv336}, 449, 987

\bibitem[\protect\citeauthoryear{{Faucher-Gigu{\`e}re}, Feldmann, Quataert,
  Kere{\v s}, Hopkins  \& Murray}{{Faucher-Gigu{\`e}re}
  et~al.}{2016}]{Faucher-Giguere2016}
{Faucher-Gigu{\`e}re} C.-A.,  Feldmann R.,  Quataert E.,  Kere{\v s} D.,
  Hopkins P.~F.,   Murray N.,  2016, \mn@doi [Monthly Notices of the Royal
  Astronomical Society: Letters] {10.1093/mnrasl/slw091}, 461, L32

\bibitem[\protect\citeauthoryear{Fielding, Quataert, McCourt  \&
  Thompson}{Fielding et~al.}{2017}]{Fielding2017}
Fielding D.,  Quataert E.,  McCourt M.,   Thompson T.~A.,  2017, \mn@doi [Mon.
  Not. R. Astron. Soc.] {10.1093/mnras/stw3326}, 466, 3810

\bibitem[\protect\citeauthoryear{Fielding, Ostriker, Bryan  \& Jermyn}{Fielding
  et~al.}{2020}]{Fielding2020}
Fielding D.~B.,  Ostriker E.~C.,  Bryan G.~L.,   Jermyn A.~S.,  2020, \mn@doi
  [The Astrophysical Journal] {10.3847/2041-8213/ab8d2c}, 894, L24

\bibitem[\protect\citeauthoryear{{Garrison-Kimmel} et~al.,}{{Garrison-Kimmel}
  et~al.}{2017}]{Garrison-Kimmel2017}
{Garrison-Kimmel} S.,  et~al., 2017, \mn@doi [Monthly Notices of the Royal
  Astronomical Society] {10.1093/mnras/stx1710}, 471, 1709

\bibitem[\protect\citeauthoryear{{Garrison-Kimmel} et~al.,}{{Garrison-Kimmel}
  et~al.}{2018b}]{Garrison-Kimmel2018}
{Garrison-Kimmel} S.,  et~al., 2018b, \mn@doi [Monthly Notices of the Royal
  Astronomical Society] {10.1093/mnras/sty2513}, 481, 4133

\bibitem[\protect\citeauthoryear{{Garrison-Kimmel} et~al.,}{{Garrison-Kimmel}
  et~al.}{2018a}]{GK18}
{Garrison-Kimmel} S.,  et~al., 2018a, \mn@doi [\mnras] {10.1093/mnras/sty2513},
  \href {https://ui.adsabs.harvard.edu/abs/2018MNRAS.481.4133G} {481, 4133}

\bibitem[\protect\citeauthoryear{{Garrison-Kimmel} et~al.,}{{Garrison-Kimmel}
  et~al.}{2019}]{Garrison-Kimmel2019a}
{Garrison-Kimmel} S.,  et~al., 2019, \mn@doi [Monthly Notices of the Royal
  Astronomical Society] {10.1093/mnras/stz1317}, 487, 1380

\bibitem[\protect\citeauthoryear{Geller \& Gurvich}{Geller \&
  Gurvich}{2018}]{Geller2018}
Geller A.~M.,  Gurvich A.,  2018, Astrophysics Source Code Library, p.
  ascl:1810.021

\bibitem[\protect\citeauthoryear{Genel, Fall, Hernquist, Vogelsberger, Snyder,
  {Rodriguez-Gomez}, Sijacki  \& Springel}{Genel et~al.}{2015}]{Genel2015}
Genel S.,  Fall S.~M.,  Hernquist L.,  Vogelsberger M.,  Snyder G.~F.,
  {Rodriguez-Gomez} V.,  Sijacki D.,   Springel V.,  2015, \mn@doi [The
  Astrophysical Journal] {10.1088/2041-8205/804/2/L40}, 804, L40

\bibitem[\protect\citeauthoryear{Gronke \& Oh}{Gronke \&
  Oh}{2020}]{Gronke2020b}
Gronke M.,  Oh S.~P.,  2020, \mn@doi [Monthly Notices of the Royal Astronomical
  Society] {10.1093/mnrasl/slaa033}, 494, L27

\bibitem[\protect\citeauthoryear{Guszejnov, Grudi{\'c}, Offner,
  {Boylan-Kolchin}, {Faucher-Gigu{\`e}re}, Wetzel, Benincasa  \&
  Loebman}{Guszejnov et~al.}{2020}]{Guszejnov2020b}
Guszejnov D.,  Grudi{\'c} M.~Y.,  Offner S. S.~R.,  {Boylan-Kolchin} M.,
  {Faucher-Gigu{\`e}re} C.-A.,  Wetzel A.,  Benincasa S.~M.,   Loebman S.,
  2020, \mn@doi [Monthly Notices of the Royal Astronomical Society]
  {10.1093/mnras/stz3527}, 492, 488

\bibitem[\protect\citeauthoryear{Hafen et~al.,}{Hafen et~al.}{2019}]{Hafen2019}
Hafen Z.,  et~al., 2019, \mn@doi [Monthly Notices of the Royal Astronomical
  Society] {10.1093/mnras/stz1773}, 488, 1248

\bibitem[\protect\citeauthoryear{Hafen et~al.,}{Hafen et~al.}{2020}]{Hafen2020}
Hafen Z.,  et~al., 2020, \mn@doi [Monthly Notices of the Royal Astronomical
  Society] {10.1093/mnras/staa902}, 494, 3581

\bibitem[\protect\citeauthoryear{Harris et~al.,}{Harris
  et~al.}{2020}]{Harris2020}
Harris C.~R.,  et~al., 2020, \mn@doi [Nature] {10.1038/s41586-020-2649-2}, 585,
  357

\bibitem[\protect\citeauthoryear{Heckman \& Thompson}{Heckman \&
  Thompson}{2019}]{Heckman2019}
Heckman T.~M.,  Thompson T.~A.,  2019, arXiv:1701.09062 [astro-ph]

\bibitem[\protect\citeauthoryear{{Hodges-Kluck}, Miller  \&
  Bregman}{{Hodges-Kluck} et~al.}{2016}]{Hodges-Kluck2016}
{Hodges-Kluck} E.~J.,  Miller M.~J.,   Bregman J.~N.,  2016, \mn@doi [The
  Astrophysical Journal] {10.3847/0004-637X/822/1/21}, 822, 21

\bibitem[\protect\citeauthoryear{Hopkins}{Hopkins}{2015}]{Hopkins2015}
Hopkins P.~F.,  2015, \mn@doi [Monthly Notices of the Royal Astronomical
  Society] {10.1093/mnras/stv195}, 450, 53

\bibitem[\protect\citeauthoryear{Hopkins}{Hopkins}{2017}]{Hopkins2017}
Hopkins P.~F.,  2017, \mn@doi [Monthly Notices of the Royal Astronomical
  Society] {10.1093/mnras/stw3306}, 466, 3387

\bibitem[\protect\citeauthoryear{Hopkins, Keres, Onorbe, {Faucher-Giguere},
  Quataert, Murray  \& Bullock}{Hopkins et~al.}{2014}]{Hopkins2014}
Hopkins P.~F.,  Keres D.,  Onorbe J.,  {Faucher-Giguere} C.-A.,  Quataert E.,
  Murray N.,   Bullock J.~S.,  2014, \mn@doi [Monthly Notices of the Royal
  Astronomical Society] {10.1093/mnras/stu1738}, 445, 581

\bibitem[\protect\citeauthoryear{Hopkins et~al.,}{Hopkins
  et~al.}{2018}]{Hopkins2018}
Hopkins P.~F.,  et~al., 2018, \mn@doi [Monthly Notices of the Royal
  Astronomical Society] {10.1093/mnras/sty1690}, 480, 800

\bibitem[\protect\citeauthoryear{Hopkins et~al.,}{Hopkins
  et~al.}{2020}]{Hopkins2020a}
Hopkins P.~F.,  et~al., 2020, \mn@doi [Monthly Notices of the Royal
  Astronomical Society] {10.1093/mnras/stz3321}, 492, 3465

\bibitem[\protect\citeauthoryear{Hopkins, Squire, Chan, Quataert, Ji, Keres  \&
  {Faucher-Giguere}}{Hopkins et~al.}{2021a}]{Hopkins2021d}
Hopkins P.~F.,  Squire J.,  Chan T.~K.,  Quataert E.,  Ji S.,  Keres D.,
  {Faucher-Giguere} C.-A.,  2021a, \mn@doi [Monthly Notices of the Royal
  Astronomical Society] {10.1093/mnras/staa3691}, 501, 4184

\bibitem[\protect\citeauthoryear{Hopkins, Chan, Squire, Quataert, Ji, Kere{\v
  s}  \& {Faucher-Gigu{\`e}re}}{Hopkins et~al.}{2021b}]{Hopkins2021e}
Hopkins P.~F.,  Chan T.~K.,  Squire J.,  Quataert E.,  Ji S.,  Kere{\v s} D.,
  {Faucher-Gigu{\`e}re} C.-A.,  2021b, \mn@doi [Monthly Notices of the Royal
  Astronomical Society] {10.1093/mnras/staa3692}, 501, 3663

\bibitem[\protect\citeauthoryear{Hopkins, Chan, Ji, Hummels, Kere{\v s},
  Quataert  \& {Faucher-Gigu{\`e}re}}{Hopkins et~al.}{2021c}]{Hopkins2021}
Hopkins P.~F.,  Chan T.~K.,  Ji S.,  Hummels C.~B.,  Kere{\v s} D.,  Quataert
  E.,   {Faucher-Gigu{\`e}re} C.-A.,  2021c, \mn@doi [Monthly Notices of the
  Royal Astronomical Society] {10.1093/mnras/staa3690}, 501, 3640

\bibitem[\protect\citeauthoryear{Hunter}{Hunter}{2007}]{Hunter2007}
Hunter J.~D.,  2007, \mn@doi [Computing in Science Engineering]
  {10.1109/MCSE.2007.55}, 9, 90

\bibitem[\protect\citeauthoryear{Huscher, Oppenheimer, Lonardi, Crain, Richings
   \& Schaye}{Huscher et~al.}{2021}]{Huscher2021}
Huscher E.,  Oppenheimer B.~D.,  Lonardi A.,  Crain R.~A.,  Richings A.~J.,
  Schaye J.,  2021, \mn@doi [Monthly Notices of the Royal Astronomical Society]
  {10.1093/mnras/staa3203}, 500, 1476

\bibitem[\protect\citeauthoryear{Ji et~al.,}{Ji et~al.}{2020}]{Ji2020}
Ji S.,  et~al., 2020, \mn@doi [Monthly Notices of the Royal Astronomical
  Society] {10.1093/mnras/staa1849}, 496, 4221

\bibitem[\protect\citeauthoryear{Joung, Putman, Bryan, Fern{\'a}ndez  \&
  Peek}{Joung et~al.}{2012}]{Joung2012}
Joung M.~R.,  Putman M.~E.,  Bryan G.~L.,  Fern{\'a}ndez X.,   Peek J. E.~G.,
  2012, \mn@doi [The Astrophysical Journal] {10.1088/0004-637X/759/2/137}, 759,
  137

\bibitem[\protect\citeauthoryear{Kassin, {de Jong}  \& Weiner}{Kassin
  et~al.}{2006}]{Kassin2006}
Kassin S.~A.,  {de Jong} R.~S.,   Weiner B.~J.,  2006, \mn@doi [The
  Astrophysical Journal] {10.1086/502959}, 643, 804

\bibitem[\protect\citeauthoryear{Kassin, Devriendt, Fall, {de Jong}, Allgood
  \& Primack}{Kassin et~al.}{2012a}]{Kassin2012}
Kassin S.~A.,  Devriendt J.,  Fall S.~M.,  {de Jong} R.~S.,  Allgood B.,
  Primack J.~R.,  2012a, \mn@doi [Monthly Notices of the Royal Astronomical
  Society] {10.1111/j.1365-2966.2012.21219.x}, 424, 502

\bibitem[\protect\citeauthoryear{Kassin et~al.,}{Kassin
  et~al.}{2012b}]{Kassin2012a}
Kassin S.~A.,  et~al., 2012b, \mn@doi [The Astrophysical Journal]
  {10.1088/0004-637X/758/2/106}, 758, 106

\bibitem[\protect\citeauthoryear{Kelly, Jenkins, Deason, Fattahi, Grand,
  Pakmor, Springel  \& Frenk}{Kelly et~al.}{2021}]{Kelly2021}
Kelly A.~J.,  Jenkins A.,  Deason A.,  Fattahi A.,  Grand R. J.~J.,  Pakmor R.,
   Springel V.,   Frenk C.~S.,  2021, arXiv:2106.08618 [astro-ph]

\bibitem[\protect\citeauthoryear{Kere{\v s} \& Hernquist}{Kere{\v s} \&
  Hernquist}{2009}]{Keres2009b}
Kere{\v s} D.,  Hernquist L.,  2009, \mn@doi [The Astrophysical Journal]
  {10.1088/0004-637X/700/1/L1}, 700, L1

\bibitem[\protect\citeauthoryear{Kere{\v s}, Katz, Weinberg  \&
  Dav{\'e}}{Kere{\v s} et~al.}{2005}]{Keres2005}
Kere{\v s} D.,  Katz N.,  Weinberg D.~H.,   Dav{\'e} R.,  2005, \mn@doi
  [Monthly Notices of the Royal Astronomical Society]
  {10.1111/j.1365-2966.2005.09451.x}, 363, 2

\bibitem[\protect\citeauthoryear{Kere{\v s}, Katz, Fardal, Dav{\'e}  \&
  Weinberg}{Kere{\v s} et~al.}{2009a}]{Keres2009a}
Kere{\v s} D.,  Katz N.,  Fardal M.,  Dav{\'e} R.,   Weinberg D.~H.,  2009a,
  \mn@doi [Monthly Notices of the Royal Astronomical Society]
  {10.1111/j.1365-2966.2009.14541.x}, 395, 160

\bibitem[\protect\citeauthoryear{Kere{\v s}, Katz, Dav{\'e}, Fardal  \&
  Weinberg}{Kere{\v s} et~al.}{2009b}]{Keres2009}
Kere{\v s} D.,  Katz N.,  Dav{\'e} R.,  Fardal M.,   Weinberg D.~H.,  2009b,
  \mn@doi [Monthly Notices of the Royal Astronomical Society]
  {10.1111/j.1365-2966.2009.14924.x}, 396, 2332

\bibitem[\protect\citeauthoryear{Kranz, Slyz  \& Rix}{Kranz
  et~al.}{2003}]{Kranz2003}
Kranz T.,  Slyz A.,   Rix H.-W.,  2003, \mn@doi [The Astrophysical Journal]
  {10.1086/367551}, 586, 143

\bibitem[\protect\citeauthoryear{Kregel, Van Der~Kruit  \& De~Grijs}{Kregel
  et~al.}{2002}]{Kregel2002}
Kregel M.,  Van Der~Kruit P.~C.,   De~Grijs R.,  2002, \mn@doi [Monthly Notices
  of the Royal Astronomical Society] {10.1046/j.1365-8711.2002.05556.x}, 334,
  646

\bibitem[\protect\citeauthoryear{Krekel, Oliveira, Pfannschmidt, Bruynooghe,
  Laugher  \& Bruhin}{Krekel et~al.}{2004}]{pytest3.4}
Krekel H.,  Oliveira B.,  Pfannschmidt R.,  Bruynooghe F.,  Laugher B.,
  Bruhin F.,  2004, pytest 3.4, \url {https://github.com/pytest-dev/pytest}

\bibitem[\protect\citeauthoryear{Krumholz, Burkhart, Forbes  \&
  Crocker}{Krumholz et~al.}{2018}]{Krumholz2018}
Krumholz M.~R.,  Burkhart B.,  Forbes J.~C.,   Crocker R.~M.,  2018, \mn@doi
  [Monthly Notices of the Royal Astronomical Society] {10.1093/mnras/sty852},
  477, 2716

\bibitem[\protect\citeauthoryear{Lam, Pitrou  \& Seibert}{Lam
  et~al.}{2015}]{Lam2015}
Lam S.~K.,  Pitrou A.,   Seibert S.,  2015, Proceedings of the Second Workshop
  on the LLVM Compiler Infrastructure in HPC - LLVM '15, pp~1--6

\bibitem[\protect\citeauthoryear{Li \& Bregman}{Li \& Bregman}{2017}]{Li2017a}
Li Y.,  Bregman J.,  2017, \mn@doi [The Astrophysical Journal]
  {10.3847/1538-4357/aa92c6}, 849, 105

\bibitem[\protect\citeauthoryear{{Li}, {Crain}  \& {Wang}}{{Li}
  et~al.}{2014}]{Li14b}
{Li} J.-T.,  {Crain} R.~A.,   {Wang} Q.~D.,  2014, \mn@doi [\mnras]
  {10.1093/mnras/stu329}, \href
  {https://ui.adsabs.harvard.edu/abs/2014MNRAS.440..859L} {440, 859}

\bibitem[\protect\citeauthoryear{Li, Bregman, Wang, Crain  \& Anderson}{Li
  et~al.}{2018}]{Li2018}
Li J.-T.,  Bregman J.~N.,  Wang Q.~D.,  Crain R.~A.,   Anderson M.~E.,  2018,
  \mn@doi [ApJ] {10.3847/2041-8213/aab2af}, 855, L24

\bibitem[\protect\citeauthoryear{Lian, Thomas, Maraston, Goddard, Comparat,
  {Gonzalez-Perez}  \& Ventura}{Lian et~al.}{2018}]{Lian2018}
Lian J.,  Thomas D.,  Maraston C.,  Goddard D.,  Comparat J.,  {Gonzalez-Perez}
  V.,   Ventura P.,  2018, \mn@doi [Monthly Notices of the Royal Astronomical
  Society] {10.1093/mnras/stx2829}, 474, 1143

\bibitem[\protect\citeauthoryear{Ma, Hopkins, Wetzel, Kirby,
  {Angl{\'e}s-Alc{\'a}zar}, {Faucher-Gigu{\`e}re}, Kere{\v s}  \& Quataert}{Ma
  et~al.}{2017}]{Ma2017a}
Ma X.,  Hopkins P.~F.,  Wetzel A.~R.,  Kirby E.~N.,  {Angl{\'e}s-Alc{\'a}zar}
  D.,  {Faucher-Gigu{\`e}re} C.-A.,  Kere{\v s} D.,   Quataert E.,  2017,
  \mn@doi [Monthly Notices of the Royal Astronomical Society]
  {10.1093/mnras/stx273}, 467, 2430

\bibitem[\protect\citeauthoryear{Maller \& Bullock}{Maller \&
  Bullock}{2004}]{Maller2004}
Maller A.~H.,  Bullock J.~S.,  2004, \mn@doi [Monthly Notices of the Royal
  Astronomical Society] {10.1111/j.1365-2966.2004.08349.x}, 355, 694

\bibitem[\protect\citeauthoryear{Mandelker, Padnos, Dekel, Birnboim, Burkert,
  Krumholz  \& Steinberg}{Mandelker et~al.}{2016}]{Mandelker2016}
Mandelker N.,  Padnos D.,  Dekel A.,  Birnboim Y.,  Burkert A.,  Krumholz
  M.~R.,   Steinberg E.,  2016, \mn@doi [Monthly Notices of the Royal
  Astronomical Society] {10.1093/mnras/stw2267}, 463, 3921

\bibitem[\protect\citeauthoryear{Mandelker, {van Dokkum}, Brodie, {van den
  Bosch}  \& Ceverino}{Mandelker et~al.}{2018}]{Mandelker2018}
Mandelker N.,  {van Dokkum} P.~G.,  Brodie J.~P.,  {van den Bosch} F.~C.,
  Ceverino D.,  2018, \mn@doi [The Astrophysical Journal]
  {10.3847/1538-4357/aaca98}, 861, 148

\bibitem[\protect\citeauthoryear{Mandelker, Nagai, Aung, Dekel, Padnos  \&
  Birnboim}{Mandelker et~al.}{2019}]{Mandelker2019a}
Mandelker N.,  Nagai D.,  Aung H.,  Dekel A.,  Padnos D.,   Birnboim Y.,  2019,
  \mn@doi [Monthly Notices of the Royal Astronomical Society]
  {10.1093/mnras/stz012}, 484, 1100

\bibitem[\protect\citeauthoryear{Mandelker, Nagai, Aung, Dekel, Birnboim  \&
  {van den Bosch}}{Mandelker et~al.}{2020}]{Mandelker2020a}
Mandelker N.,  Nagai D.,  Aung H.,  Dekel A.,  Birnboim Y.,   {van den Bosch}
  F.~C.,  2020, \mn@doi [Monthly Notices of the Royal Astronomical Society]
  {10.1093/mnras/staa812}, 494, 2641

\bibitem[\protect\citeauthoryear{Mandelker, van~den Bosch, Springel, {van de
  Voort}, Burchett, Butsky, Nagai  \& Oh}{Mandelker
  et~al.}{2021}]{Mandelker2021}
Mandelker N.,  van~den Bosch F.~C.,  Springel V.,  {van de Voort} F.,  Burchett
  J.~N.,  Butsky I.~S.,  Nagai D.,   Oh S.~P.,  2021, arXiv:2107.03395
  [astro-ph]

\bibitem[\protect\citeauthoryear{Martin et~al.,}{Martin
  et~al.}{2019}]{Martin2019a}
Martin D.~C.,  et~al., 2019, arXiv

\bibitem[\protect\citeauthoryear{Mathews \& Bregman}{Mathews \&
  Bregman}{1978}]{Mathews1978}
Mathews W.~G.,  Bregman J.~N.,  1978, \mn@doi [The Astrophysical Journal]
  {10.1086/156379}, 224, 308

\bibitem[\protect\citeauthoryear{Mayor \& Vigroux}{Mayor \&
  Vigroux}{1981}]{Mayor1981}
Mayor M.,  Vigroux L.,  1981, Astronomy and Astrophysics, vol. 98, no. 1, May
  1981, p. 1-8., 98, 1

\bibitem[\protect\citeauthoryear{McCourt, Sharma, Quataert  \& Parrish}{McCourt
  et~al.}{2012}]{Mccourt2012}
McCourt M.,  Sharma P.,  Quataert E.,   Parrish I.~J.,  2012, \mn@doi [Monthly
  Notices of the Royal Astronomical Society]
  {10.1111/j.1365-2966.2011.19972.x}, 419, 3319

\bibitem[\protect\citeauthoryear{McCourt, Oh, O'Leary  \& Madigan}{McCourt
  et~al.}{2018}]{McCourt2018}
McCourt M.,  Oh S.~P.,  O'Leary R.~M.,   Madigan A.-M.,  2018, \mn@doi [Monthly
  Notices of the Royal Astronomical Society] {10.1093/mnras/stx2687}, 473, 5407

\bibitem[\protect\citeauthoryear{{McDonald}, {Gaspari}, {McNamara}  \&
  {Tremblay}}{{McDonald} et~al.}{2018}]{McDonald18}
{McDonald} M.,  {Gaspari} M.,  {McNamara} B.~R.,   {Tremblay} G.~R.,  2018,
  \mn@doi [\apj] {10.3847/1538-4357/aabace}, \href
  {https://ui.adsabs.harvard.edu/abs/2018ApJ...858...45M} {858, 45}

\bibitem[\protect\citeauthoryear{McKinney}{McKinney}{2010}]{McKinney2010}
McKinney W.,  2010, in Python in {{Science Conference}}. {Austin, Texas}, pp
  56--61, \mn@doi{10.25080/Majora-92bf1922-00a}

\bibitem[\protect\citeauthoryear{McNamara \& Nulsen}{McNamara \&
  Nulsen}{2007}]{McNamara2007}
McNamara B.~R.,  Nulsen P. E.~J.,  2007, \mn@doi [Annual Review of Astronomy
  and Astrophysics] {10.1146/annurev.astro.45.051806.110625}, 45, 117

\bibitem[\protect\citeauthoryear{{Mo}, {Mao}  \& {White}}{{Mo}
  et~al.}{1998}]{MMW98}
{Mo} H.~J.,  {Mao} S.,   {White} S. D.~M.,  1998, \mn@doi [\mnras]
  {10.1046/j.1365-8711.1998.01227.x}, \href
  {https://ui.adsabs.harvard.edu/abs/1998MNRAS.295..319M} {295, 319}

\bibitem[\protect\citeauthoryear{{Moffett} et~al.,}{{Moffett}
  et~al.}{2016}]{Moffett16}
{Moffett} A.~J.,  et~al., 2016, \mn@doi [\mnras] {10.1093/mnras/stv2883}, \href
  {https://ui.adsabs.harvard.edu/abs/2016MNRAS.457.1308M} {457, 1308}

\bibitem[\protect\citeauthoryear{Murante, Calabrese, De~Lucia, Monaco, Borgani
  \& Dolag}{Murante et~al.}{2012}]{Murante2012}
Murante G.,  Calabrese M.,  De~Lucia G.,  Monaco P.,  Borgani S.,   Dolag K.,
  2012, \mn@doi [The Astrophysical Journal Letters]
  {10.1088/2041-8205/749/2/L34}, 749, L34

\bibitem[\protect\citeauthoryear{Muratov, Kere{\v s}, {Faucher-Gigu{\`e}re},
  Hopkins, Quataert  \& Murray}{Muratov et~al.}{2015}]{Muratov2015}
Muratov A.~L.,  Kere{\v s} D.,  {Faucher-Gigu{\`e}re} C.-A.,  Hopkins P.~F.,
  Quataert E.,   Murray N.,  2015, \mn@doi [Monthly Notices of the Royal
  Astronomical Society] {10.1093/mnras/stv2126}, 454, 2691

\bibitem[\protect\citeauthoryear{Muratov et~al.,}{Muratov
  et~al.}{2017}]{Muratov2017}
Muratov A.~L.,  et~al., 2017, \mn@doi [Monthly Notices of the Royal
  Astronomical Society] {10.1093/mnras/stx667}, 468, 4170

\bibitem[\protect\citeauthoryear{Nelson, Vogelsberger, Genel, Sijacki, Kere{\v
  s}, Springel  \& Hernquist}{Nelson et~al.}{2013}]{Nelson2013}
Nelson D.,  Vogelsberger M.,  Genel S.,  Sijacki D.,  Kere{\v s} D.,  Springel
  V.,   Hernquist L.,  2013, \mn@doi [Monthly Notices of the Royal Astronomical
  Society] {10.1093/mnras/sts595}, 429, 3353

\bibitem[\protect\citeauthoryear{Nelson, Genel, Pillepich, Vogelsberger,
  Springel  \& Hernquist}{Nelson et~al.}{2016}]{Nelson2016}
Nelson D.,  Genel S.,  Pillepich A.,  Vogelsberger M.,  Springel V.,
  Hernquist L.,  2016, \mn@doi [Monthly Notices of the Royal Astronomical
  Society] {10.1093/mnras/stw1191}, 460, 2881

\bibitem[\protect\citeauthoryear{Oppenheimer}{Oppenheimer}{2018}]{Oppenheimer2018}
Oppenheimer B.~D.,  2018, \mn@doi [Monthly Notices of the Royal Astronomical
  Society] {10.1093/mnras/sty1918}, 480, 2963

\bibitem[\protect\citeauthoryear{Pandya et~al.,}{Pandya
  et~al.}{2021}]{Pandya2021}
Pandya V.,  et~al., 2021, arXiv

\bibitem[\protect\citeauthoryear{{Peebles}}{{Peebles}}{1969}]{Peebles69}
{Peebles} P.~J.~E.,  1969, \mn@doi [\apj] {10.1086/149876}, \href
  {https://ui.adsabs.harvard.edu/abs/1969ApJ...155..393P} {155, 393}

\bibitem[\protect\citeauthoryear{Peebles}{Peebles}{2020}]{Peebles2020}
Peebles P. J.~E.,  2020, \mn@doi [Monthly Notices of the Royal Astronomical
  Society] {10.1093/mnras/staa2649}, 498, 4386

\bibitem[\protect\citeauthoryear{Pezzulli \& Fraternali}{Pezzulli \&
  Fraternali}{2016}]{Pezzulli2016a}
Pezzulli G.,  Fraternali F.,  2016, \mn@doi [Monthly Notices of the Royal
  Astronomical Society] {10.1093/mnras/stv2397}, 455, 2308

\bibitem[\protect\citeauthoryear{Pezzulli, Fraternali  \& Binney}{Pezzulli
  et~al.}{2017}]{Pezzulli2017}
Pezzulli G.,  Fraternali F.,   Binney J.,  2017, \mn@doi [Monthly Notices of
  the Royal Astronomical Society] {10.1093/mnras/stx029}, 467, 311

\bibitem[\protect\citeauthoryear{{Planck Collaboration} et~al.,}{{Planck
  Collaboration} et~al.}{2018}]{PlanckCollaboration2018}
{Planck Collaboration} et~al., 2018, arXiv:1807.06209

\bibitem[\protect\citeauthoryear{Quataert, Thompson  \& Jiang}{Quataert
  et~al.}{2021a}]{Quataert2021a}
Quataert E.,  Thompson T.~A.,   Jiang Y.-F.,  2021a, arXiv:2102.05696
  [astro-ph]

\bibitem[\protect\citeauthoryear{Quataert, Jiang  \& Thompson}{Quataert
  et~al.}{2021b}]{Quataert2021}
Quataert E.,  Jiang Y.-F.,   Thompson T.~A.,  2021b, arXiv:2106.08404
  [astro-ph]

\bibitem[\protect\citeauthoryear{Reback et~al.,}{Reback
  et~al.}{2020}]{Reback2020}
Reback J.,  et~al., 2020, Pandas-Dev/Pandas: {{Pandas}} 1.0.3, Zenodo,
  \mn@doi{10.5281/zenodo.3715232}

\bibitem[\protect\citeauthoryear{Rohr et~al.,}{Rohr et~al.}{2021}]{Rohr2021}
Rohr E.,  et~al., 2021, arXiv:2112.05159 [astro-ph]

\bibitem[\protect\citeauthoryear{Sales, Navarro, Theuns, Schaye, White, Frenk,
  Crain  \& Dalla~Vecchia}{Sales et~al.}{2012}]{Sales2012}
Sales L.~V.,  Navarro J.~F.,  Theuns T.,  Schaye J.,  White S. D.~M.,  Frenk
  C.~S.,  Crain R.~A.,   Dalla~Vecchia C.,  2012, \mn@doi [Monthly Notices of
  the Royal Astronomical Society] {10.1111/j.1365-2966.2012.20975.x}, 423, 1544

\bibitem[\protect\citeauthoryear{Samuel et~al.,}{Samuel
  et~al.}{2020}]{Samuel2020}
Samuel J.,  et~al., 2020, \mn@doi [Monthly Notices of the Royal Astronomical
  Society] {10.1093/mnras/stz3054}, 491, 1471

\bibitem[\protect\citeauthoryear{Sanderson et~al.,}{Sanderson
  et~al.}{2020}]{Sanderson2020}
Sanderson R.~E.,  et~al., 2020, \mn@doi [The Astrophysical Journal Supplement
  Series] {10.3847/1538-4365/ab5b9d}, 246, 6

\bibitem[\protect\citeauthoryear{Sharma, Mccourt, Quataert  \& Parrish}{Sharma
  et~al.}{2012}]{Sharma2012}
Sharma P.,  Mccourt M.,  Quataert E.,   Parrish I.~J.,  2012, \mn@doi [Monthly
  Notices of the Royal Astronomical Society]
  {10.1111/j.1365-2966.2011.20246.x}, 420, 3174

\bibitem[\protect\citeauthoryear{Simons, Kassin, Weiner, Heckman, Lee, Lotz,
  Peth  \& Tchernyshyov}{Simons et~al.}{2015}]{Simons2015}
Simons R.~C.,  Kassin S.~A.,  Weiner B.~J.,  Heckman T.~M.,  Lee J.~C.,  Lotz
  J.~M.,  Peth M.,   Tchernyshyov K.,  2015, \mn@doi [Monthly Notices of the
  Royal Astronomical Society] {10.1093/mnras/stv1298}, 452, 986

\bibitem[\protect\citeauthoryear{Simons et~al.,}{Simons
  et~al.}{2017}]{Simons2017}
Simons R.~C.,  et~al., 2017, \mn@doi [The Astrophysical Journal]
  {10.3847/1538-4357/aa740c}, 843, 46

\bibitem[\protect\citeauthoryear{Sormani, Sobacchi, Pezzulli, Binney  \&
  Klessen}{Sormani et~al.}{2018}]{Sormani2018}
Sormani M.~C.,  Sobacchi E.,  Pezzulli G.,  Binney J.,   Klessen R.~S.,  2018,
  \mn@doi [Monthly Notices of the Royal Astronomical Society]
  {10.1093/mnras/sty2500}, 481, 3370

\bibitem[\protect\citeauthoryear{Stern, Fielding, {Faucher-Gigu{\`e}re}  \&
  Quataert}{Stern et~al.}{2019}]{Stern2019}
Stern J.,  Fielding D.,  {Faucher-Gigu{\`e}re} C.-A.,   Quataert E.,  2019,
  \mn@doi [Monthly Notices of the Royal Astronomical Society]
  {10.1093/mnras/stz1859}, 488, 2549

\bibitem[\protect\citeauthoryear{Stern, Fielding, {Faucher-Gigu{\`e}re}  \&
  Quataert}{Stern et~al.}{2020}]{Stern2020}
Stern J.,  Fielding D.,  {Faucher-Gigu{\`e}re} C.-A.,   Quataert E.,  2020,
  \mn@doi [Monthly Notices of the Royal Astronomical Society]
  {10.1093/mnras/staa198}, 492, 6042

\bibitem[\protect\citeauthoryear{Stern et~al.,}{Stern
  et~al.}{2021a}]{Stern2021}
Stern J.,  et~al., 2021a, \mn@doi [ApJ] {10.3847/1538-4357/abd776}, 911, 88

\bibitem[\protect\citeauthoryear{Stern et~al.,}{Stern
  et~al.}{2021b}]{Stern2021a}
Stern J.,  et~al., 2021b, \mn@doi [Monthly Notices of the Royal Astronomical
  Society] {10.1093/mnras/stab2240}, 507, 2869

\bibitem[\protect\citeauthoryear{Stevens, Lagos, Contreras, Croton, Padilla,
  Schaller, Schaye  \& Theuns}{Stevens et~al.}{2017}]{Stevens2017}
Stevens A. R.~H.,  Lagos C. d.~P.,  Contreras S.,  Croton D.~J.,  Padilla
  N.~D.,  Schaller M.,  Schaye J.,   Theuns T.,  2017, \mn@doi [Monthly Notices
  of the Royal Astronomical Society] {10.1093/mnras/stx243}, 467, 2066

\bibitem[\protect\citeauthoryear{Stewart, Kaufmann, Bullock, Barton, Maller,
  Diemand  \& Wadsley}{Stewart et~al.}{2011}]{Stewart2011a}
Stewart K.~R.,  Kaufmann T.,  Bullock J.~S.,  Barton E.~J.,  Maller A.~H.,
  Diemand J.,   Wadsley J.,  2011, \mn@doi [The Astrophysical Journal]
  {10.1088/0004-637X/738/1/39}, 738, 39

\bibitem[\protect\citeauthoryear{Stewart, Brooks, Bullock, Maller, Diemand,
  Wadsley  \& Moustakas}{Stewart et~al.}{2013}]{Stewart2013}
Stewart K.~R.,  Brooks A.~M.,  Bullock J.~S.,  Maller A.~H.,  Diemand J.,
  Wadsley J.,   Moustakas L.~A.,  2013, \mn@doi [The Astrophysical Journal]
  {10.1088/0004-637X/769/1/74}, 769, 74

\bibitem[\protect\citeauthoryear{Stewart et~al.,}{Stewart
  et~al.}{2017}]{Stewart2017}
Stewart K.~R.,  et~al., 2017, \mn@doi [The Astrophysical Journal]
  {10.3847/1538-4357/aa6dff}, 843, 47

\bibitem[\protect\citeauthoryear{Thompson, Quataert  \& Murray}{Thompson
  et~al.}{2005}]{Thompson2005}
Thompson T.~A.,  Quataert E.,   Murray N.,  2005, \mn@doi [The Astrophysical
  Journal] {10.1086/431923}, 630, 167

\bibitem[\protect\citeauthoryear{Trapp et~al.,}{Trapp et~al.}{2021}]{Trapp2021}
Trapp C.,  et~al., 2021, arXiv:2105.11472 [astro-ph]

\bibitem[\protect\citeauthoryear{{\"U}bler, Naab, Oser, Aumer, Sales  \&
  White}{{\"U}bler et~al.}{2014}]{Ubler2014}
{\"U}bler H.,  Naab T.,  Oser L.,  Aumer M.,  Sales L.~V.,   White S. D.~M.,
  2014, \mn@doi [Monthly Notices of the Royal Astronomical Society]
  {10.1093/mnras/stu1275}, 443, 2092

\bibitem[\protect\citeauthoryear{{Van de Voort} \& Schaye}{{Van de Voort} \&
  Schaye}{2012}]{VandeVoort2012a}
{Van de Voort} F.,  Schaye J.,  2012, \mn@doi [Monthly Notices of the Royal
  Astronomical Society] {10.1111/j.1365-2966.2012.20949.x}, 423, 2991

\bibitem[\protect\citeauthoryear{{Van de Voort}, Schaye, Booth  \&
  Dalla~Vecchia}{{Van de Voort} et~al.}{2011}]{VandeVoort2011}
{Van de Voort} F.,  Schaye J.,  Booth C.~M.,   Dalla~Vecchia C.,  2011, \mn@doi
  [Monthly Notices of the Royal Astronomical Society]
  {10.1111/j.1365-2966.2011.18896.x}, 415, 2782

\bibitem[\protect\citeauthoryear{Virtanen et~al.,}{Virtanen
  et~al.}{2020}]{Virtanen2020}
Virtanen P.,  et~al., 2020, \mn@doi [Nat Methods] {10.1038/s41592-019-0686-2},
  17, 261

\bibitem[\protect\citeauthoryear{Voit}{Voit}{2021}]{Voit2021}
Voit G.~M.,  2021, \mn@doi [The Astrophysical Journal]
  {10.3847/2041-8213/abe11f}, 908, L16

\bibitem[\protect\citeauthoryear{Voit, Donahue, O'Shea, Bryan, Sun  \&
  Werner}{Voit et~al.}{2015}]{Voit2015}
Voit G.~M.,  Donahue M.,  O'Shea B.~W.,  Bryan G.~L.,  Sun M.,   Werner N.,
  2015, \mn@doi [The Astrophysical Journal] {10.1088/2041-8205/803/2/L21}, 803,
  L21

\bibitem[\protect\citeauthoryear{Voit, Meece, Li, O'Shea, Bryan  \&
  Donahue}{Voit et~al.}{2017}]{Voit2017}
Voit G.~M.,  Meece G.,  Li Y.,  O'Shea B.~W.,  Bryan G.~L.,   Donahue M.,
  2017, \mn@doi [The Astrophysical Journal] {10.3847/1538-4357/aa7d04}, 845, 80

\bibitem[\protect\citeauthoryear{{Wellons} et~al.,}{{Wellons}
  et~al.}{2022}]{Wellons22}
{Wellons} S.,  et~al., 2022, arXiv e-prints, \href
  {https://ui.adsabs.harvard.edu/abs/2022arXiv220306201W} {p. arXiv:2203.06201}

\bibitem[\protect\citeauthoryear{Wetzel, Hopkins, Kim, {Faucher-Gigu{\`e}re},
  Kere{\v s}  \& Quataert}{Wetzel et~al.}{2016}]{Wetzel2016}
Wetzel A.~R.,  Hopkins P.~F.,  Kim J.-h.,  {Faucher-Gigu{\`e}re} C.-A.,
  Kere{\v s} D.,   Quataert E.,  2016, \mn@doi [The Astrophysical Journal]
  {10.3847/2041-8205/827/2/L23}, 827, L23

\bibitem[\protect\citeauthoryear{Wheeler et~al.,}{Wheeler
  et~al.}{2016}]{Wheeler16}
Wheeler C.,  et~al., 2016, \mn@doi [Monthly Notices of the Royal Astronomical
  Society] {10.1093/mnras/stw2583}, 465, 2420

\bibitem[\protect\citeauthoryear{White \& Frenk}{White \&
  Frenk}{1991}]{White1991}
White S. D.~M.,  Frenk C.~S.,  1991, \mn@doi [The Astrophysical Journal]
  {10.1086/170483}, 379, 52

\bibitem[\protect\citeauthoryear{White \& Rees}{White \&
  Rees}{1978}]{White1978}
White S. D.~M.,  Rees M.~J.,  1978, \mn@doi [Monthly Notices of the Royal
  Astronomical Society] {10.1093/mnras/183.3.341}, 183, 341

\bibitem[\protect\citeauthoryear{Yu et~al.,}{Yu et~al.}{2021}]{Yu2021}
Yu S.,  et~al., 2021, arXiv e-prints, 2103, arXiv:2103.03888

\bibitem[\protect\citeauthoryear{{van de Voort}, Schaye, Booth, Haas  \&
  Dalla~Vecchia}{{van de Voort} et~al.}{2011}]{VanDeVoort2011a}
{van de Voort} F.,  Schaye J.,  Booth C.~M.,  Haas M.~R.,   Dalla~Vecchia C.,
  2011, \mn@doi [Monthly Notices of the Royal Astronomical Society]
  {10.1111/j.1365-2966.2011.18565.x}, 414, 2458

\bibitem[\protect\citeauthoryear{{van der Wel} et~al.,}{{van der Wel}
  et~al.}{2014}]{vanderWel14}
{van der Wel} A.,  et~al., 2014, \mn@doi [\apjl] {10.1088/2041-8205/792/1/L6},
  \href {https://ui.adsabs.harvard.edu/abs/2014ApJ...792L...6V} {792, L6}

\makeatother
\end{thebibliography}



\appendix

\section{Observationally-motivated thin disk fraction}
\label{s: appendix-sloan thin disk fraction}

\begin{figure}
    \centering
    \includegraphics[width=\columnwidth]{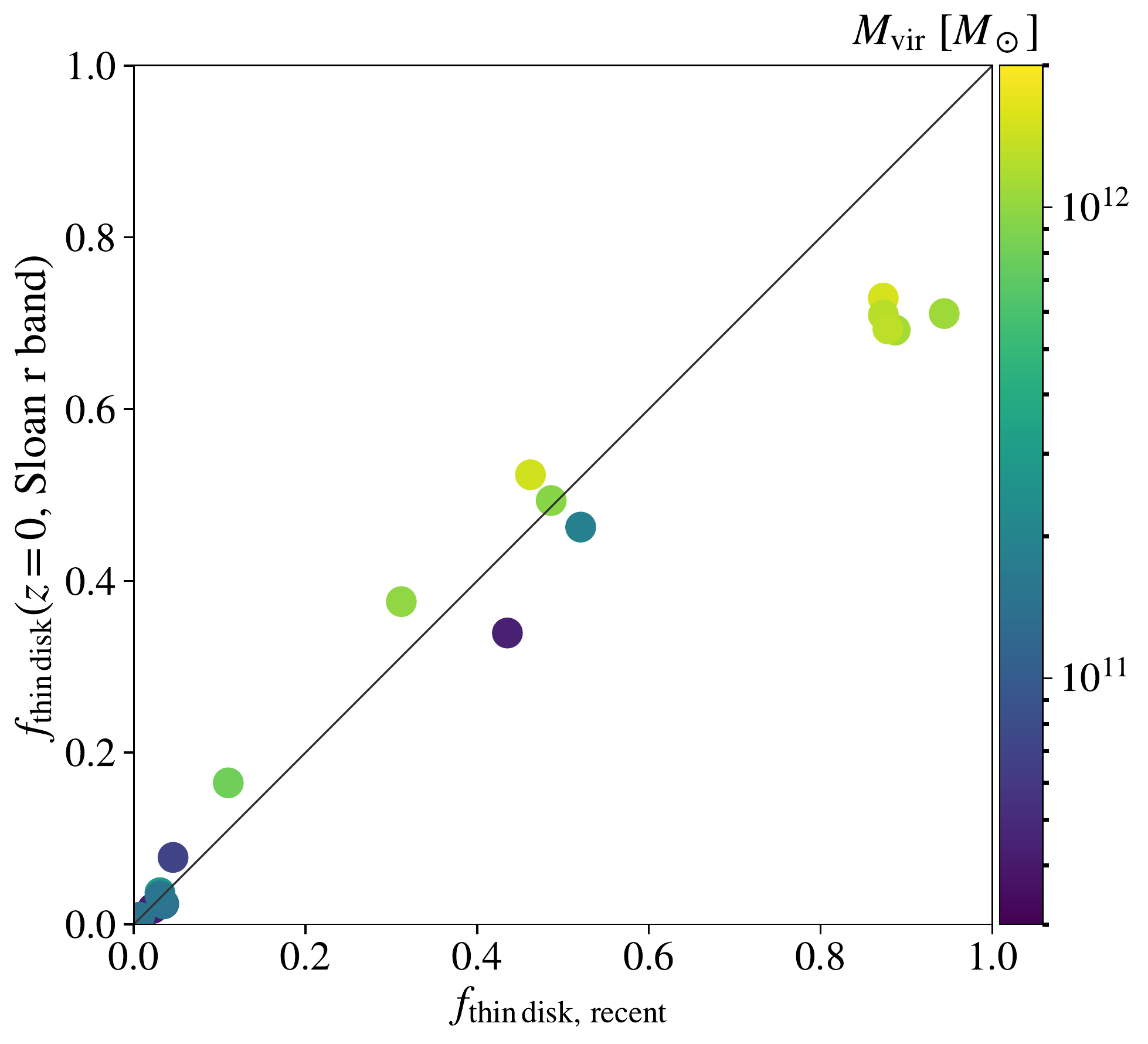}
    \caption{
    Fraction of recent stars in a thin disk versus Sloan-r-band-weighted thin disk fraction, in our FIRE sample. The value of $\fthin$ closely tracks the observationally-motivated luminosity-weighted thin disk fraction.
    }
    \label{f: thin disk v thin disk}
\end{figure}

Fig.~\ref{f: thin disk v thin disk} shows the relationship between the fraction of stars with $j_z/j_c(E)>0.8$ at $z=0$ ($\fthin$) and the same fraction of stars weighted by their Sloan $r$ band luminosity-weighted.
The figure shows that $\fthin$, the thin disk metric used throughout our analysis, is closely related to the observationally-motivated luminosity-weighted thin disk fraction.

\section{Sample validation}
\label{s: appendix-sample validation}

\begin{figure}
    \centering
    \includegraphics[width=\columnwidth]{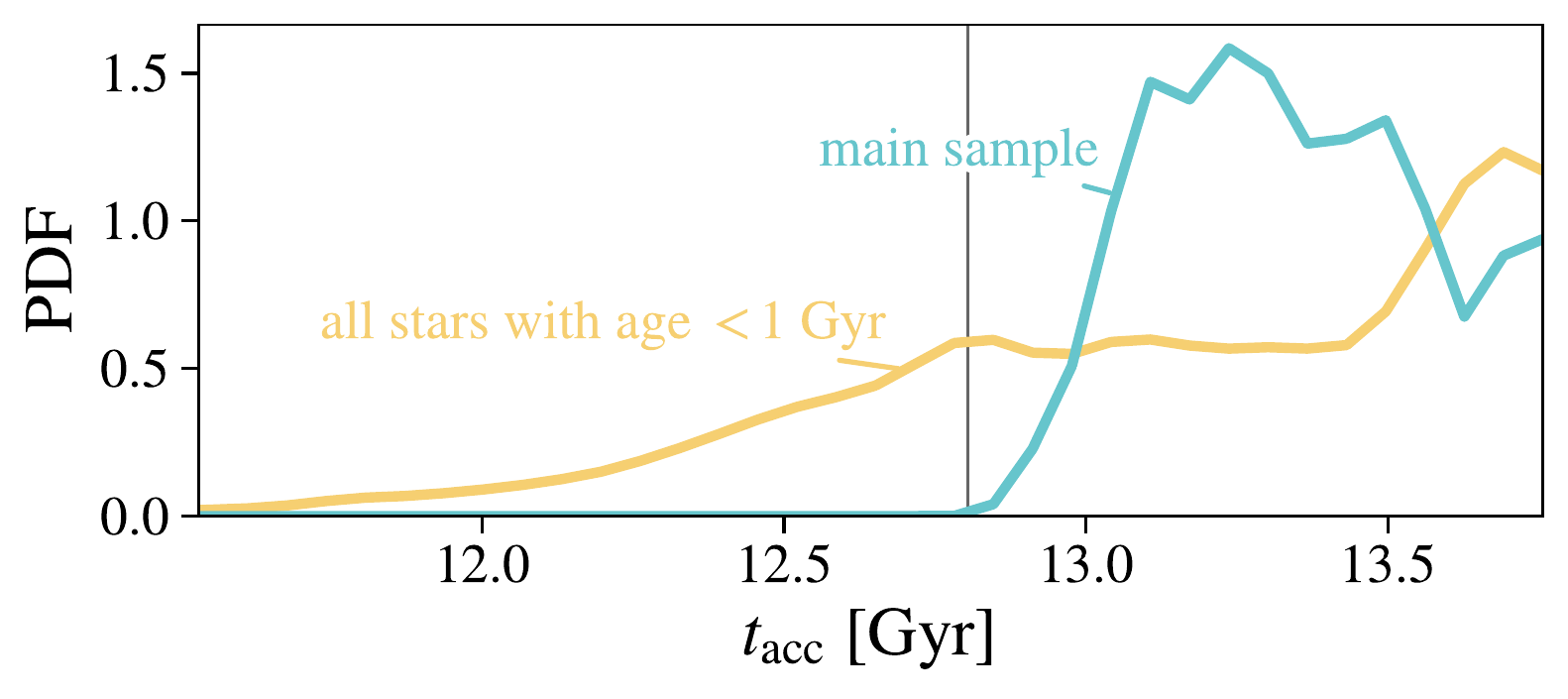}
    \caption{
    Accretion time distribution for all accreting gas in the last Gyr (green, corresponding to the sample analyzed in the main text), compared to accretion time for all gas which forms stars in the last Gyr (yellow), in the \texttt{m12i} simulation.
    Of stars born in the last Gyr, $> 70\%$ ($> 98\%$) are formed from gas accreted within the last one Gyr (two Gyr), indicating that the main sample is representative of all gas which formed recent stars.
    }
    \label{f: sample validation -- tacc}
\end{figure}

\begin{figure}
    \centering
    \includegraphics[width=\columnwidth]{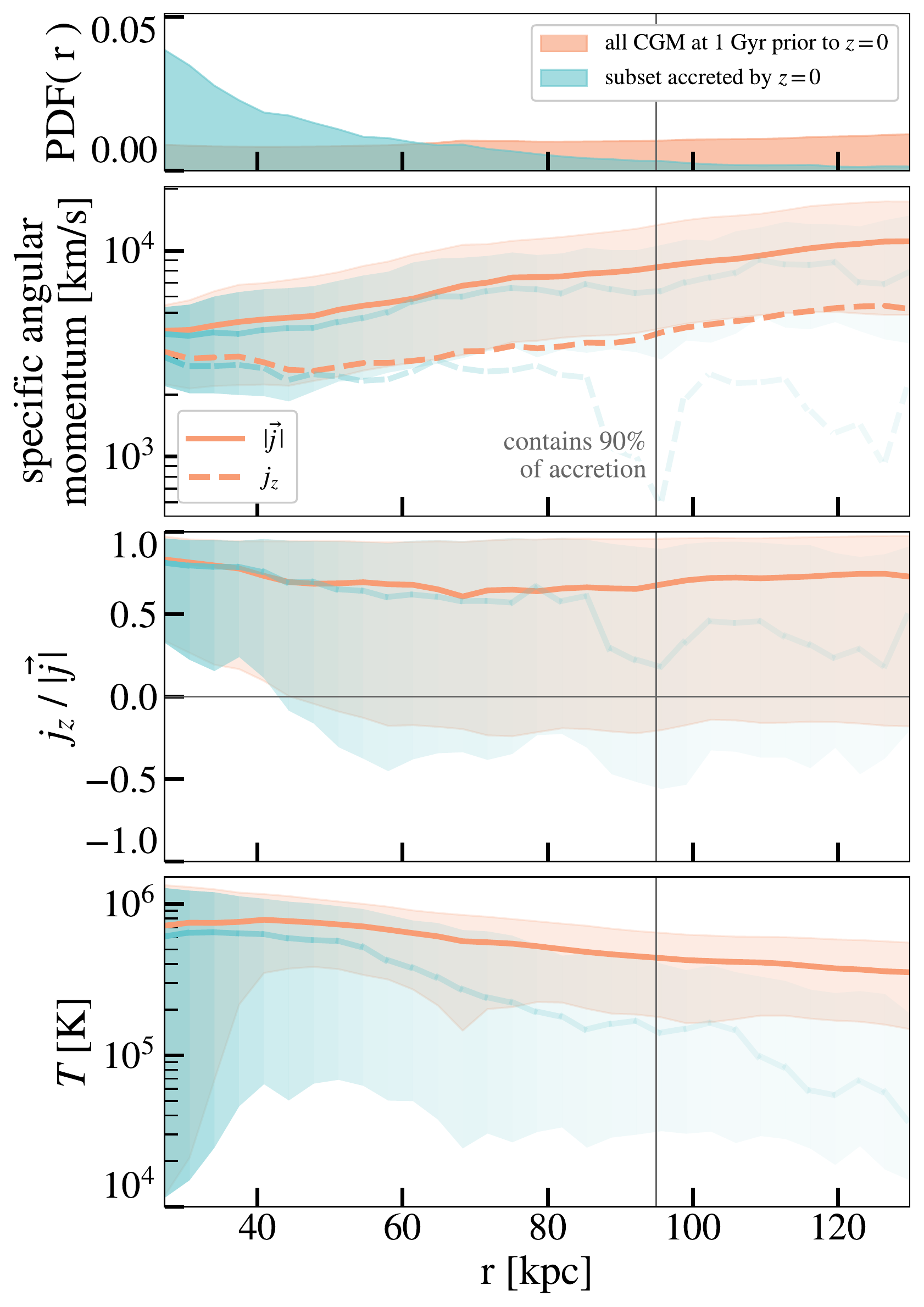}
    \caption{
    \textbf{Topmost panel:}
    Radial distribution of gas in the CGM of \texttt{m12i} at 1 Gyr prior to $z=0$, comparing all gas (orange) to the subset accreted onto the galaxy (green, corresponding to the sample analyzed in the main text).
    Accreted gas mainly originates in the inner CGM. 
    \textbf{Bottom panels:}
    From top to bottom, specific angular momentum, angular momentum alignment, and temperature for both samples.
    Solid line shows the median, while colored regions cover 16th--84th percentiles.
    Accreted gas has similar specific angular momentum and alignment as all gas at the same radii, especially at $r<50$ kpc which dominates the accretion. 
    The median temperature is also consistent between the two samples at $r<50$ kpc, though the distribution extends to lower temperatures in the accreted sample. 
    }
    \label{f: sample validation -- spatial}
\end{figure}

Our main sample of tracked particles for each simulation (\S\ref{s: methods -- analysis}) selects particles which accrete from the CGM in the last Gyr prior to $z=0$. 
In this appendix we check the extent to which our sample is representative of all gas that fuels  star formation at $z\sim0$, and the extent to which our sample is representative of all circumgalactic gas.
We use the simulation \texttt{m12i} as a test case.

Fig.~\ref{f: sample validation -- tacc} shows the age of the universe at which particles in our main sample accrete onto the main galaxy, compared to the accretion  time for all gas which forms stars in the last Gyr.
To calculate $\tacc$ for all recently-formed stars we track the history of $10^5$ randomly-selected star particles that formed within the last Gyr, and set $\tacc$ to the time of the last accretion event.
The $\tacc$ distributions for the main sample and the sample probing all recently-formed stars largely overlap, indicating our selection method is representative of star forming gas in the last Gyr of the simulation. 

Fig.~\ref{f: sample validation -- spatial} compares the properties of all gas in the CGM with the subset of gas which accretes onto the galaxy, where the latter corresponds to the main sample of tracked particles. 
The top panel shows that most of the accreted gas originates from the inner CGM, likely since cooling times in the hot phase are shorter at small radii.
The distributions of specific angular momentum (2nd panel) and angular momentum alignment (3rd panel) are 
similar between the accreted gas and general CGM gas. The bottom panel shows that the median temperature of accreted gas and all CGM gas are similar at $r<50$ kpc, at which most of the accretion originates, with a tail extending to lower temperatures in the accreted gas. At larger radii the accreted gas is cooler than general CGM gas, suggesting that from these radii accretion is via cooling of clumps rather than via inflow of the hot medium. Similar results are found in 
\cite{Hafen2020}.

\section{Instantaneous mass flow}
\label{s: appendix-mass flow}

\begin{figure*}
    \centering
    \includegraphics[width=\textwidth]{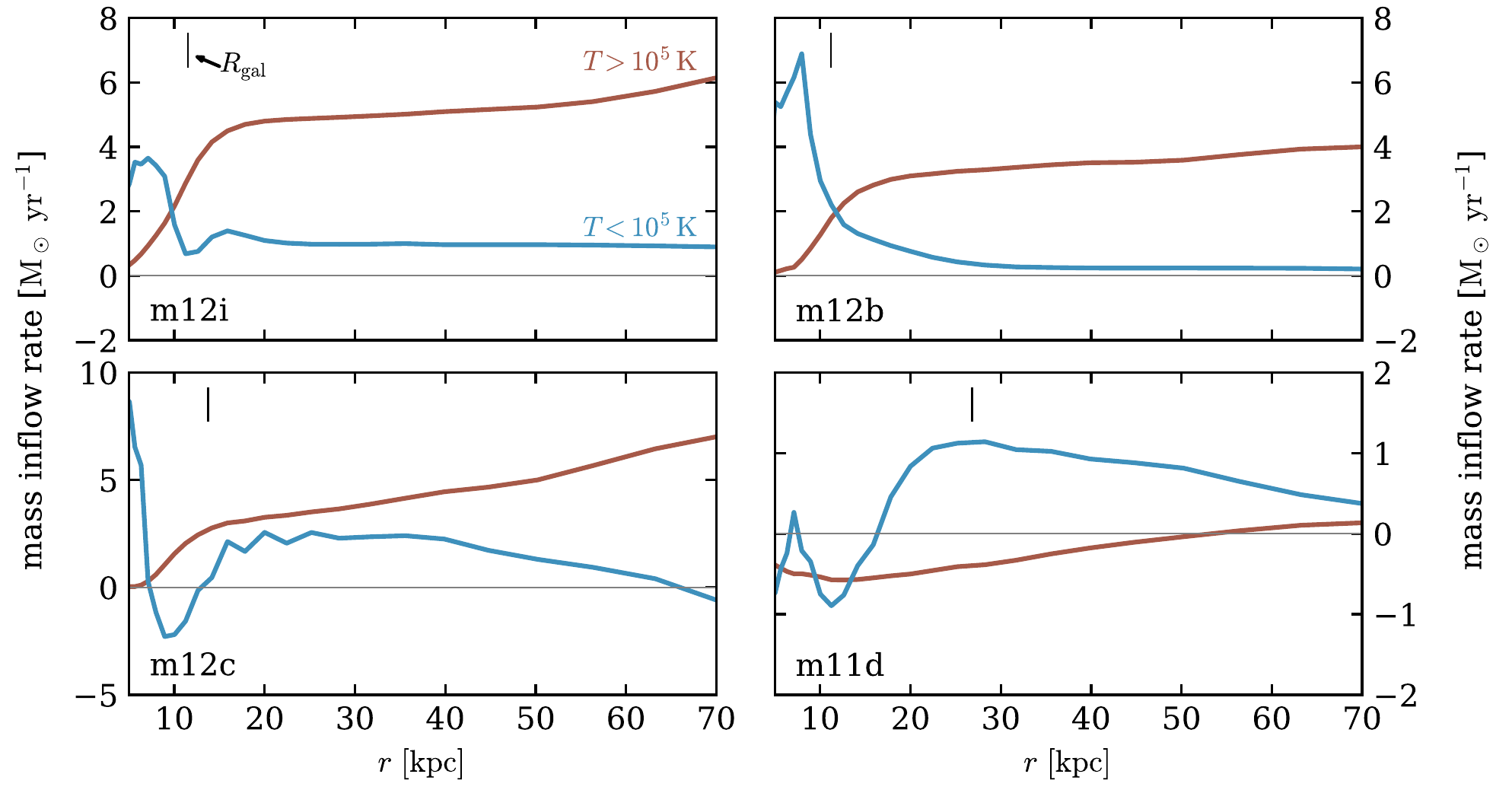}
    \caption{
    Average mass inflow rate versus gas temperature, during the last Gyr prior to $z=0$ in the same haloes as in Fig.~\ref{f: stars}.
    The galaxy radius $r_{\rm gal}=4 r_{\star,0.5}$ is marked in the panels. 
    Hot accretion ($T>10^5$ K) dominates the mass inflow onto thin-disk galaxies at $r \gtrsim r_{\rm gal} \sim 10-20$ kpc.
    Cold accretion ($T<10^5$ K) dominates the inflow onto the  
    irregular galaxy.
    }
    \label{f: Mdot}
\end{figure*}

In this appendix we analyze if inflow is dominated by hot or cold gas, which allows us to distinguish a hot inflow from the accretion of cool clouds (formed, e.g., via instabilities).
Fig.~\ref{f: Mdot} shows the mass inflow rate versus radius for hot gas (red curves; $T>10^5$ K) and for cool gas (blue curves; $T < 10^5\,{\rm K}$),  for the four simulations shown in Fig.~\ref{f: stars}. 
We calculate the mass inflow rate at a given radius as
\begin{equation}
     \Mdot(r) = \frac{\int_{\rm shell} v_r dm}{\Delta r} = \frac{M_{\rm shell}}{\Delta r} \langle v_r\rangle_{\rm mass\, weighted}
     \label{e: Mdot}
\end{equation}
where $\Delta r=0.05\,{\rm dex}$ is the shell thickness, and the integration is done on all particles with centers within the shell which satisfy the appropriate temperature cut. 

In the three MW-mass galaxies with a large thin disk fraction, at halo scales of $r>20$ kpc the inflow is dominated by hot gas, where in \texttt{m12i} and \texttt{m12b} $\Mdot$ of the hot gas is larger by a factor of $\gtrsim 4\times$ than that of cool gas.
The greater mass flux of hot gas indicates that the dominant form of accretion is an inflowing hot phase, rather than cold streams or cool clouds precipitating from the hot phase.
In contrast, in the lower mass galaxy shown in the bottom right the inflow is dominated by cool gas, while the hot gas is outflowing out to $\approx50$ kpc.
Fig.~\ref{f: Mdot} also shows that the inflow rate of hot gas in the MW-mass galaxies drops within $r_{\rm gal}$, reflecting the cooling of the hot inflow at the galaxy-halo interface, as shown in \S\ref{s: characteristics -- cools}.

\section{Notes on individual galaxies}

\label{s: appendix-individual}

\begin{figure*}\includegraphics[width=\textwidth]{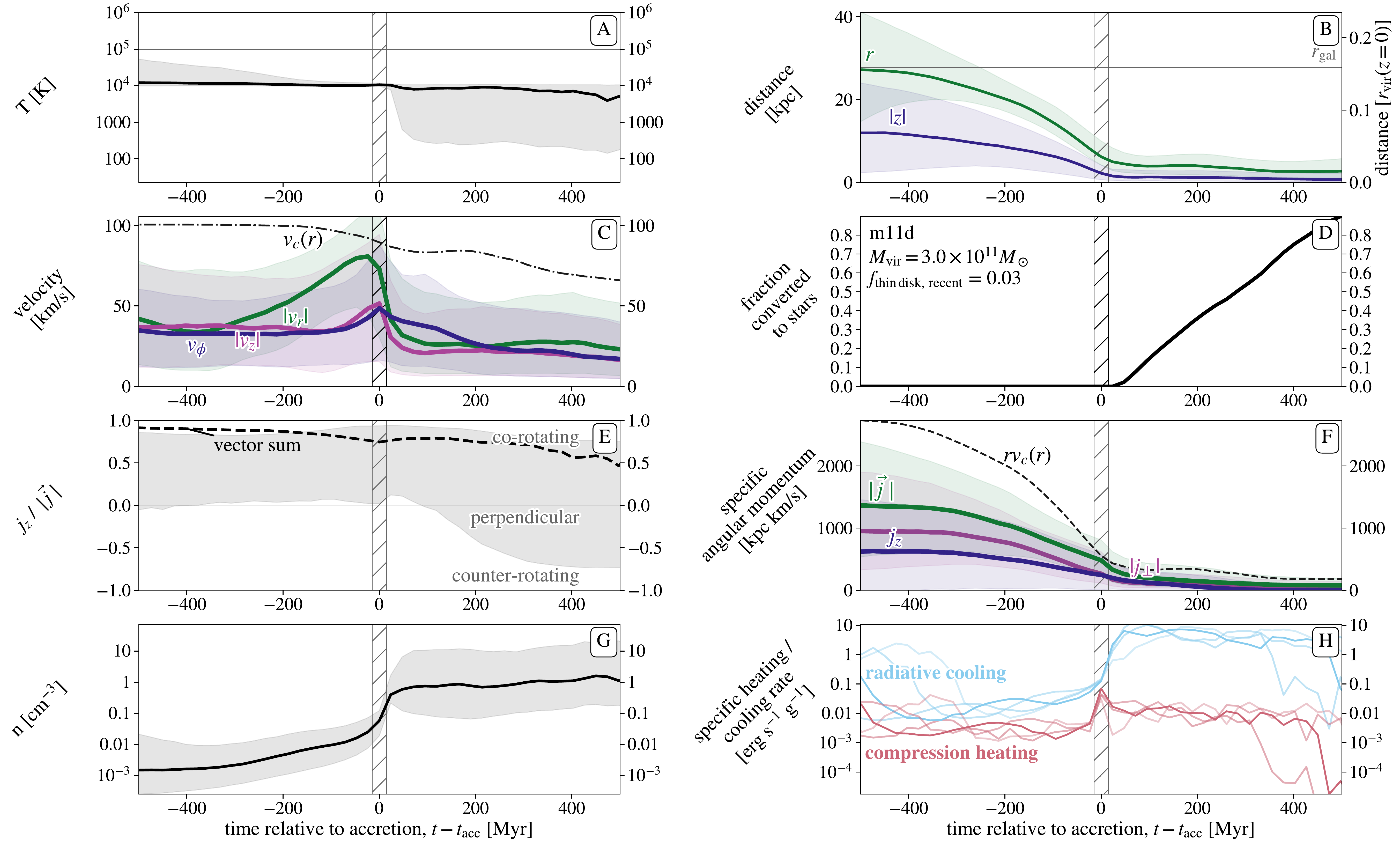}
\caption{
Temperature and geometry of gas accreting onto a $z\sim0$ galaxy in FIRE with a thin disk fraction $\fthin=0.03$ versus time relative to accretion ($t - \tacc$).
In each panel solid lines and shaded regions mark the medians and 16th to 84th percentile ranges of all particles accreted within 1 Gyr prior to $z=0$.
\textbf{A:} Temperature.
\textbf{B:} 3D distance from halo center.
\textbf{C:} Velocity components of accretion (colored lines and band), relative to circular velocity at the median radius (dash-dotted line).
\textbf{D:} Fraction of gas converted into stars.
\textbf{E:} The ratio of $j_z / \vert \vec j \vert$, the cosine of the angle between the accreting gas and the galaxy angular momentum.
The dashed line shows this ratio for the total angular momentum of all accreted particles.
\textbf{F:} The magnitude of the specific angular momentum of particles ($\vert\vec{j}\vert$, green), the component of angular momentum aligned with the galaxy disk ($j_z$; purple), and the perpendicular component ($j_{\perp} = \vert \vec{j} - j_z \hat{z} \vert$; pink).
The dashed line shows the angular momentum necessary for rotational support.
\textbf{G:} Baryon number density.
\textbf{H:} Energy loss from radiative cooling (blue) and heating from $PdV$ work on the gas particles (red).
}
\label{f: m11d}
\end{figure*}

\begin{figure*}\includegraphics[width=\textwidth]{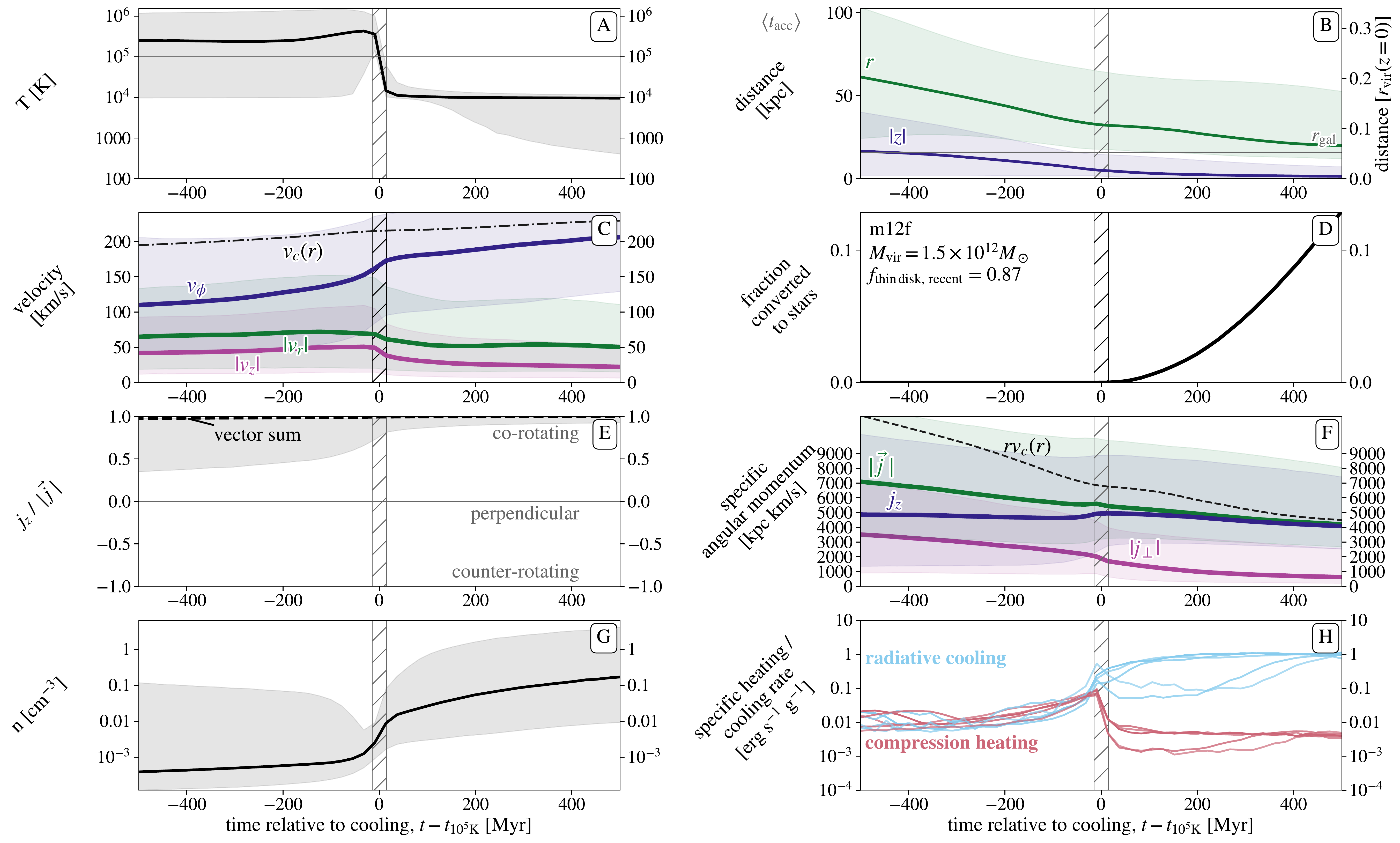}
\caption{
Same as Figure~\ref{f: m11d} but for accretion onto a $z\sim0$ galaxy in FIRE with a thin disk fraction $\fthin=0.87$ and versus time relative to the final cooling time ($t - \tcools$).
This galaxy (\texttt{m12f}) has qualitatively similar properties to other thin disk galaxies relative to $t-\tcools$, as seen via comparison with Figs~\ref{f: before and after A},~\ref{f: before and after B}, and~\ref{f: before and after combined}.
}
\label{f: m12f-tcools}
\end{figure*}

\begin{figure*}\includegraphics[width=\textwidth]{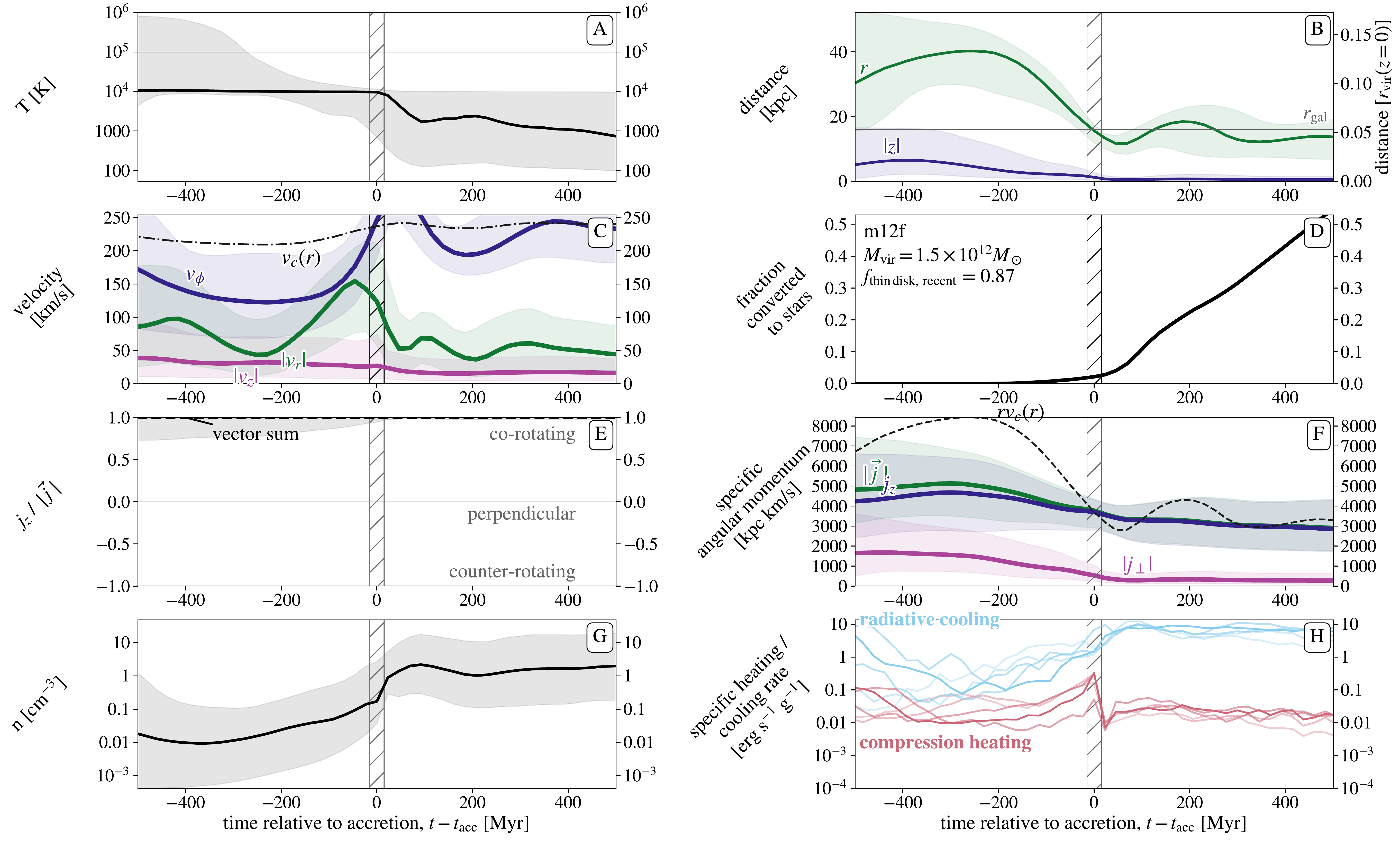}
\caption{
Same as Fig.~\ref{f: m12f-tcools}, but with time relative to $\tacc$ in the horizontal axis. 
In this galaxy the properties of accreting gas relative to $\tacc$ are qualitatively different from those relative to $\tcools$ shown in Fig.~\ref{f: m12f-tcools}, likely due to the relatively high specific angular momentum of the accreting gas (see text). 
}
\label{f: m12f-tacc}
\end{figure*}

We show the equivalent of Figs~\ref{f: before and after A} and~\ref{f: before and after B} for a representative irregular galaxy, \texttt{m11d}.
Panel E shows that prior to $\tacc$ the accreting gas has a very broad and unchanging $j_z/\vert \vec j \vert$ distribution, in contrast with the narrowing angular momentum distribution of thin disk galaxies shown in the main text, and despite the total angular momentum being mostly aligned with the galaxy $(\Sigma \vec{j})_z / \vert \vec \Sigma\vec{j} \vert \sim 0.8-0.9$.
The low thin disk fraction of $0.03$ in this galaxy is thus  consistent with angular momentum coherence in accreted gas being necessary for thin disk formation. Other properties of the accretion onto m11d are apparent also in the sample average onto irregular galaxies (Fig.~\ref{f: before and after combined}) and are discussed there.

Figures~\ref{f: m12f-tcools} and~\ref{f: m12f-tacc} show properties of accretion onto \texttt{m12f} (which was excluded from the averages in Fig.~\ref{f: before and after combined}) relative to $\tcools$ and $\tacc$ respectively.
Figure~\ref{f: m12f-tcools} shows that relative to $\tcools$ the accreting gas has the same key characteristics as accretion onto other thin disk galaxies --- inflow is hot in the CGM, and cooling is simultaneous with flattening, occuring when angular momentum support becomes significant.
Also, panels E and H show that angular momentum coherence increases prior to cooling, while radiative cooling in the hot gas is offset by compression heating.
By comparison, Fig.~\ref{f: m12f-tacc} shows that relative to $\tacc$ many of the properties of \texttt{m12f} are qualitatively different from those relative to $\tcools$, in contrast with other thin disk galaxies.
We suspect this difference is due to the relatively large specific angular momentum of accreted gas, which implies that the radius at which the accretion circularizes and cools is substantially larger than the galaxy radius, $\Rcirc \approx 2 r_{\rm gal}$ (see panel B in Fig.~\ref{f: m12f-tcools}).
As a result, gas reaches $r_{\rm gal}$ from within the disk, and quantities measured versus $\tacc$ track the evolution of gas accretion within the disk, rather than the evolution of accretion in the CGM. 

We also briefly discuss noticeable features of accretion onto additional specific galaxies.
The haloes with intermediate thin disk fractions $(0.1 \lesssim \fthin < 0.6)$ also have intermediate levels of change in accretion alignment upon cooling (Fig.~\ref{f: prevalence}). 
Four of these six haloes, \texttt{m12z}, \texttt{m12r}, \texttt{m12w}, and \texttt{m12m}, have $M_{\rm vir} \sim 10^{12} M_\odot$.
\texttt{m12z} has a chaotic halo with a number of ongoing mergers.
\texttt{m12r} has some ongoing thin disk formation, but is undergoing a major merger during the last Gyr that dominates the accreting gas supply and disrupts the galaxy structure.
This merger is relatively well-aligned with the disk, with $f_{\rm aligned} \approx 0.3-0.4$, and the aligned mass fraction does not change significantly as the gas cools past $\tcools$.
\texttt{m12w} is a galaxy where its inner CGM is only just virializing at $z=0$~\citep{Yu2021}, and the majority of the gas accretes with angular momentum perpendicular to the galaxy angular momentum.
Of these four the one with the highest thin disk fraction, \texttt{m12m}, 
has a prominent stellar bar.
\texttt{m12m} may be evidence that if rotating cooling flows are a condition for disk formation, they are a necessary-but-not-sufficient condition (\S\ref{s: why CFs thin disks}).
\texttt{m12m} is one of the few MW-mass halo simulations that does not include metal diffusion, which may also play a role in its morphology.
\bsp	
\label{lastpage}
\end{document}